\documentclass[english,aps,preprintnumbers,amsmath,amssymb,nofootinbib,twocolumn,10pt,prd]{revtex4-1} 

\usepackage[utf8]{inputenc}
\usepackage[T1]{fontenc}

\usepackage{ifthen}
\usepackage{forloop}
\usepackage{babel}
\usepackage{graphicx}
\usepackage{psfrag}
\usepackage{color}
\usepackage{dsfont}
\usepackage{mathrsfs}
\usepackage{mathtools} % allows positioning option inside matrix
\usepackage{slashed}

\usepackage{mathbbol}

\usepackage{ulem}

\usepackage{color}
\newcommand{\cen}{\operatorname{Cen}}
\renewcommand{\Re}{\operatorname{Re}}
\renewcommand{\Im}{\operatorname{Im}}
\newcommand{\metric}{\ensuremath{g}}
\newcommand{\flucmet}{\ensuremath{h}}
\newcommand{\spinmetric}{\ensuremath{h}}
\newcommand{\vielbein}{\ensuremath{e}}
\newcommand{\deltaA}{\ensuremath{{\delta_{\mathrm{A}}}}}
\newcommand{\deltaS}{\ensuremath{{\delta_{\mathrm{S}}}}}
\newcommand{\DLC}{\ensuremath{{D_{(\mathrm{LC})}}}}

\newcommand{\minsb}{\ensuremath{\mathrm{SB}_{\text{min}}}}
\newcommand{\psibar}{\ensuremath{\bar{\psi}}}
\newcommand{\flatgamma}{{\ensuremath{\gamma_{\ttm{(\text{f})}}}}}

\newcommand{\christoffel}[1]{\text{\footnotesize$\left\{ \begin{matrix} #1 \end{matrix} \right\}$}}
\newcounter{pointcounter}
\newcommand{\point}[1][\empty]{
  \ifthenelse
    {\equal{#1}{\empty}}
    {\ensuremath{\, \phantom{\cdot} \,}}
    {\setcounter{pointcounter}{1} \forloop{pointcounter}{0}{\value{pointcounter} < #1}{\ensuremath{\, \phantom{\cdot} \,}}}
}
\newcommand{\Eqref}[1]{Eq.~\eqref{#1}}
\newcommand{\ttm}[1]{\ensuremath{\text{\tiny{$#1$}}}}
\newcommand{\euler}{\mathrm{e}}
\newcommand{\cplx}{\mri}
\newcommand{\tr}{\operatorname{tr}}

\newcommand{\abs}[1]{\ensuremath{\left\vert#1\right\vert}}
\newcommand{\regint}[1]{\int \!\!\! {}_{{}_{{}_{{}_{{}_\text{\small{\ensuremath{#1}}}}}}}}
\newcommand{\mcA}{\ensuremath{\mathcal{A}}}
\newcommand{\mcB}{\ensuremath{\mathcal{B}}}
\newcommand{\mcC}{\ensuremath{\mathcal{C}}}
\newcommand{\mcD}{\ensuremath{\mathcal{D}}}

\newcommand{\mcS}{\ensuremath{\mathcal{S}}}

\newcommand{\mrd}{\ensuremath{\mathrm{d}}}

\newcommand{\mri}{\ensuremath{\mathrm{i}}}
\newcommand{\mrI}{\ensuremath{\mathrm{I}}}

\newcommand{\mrT}{\ensuremath{\mathrm{T}}}
\newcommand{\mfrL}{\ensuremath{\mathfrak{L}}}

\newcommand{\mfrS}{\ensuremath{\mathfrak{S}}}
\newcommand{\C}{\ensuremath{\mathds{C}}}
\newcommand{\N}{\ensuremath{\mathds{N}}}
\newcommand{\R}{\ensuremath{\mathds{R}}}
\newcommand{\Z}{\ensuremath{\mathds{Z}}}
\newcommand{\rmi}{\ensuremath{(\mathrm{i})}}
\newcommand{\rmii}{\ensuremath{(\mathrm{ii})}}
\newcommand{\rmiii}{\ensuremath{(\mathrm{iii})}}
\newcommand{\rmiv}{\ensuremath{(\mathrm{iv})}}

\makeatother

\begin{document}

\title{Spin-base invariance of Fermions in arbitrary dimensions %\\
%***\\
%Metric-based description of fermions in curved spacetime
}
%\date{\today}   
\author{Stefan Lippoldt}
\affiliation{Theoretisch-Physikalisches Institut, Friedrich-Schiller-Universit\"at Jena, 
Max-Wien-Platz 1, D-07743 Jena, Germany}                                        

\begin{abstract}

The concept of spin-base invariance is extended to arbitrary integer dimension $d \geq 2$.
Explicit formulas for the spin connection as a function of the Dirac matrices are found.
We disclose the hidden spin-base invariance of the vielbein formalism and give a detailed motivation for this symmetry from first principles.
The common Lorentz symmetric gauge for the vielbein is constructed for the Dirac matrices, even for metrics which are not linearly connected. Under certain criteria, it constitutes the simplest possible gauge, demonstrating why this gauge is so useful.

\end{abstract}

\maketitle

%-----------------------------------------------------------------------------
\section{Introduction}\label{sec:intro}
%-----------------------------------------------------------------------------

Gravitation is unique in many ways.
Especially, the quantization of the general theory of relativity seems to be rather different from the quantization of the standard model of particles and forces \cite{Ashtekar:2014kba}.
Currently, there is no consensus about the ultraviolet completion of gravity on the theoretical side.
Of course there are many approaches \cite{Ashtekar:2004eh,Weinberg:1980gg,Reuter:1996cp,Harst:2012ni,Plebanski:1977zz,Perez:2004hj,Bombelli:1987aa,Ambjorn:1998xu,Horava:2009uw,Ambjorn:2010hu} using very different fundamental degrees of freedom.
Whatever the correct ultraviolet description of gravity may be, it has to accommodate the other interactions and matter degrees of freedom -- in particular fermions \cite{Percacci:2002ie,Percacci:2003jz,Eichhorn:2011pc,Dona:2013qba,Dona:2014pla}.
The two most common variables for the gravitational field are the metric $\metric_{\mu \nu}$ and the vielbein $\vielbein_{\mu}^{\point a}$ related by
\begin{align}
 \metric_{\mu \nu} = \vielbein_{\mu}^{\point a} \vielbein_{\nu}^{\point b} \eta_{a b} \text{.}
\end{align}
In order to decide which of these classically (infrared) equivalent parametrizations is realized in nature, we need to make predictions for their quantum (ultraviolet) behavior and compare these to experiments.
Unfortunately so far we have only experimental access to the non-quantum regime of gravity.
Hence, even if we had a complete theory of gravity, we were not able to decide whether classically equivalent theories are describing our world properly also for high energies or not.

In this work, we want to review a common line of reasoning that suggests that the mere existence of fermions should give preference to vielbein based theories of gravity.
It goes as follows:
As gravity is encoded in the spacetime curvature and matter is fermionic we need to describe fermions in curved spacetimes.
According to textbook knowledge the coupling of fermions to curved spacetimes makes the introduction of a vielbein necessary \cite{Weyl:1929,Fock:1929,DeWitt:1965jb,Buchbinder:1992rb}.
Since the metric can be constructed from the vielbein, it is now tempting to argue that the vielbein language is at least better adapted to the description of fermions.

Surprisingly the common practice is to first write the action in terms of the vielbein, and then reexpress the vielbein as a function of the metric with the help of some gauge-fixing condition.
While this is perfectly valid for classical dynamics, this is somewhat irritating for a quantum theory.
If the vielbein was a fundamental variable, then the path integral measure $\mcD \vielbein$ should be defined in terms of the vielbein degrees of freedom.
If so, one would have to take into account a nontrivial Jacobian coming from the variable transformation $\mcD \vielbein$ to $\mcD \metric$ (not to be confused with the Faddeev-Popov determinant from the gauge fixing).
This Jacobian is usually disregarded.
Meanwhile, there are indications that a pure vielbein quantization will have at least quantitative differences compared to the case where one reexpresses the vielbein as a function of the metric \cite{Harst:2012ni}.

In fact, it is by no means obvious that one has to introduce a vielbein in order to describe fermions in curved spacetimes at all.
In this work, we demonstrate that the introduction of a vielbein (or something similar) can be avoided completely in a very natural way.
%It turns out that the metric degrees of freedom are perfectly adapted for the description of fermions in curved spacetimes.
In the following we aim at working out the ideas from Schr\"{o}dinger \cite{Schroedinger:1932}, Bargmann \cite{Bargmann:1932}, Finster \cite{Finster:1997gn} and Weldon \cite{Weldon:2000fr} on a spin-base invariant formulation of fermions on curved space.
Especially we extend our earlier work \cite{Gies:2013noa} to arbitrary integer dimensions $d \geq 2$.%
\footnote{
The one dimensional case is structurally different from all other dimensions. This is mainly because in irreducible representation the Dirac matrices do not satisfy $\tr \gamma_{\mu} = 0$.
}
Even though the spin-base invariant formalism has quite some advantages compared to its vielbein counterpart, it is rarely used in the literature \cite{Kofink:1949,Brill:1957fx,Unruh:1974bw,Finster:1998ws,Casals:2012es,Gies:2013dca,Christiansen:2015}.

Let us start our considerations with the following observation concerning the Dirac matrices $\gamma_{\mu}$.
The Clifford algebra
\begin{align}
 \{ \gamma_{\mu} , \gamma_{\nu} \} = 2 \metric_{\mu \nu} \mrI
\end{align}
by construction is present in any description of Dirac fermions.
%We restrict ourselves to irreducible representation $d_{\gamma} = 2^{\lfloor d/2 \rfloor}$.
We treat this relation as a fundamental equation, valid in the classical as well as the quantum regime.
%
%\colS{We argue that the Clifford algebra and accordingly spinors enjoy an $\mathrm{SL}(d_{\gamma}, \C)$ symmetry (spin-base invariance) and}
%
This suggests to construct everything we need for the description of fermions in curved spacetimes in terms of the Dirac matrices $\gamma_{\mu}$.
Since the metric is also determined by the $\gamma_{\mu}$ it is tempting to use them as the fundamental variables of gravity.
If we now aim at a functional integral over the Dirac matrices the metric arises naturally as the only relevant degree of freedom \cite{Gies:2013noa}.
To see this we have to keep in mind, that we cannot integrate over arbitrary $\gamma_{\mu}$, but they have to satisfy a Clifford algebra at every spacetime point.
The most general infinitesimal variation $\delta \gamma_{\mu}$ of the Dirac matrices (one integration step within a path integral) can be decomposed as
\begin{align}\label{eq:Weldon_Theorem_main}
 \delta \gamma_{\mu} = \frac{1}{2} (\delta \metric_{\mu \nu}) \gamma^{\nu} + [ \delta \mcS_{\gamma} , \gamma_{\mu} ] , \quad \tr \delta \mcS_{\gamma} = 0 \text{.}
\end{align}
In $d=4$, this has been shown by Weldon \cite{Weldon:2000fr}.
A general proof for arbitrary integer $d \geq 2$ is given in App. \ref{App:Weldon_theorem}.
Here $\delta \metric_{\mu \nu}$ corresponds to a metric fluctuation and $\delta \mcS_{\gamma}$ to a spin-base fluctuation.%
\footnote{
A spin-base fluctuation $\delta \mcS_{\gamma}$ corresponds to an element of the Lie algebra $\mathfrak{sl}(d_{\gamma} , \C)$ of the group of spin-base transformations $\mathrm{SL}(d_{\gamma} , \C)$.
}
Note, that this is a one-to-one mapping.
In other words, given an allowed variation of the Dirac matrices $\delta \gamma_{\mu}$ (compatible with the Clifford algebra), then there is a unique metric fluctuation $\delta \metric_{\mu \nu}$ and a unique spin-base fluctuation $\delta \mcS_{\gamma}$, satisfying \Eqref{eq:Weldon_Theorem_main}.
On the other hand for an arbitrary metric fluctuation $\delta \metric_{\mu \nu}$ and an arbitrary spin-base fluctuation $\delta \mcS_{\gamma}$ we can calculate the corresponding Dirac matrix fluctuation from \Eqref{eq:Weldon_Theorem_main}.
Hence, we can give the restricted integral over Dirac matrices compatible with the Clifford algebra a meaning by an unrestricted integral over metrics and spin-bases.
As we will argue in favor of spin-base invariance, the integration over spin-bases turns out to be just a trivial normalization constant for the path integral \cite{Gies:2013noa}, leaving us with a pure metric quantization.
We stress that it is more complicated and inconvenient to integrate over Dirac matrices in terms of vielbeins.
This is mainly because the vielbein alone does not cover all possible Dirac matrices.
Hence we need some additional quantity to integrate over.
It turns out that this additional quantity does not form a group.
Whereas for the metric decomposition this additional quantity is the integration over the spin-base transformations, and hence forms a group.
Details are found in App. \ref{App:impossible_vielbein}.

In this way the above-mentioned common treatment of the vielbein as a function of the metric (without keeping the Jacobian) becomes fully justified.
In order to perform the path integral we can choose a gauge for the spin base, and reexpress the Dirac matrices as a function of the metric.
This procedure leads to the exact same results as one finds for the reexpression of the vielbein as a function of the metric, without keeping the aforementioned Jacobian.

Since the vielbein is not needed at any step in our analysis we do not recapitulate the well known formulas of the vielbein formalism.
They can be found e.g.~in \cite{Weyl:1929,Fock:1929,DeWitt:1965jb,Buchbinder:1992rb,Watanabe:2004nt}.
But we encourage the reader to compare all the results with the standard vielbein formalism to find that we are covering the vielbein formalism completely.
Additionally we will comment on the relation to the vielbein formalism at the appropriate points.
% 
% All the matter we know about is composed of fermions.
% Therefore the description of matter is the description of fermions.
% Since the spacetime we are living in is curved we need to describe fermions in curved spacetimes.
% If we also accept that our world is quantum we also have to think about the respective fundamental degrees of freedom.
% For the matter sector these degrees are usually captured by Dirac spinors.
% The electro-weak and the strong interaction are mediated via gauge fields.
% But the big question is what the fundamental degrees of freedom for gravity are.
% Unfortunately there are barely any experimental guidelines whereas there are many possible candidates from the theoretical side.
% The two most prominent variables for gravity are the metric $\metric_{\mu \nu}$ and the tetrad $\vielbein_{\mu}^{\point a}$.
% On the classical level these degrees of freedom lead together with the Einstein-Hilbert action to the Einstein field equations.
% Once these theories are quantized they are expected to differ at least quantitatively \cite{Harst:2012ni}.
% One can also include torsion into the connection and treat these as additional variables which discloses us an even bigger variety of possible theories of gravity.

The paper is organized as follows.
In Sect.~\ref{sec:How_spinors_transform}, we give a detailed motivation of spin-base invariance.
Particularly we find a hidden spin-base invariance within the vielbein formalism.
%Our approach results in a more general description of Dirac fermions in curved spacetimes compared to the usual vielbein formalism.
In order to be as comprehensible as possible we summarize our mathematical assumptions in Sect.~\ref{sec:general_requirements}.
Sect.~\ref{sec:spin_connection} is devoted to the analysis of the spin metric and spin connection properties.
The constraints of the spin torsion and an action suggested from the field strength are discussed in Sect.~\ref{sec:spin_torsion_and_the_gauge_field}.
We construct the simplest choice of Dirac matrices for a given set of background metric and full metric in Sect.~\ref{sec:lorentz_symmetric_gauge}.
It turns out that this is exactly the Dirac matrix analog of the well known Lorentz symmetric gauge for the vielbein.
Conclusions are drawn in Sect.~\ref{sec:conc}.
We prove the Weldon theorem for arbitrary integer dimensions $d \geq 2$ in App.~\ref{App:Weldon_theorem}.
In App.~\ref{App:impossible_vielbein} we show that it is complicated and inconvenient to integrate over Dirac matrices in terms of vielbeins.
App.~\ref{App:minimal_spin_base_group} is devoted to the construction of the minimal group ensuring full spin-base invariance.
Some important identities for the Dirac matrices are derived in App.~\ref{App:Special_relations_gamma_matrices} and \ref{App:Special_relations_gamma_matrices_II}.
App.~\ref{App:Existence_of_the_spin_connection} shows the existence and uniqueness of the canonical part of the spin connection.
The existence and uniqueness (up to a sign) of the spin metric is shown in App.~\ref{App:spin_metric}.

%-----------------------------------------------------------------------------
\section{How spinors transform under coordinate transformations}
\label{sec:How_spinors_transform}
%-----------------------------------------------------------------------------

In this section we give a motivation for the spin base invariant formalism.
We aim at describing fermions in a curved spacetime with $1$ timelike and $d-1$ spacelike directions. In our conventions the signature of the metric reads $(-,+,\ldots,+)$. The Dirac structure is introduced with the Clifford algebra
\begin{align}
 \{ \gamma_{\mu} , \gamma_{\nu} \} = 2 \metric_{\mu \nu} \mrI \text{.}
\end{align}
Here $\metric_{\mu \nu}$ is the spacetime metric with the greek spacetime indices $\mu, \nu, \ldots$ running from $0$ to $d-1$.
The $\gamma_{\mu}$ are complex $d_{\gamma} \times d_{\gamma}$ matrices, where $d_{\gamma} = 2^{\lfloor d/2 \rfloor}$ and $\mrI$ is the unit matrix.
It is important to note that the Clifford algebra enjoys an invariance with respect to similarity transformations $\gamma_{\mu} \to \mcS \gamma_{\mu} \mcS^{-1}$, where $\mcS \in \mathrm{SL}(d_{\gamma} , \C)$ \cite{Pauli:1936,Cornwell:1989bx}.

Fermions are then represented as vectors $\psi$ in Dirac space with $d_{\gamma}$ components. The corresponding dual vectors $\psibar$ are denoted with a bar.
The dual vector $\psibar$ is related to the vector $\psi$ via the spin metric $\spinmetric$
\begin{align}
 \psibar = \psi^{\dagger} \spinmetric \text{.}
\end{align}
We will give a precise definition of the spin metric later.
For the moment it suffices to know that we need a spin metric in order to define a product between two fermionic fields $\psi$ and $\chi$ which results in a scalar with respect to coordinate transformations
\begin{align}\label{eq:spin_scalar}
 \psibar \chi = \psi^{\dagger} \spinmetric \chi \text{.}
\end{align}
Additionally we require this spin metric to not introduce any scale and therefore demand
\begin{align}\label{eq:spin_metric_det_1}
 \abs{\det \spinmetric} = 1 \text{.}
\end{align}
As is well known, in flat spacetimes we can choose Cartesian coordinates and the Dirac matrices in Dirac representation \cite{Cornwell:1989bx}.
There the spin metric turns out to be $\spinmetric = \gamma^{0}$.
In other representations of the Dirac matrices the spin metric is in general not equal to $\gamma^{0}$.
We will see how this comes about later on.

First we have to understand what fermions are.
From the view point of a theoretical physicist this means that we need to know how they transform under which symmetry group.
Since we deal with curved spacetimes we have to know how the fermionic fields behave under coordinate transformations.
To this end one usually looks at $\psibar \gamma^{\mu} \psi$ and demands that this object transforms like a usual contravariant spacetime vector since the complete Dirac structure is eliminated
\begin{align}\label{eq:lorentz_transformation_current}
 \psibar \gamma^{\mu} \psi \rightarrow \frac{\partial x'^{\mu}}{\partial x^{\rho}} \psibar \gamma^{\rho} \psi = \psibar \frac{\partial x'^{\mu}}{\partial x^{\rho}} \gamma^{\rho} \psi \text{.}
\end{align}
In flat spacetimes where one usually restricts oneself to Lorentz transformations $\Lambda^{a}_{\point b}$ as coordinate transformations we are used to a nice property of the flat Dirac matrices $\flatgamma^{a}$, namely
\begin{align}\label{eq:lorentz_transformation_flatgamma}
 \mcS_{\text{Lor}} \flatgamma^{a} \mcS_{\text{Lor}}^{-1} = {\Lambda_{\mcS}}^{a}_{\point b} \flatgamma^{b} \text{,}
\end{align}
where $\mcS_{\text{Lor}} \in \mathrm{Spin}(d - 1, 1)$ and ${\Lambda_{\mcS}}^{a}_{\point b} \in \mathrm{SO}(d - 1 , 1)$ is the corresponding Lorentz transformation \cite{Cornwell:1989bx}.
And therefore we can write
\begin{align}
 \psibar \flatgamma^{a} \psi \rightarrow \psibar' \flatgamma^{a} \psi' \stackrel{!}{=} {\Lambda_{\mcS}}^{a}_{\point b} \psibar \flatgamma^{b} \psi = \psibar \mcS_{\text{Lor}} \flatgamma^{a} \mcS_{\text{Lor}}^{-1} \psi \text{,}
\end{align}
which suggests that spinors transform under Lorentz transformations according to
\begin{align}\label{eq:lorentz_transformation_psi}
 \psi \rightarrow \psi' = \mcS_{\text{Lor}}^{-1} \psi, \quad \psibar \rightarrow \psibar' = \psibar \mcS_{\text{Lor}} \text{.}
\end{align}
But this is rather a group theoretical accident for Lorentz transformations than a rule for general coordinate transformations which enjoy no such relation.
One simple counterexample is the stretching of one of the axis, $x^{3} \rightarrow x'^{3} = \frac{1}{\alpha} x^{3}$.
Then the Minkowski metric in $d=4$ spacetime dimensions changes to
\begin{align}
 (\eta_{a b}) = \operatorname{diag}(-1, 1, 1, 1) \rightarrow (\eta_{a b}') = \operatorname{diag}( -1, 1, 1, \alpha^2 )
\end{align}
and therefore the transformed Dirac matrix ${\gamma_{(\text{f})}'}^{3}$ would have to square to $\alpha^2 \mrI$.
But this cannot be achieved via a similarity transformation, since $\big( \mcS {\gamma_{(\text{f})}}^{3} \mcS^{-1} \big)^{2} = \mrI$ for all $\mcS \in \mathrm{SL}(d_{\gamma} , \C)$.
This example illustrates why it is in general not possible to pass on a coordinate transformation to a similarity transformation and why we should start rethinking. We will give an intuitive introduction to spin-base invariance in the following.
%There is even another hint that something strange is going on. If we look closer at \Eqref{eq:lorentz_transformation_flatgamma}, we find that $\mcS_{\text{Lor}}(\Lambda)$ is by far not unique.
%Especially we can always take $\mcS_{\text{Lor}}$ with the opposite sign $- \mcS_{\text{Lor}}$.
%This possibility of the sign change makes the prescription for the behavior of the fermion fields under a given Lorentz transformation ambigous.
%We will resolve this ambiguity in the following.

If we perform a general coordinate transformation we have to transform the metric, unlike for Lorentz transformations, in a non trivial way
\begin{align}
 \metric_{\mu \nu} \rightarrow \metric'_{\mu \nu} = \frac{\partial x^{\rho}}{\partial x'^{\mu}} \frac{\partial x^{\lambda}}{\partial x'^{\nu}} \metric_{\rho \lambda} \text{.}
\end{align}
Therefore we also have to transform the Dirac matrices non trivially $\gamma_{\mu} \to \gamma'_{\mu}$.
%Unfortunately we do not know of what form the transformation law is.
Taking the Clifford algebra as a guideline, we find
\begin{align}
 \{ \gamma'_{\mu} , \gamma'_{\nu} \} = 2 \metric'_{\mu \nu} \mrI = 2 \frac{\partial x^{\rho}}{\partial x'^{\mu}} \frac{\partial x^{\lambda}}{\partial x'^{\nu}} \metric_{\rho \lambda} \mrI = \! \left\{ \! \frac{\partial x^{\rho}}{\partial x'^{\mu}} \gamma_{\rho} , \frac{\partial x^{\lambda}}{\partial x'^{\nu}} \gamma_{\lambda} \! \right\} \text{.}
\end{align}
This equation implies that
\begin{align}\label{eq:coord_transf_gamma}
 \gamma_{\mu} \rightarrow \gamma'_{\mu} = \left\{
 \begin{matrix*}[l]
  \frac{\partial x^{\rho}}{\partial x'^{\mu}} \mcS \gamma_{\rho} \mcS^{-1} \vphantom{A^{A^{A}}} & , d \text{ even}\\
  \pm \frac{\partial x^{\rho}}{\partial x'^{\mu}} \mcS \gamma_{\rho} \mcS^{-1} \vphantom{A^{A^{A}}} & , d \text{ odd}
 \end{matrix*}
 \right.
\end{align}
where $\mcS \in \mathrm{SL}( d_{\gamma}, \C)$ is arbitrary. %
% \footnote{
% $\mathrm{SL}( d_{\gamma} , \C )$ as a set in $\mathrm{Mat}(d_{\gamma} \times d_{\gamma} , \C)$ is the unique set $\setD \subset \mathrm{Mat}(d_{\gamma} \times d_{\gamma} , \C)$ which satisfies the following three conditions: $\rmi$ Every set of Dirac matrices compatible with a given metric is connected via \Eqref{eq:coord_transf_gamma}. $\rmii$ The set $\setD$ forms a group with the usual matrix multiplication as the group product. $\rmiii$ It is the set $\setD$ (compatible with $\rmi$ and $\rmii$) which minimizes the set $\{ \mcS \in \setD : \mcS \gamma_{\mu} \mcS^{-1} = \gamma_{\mu} \}$.
% }
The proof of this relation uses that every irreducible representation of the Clifford algebra for a given metric is connected to each other via a similarity transformation and in odd dimensions if necessary via an additional sign change since there are two connected components \cite{Pauli:1936,Cornwell:1989bx}.
This sign flip has to be global if we want the Dirac matrices to be differentiable.

We can rephrase our finding \Eqref{eq:coord_transf_gamma} by saying that a coordinate transformation for Dirac matrices is a combination of the usual transformation of the vector part $\frac{\partial x^{\rho}}{\partial x'^{\mu}}$, a similarity transformation $\mcS \in \mathrm{SL} ( d_{\gamma}, \C)$ and if necessary a sign flip.
But since we still have a solution to the Clifford algebra if we perform a similarity transformation or a sign flip on the Dirac matrices we should distinguish two kinds of coordinate transformations \cite{Gies:2013noa,Gies:2015cka}.

First we have the usual spacetime coordinate transformations
\begin{align}
 \gamma_{\mu} \rightarrow \gamma'_{\mu} = \frac{\partial x^{\rho}}{\partial x'^{\mu}} \gamma_{\rho} \text{.}
\end{align}
These transformations change the spacetime coordinate bases and are called diffeomorphisms.
Second we have the similarity transformations $\mcS \in \mathrm{SL}(d_{\gamma} , \C)$ and in odd dimensions also the sign flip, which are the Dirac (or spin) coordinate transformations,
\begin{align}\label{eq:spinbase_transformation_gamma}
 \gamma_{\mu} \rightarrow \gamma'_{\mu} = \left\{
 \begin{matrix*}[l]
  \mcS \gamma_{\mu} \mcS^{-1} & , d \text{ even} \\
  \pm \mcS \gamma_{\mu} \mcS^{-1} & , d \text{ odd}
 \end{matrix*}
 \right. \text{.}
\end{align}
They change the spin bases and therefore we will call them spin-base transformations in the following.

At the moment the choice of $\mathrm{SL}(d_{\gamma} , \C)$
as the transformation group%
\footnote{
In fact we are dealing with the fundamental representation of $\mathrm{SL}(d_{\gamma} , \C)$ and not the group itself. But we will keep this terminology in the following for simplicity, as we are working with the representations of the groups exclusively throughout this paper. By fundamental representation we mean the defining matrix representation of $\mathrm{SL}(d_{\gamma} , \C)$, which is $\{ \mcS \in \mathrm{Mat}(d_{\gamma} \times d_{\gamma} , \C) : \det \mcS = 1 \}$ together with the matrix multiplication as the group law.
}
for the spin-base transformations seems a little arbitrary.
For example we could also take $\mathrm{GL}(d_{\gamma} , \C)$ or $\mathrm{SL}(d_{\gamma} , \C) / \Z_{d_{\gamma}}$.
But it turns out that $\mathrm{SL}(d_{\gamma} , \C)$ is special.
In order to formalize this choice, we have to clarify what we need from the spin-base transformations.

First of all we are dealing with different choices for a spin-base coordinate system, therefore we need a group $\minsb$ to connect these.
As the different spin bases are connected via similarity transformations, this group should be a subgroup of $\mathrm{GL}(d_{\gamma} , \C)$, with the usual matrix multiplication as the group law, $\minsb \leq \mathrm{GL}(d_{\gamma} , \C)$.
Next we have to ensure that we do not miss any spin base, i.e. every two sets $\gamma_{\mu}$ and $\gamma'_{\mu}$ compatible with the Clifford algebra for a given metric have to be connected via \Eqref{eq:spinbase_transformation_gamma} where $\mcS \in \minsb$.
And finally we want to keep $\minsb$ minimal in order not to artificially inflate the symmetry.
In other words we have to minimize the cardinality of the set $\{ \mcS \in \minsb : \mcS \gamma_{\mu} \mcS^{-1} = \gamma_{\mu} \}$.
In App. \ref{App:minimal_spin_base_group} it is shown that $\minsb = \mathrm{SL}(d_{\gamma} , \C)$ is the unique
group satisfying the preceding conditions.

A general coordinate transformation of the Dirac matrices is therefore given by an independent change of the spacetime base and the spin base.
Here independent means that we can in principle perform one of them without the other, as long as we stay on one fixed patch of the manifold.
But we have to keep in mind that there might be some topological obstructions similar to those encountered in the vielbein formalism.
There it can happen that one has to change the orthonormal frame while changing the patch on the manifold.
For the vielbein this is already true on the 2-sphere due to the Poincar\'{e}-Brouwer (hairy-ball) theorem.
The Dirac matrices on the other hand do have a global spin base on the 2-sphere, rendering the complete decoupling of spacetime coordinates and spin bases obvious.
A detailed analysis of the situation on the 2-sphere is given in \cite{Gies:2015cka}.
Whether a global spin base exists on all metrizable manifolds is unclear so far.

Now we can turn back to the question how the fermionic fields behave under spacetime coordinate transformations and spin-base transformations.
For the description of dynamics we need a kinetic fermion term.
If we additionally want to have covariance we need this term to be invariant under all types of coordinate transformations.
We assume the kinetic term to be of the form $\psibar \slashed{\nabla} \psi$ where $\slashed{\nabla} = \gamma^{\mu} \nabla_{\mu}$ is the Dirac operator with $\nabla_{\mu}$ the covariant derivative.
Again we postpone the precise definition of $\nabla_{\mu}$, but for the moment it is sufficient to know that this derivative has to have two important properties.
First if $\psi$ is a fermionic Dirac spinor, then $\nabla_{\mu} \psi$ is also a fermionic Dirac spinor, i.e.~it transforms in the same way under spin-base transformations.
And second if $\psi$ is a spacetime tensor, then $\nabla_{\mu} \psi$ is a spacetime tensor of one rank higher, i.e.~the additional spacetime index $\mu$ transforms like a covariant vector index under spacetime coordinate transformations.
At the moment we do not assume anything about the tensorial rank of $\psi$.

Since $\nabla_{\mu} \psi$ acts exactly like $\psi$ under spin-base transformations and as a tensor of one rank higher than $\psi$ under spacetime coordinate transformations, we can investigate $\psibar \gamma^{\mu} \psi$ instead of the original kinetic operator, demanding that it transforms like a scalar under spin-base transformations and as a contravariant vector under spacetime coordinate transformations.

% Note however, that we are restricting the discussion to just one family of fermions.
% We know that there are many different kinds of fermions in our universe, and in principle any of them could experience the gravitational force in a different way.
% But all experiments done so far confirmed that every known form of matter (fermions) is coupled to gravity in the exact same way.
% Therefore the restriction to one family of fermions is justified.

The discussion straightforwardly generalizes to fermions with further internal (flavor, color) symmetries.
As we are dealing with complex degrees of freedom, we expect to find a $\mathrm{U}(1)$ symmetry for $\psibar \gamma^{\mu} \psi$. 
If we dealt with $N$ families of fermions we would find a $\mathrm{U}(N)$ symmetry, similar to the gauge symmetries of the standard model of particle physics.
We are going to ignore these symmetries, as we could always regain them, by adding a gauge field respectively with an appropriate charge to the covariant derivative, cf. \cite{Gies:2013noa}.

The even and the odd dimensional case are structurally very different therefore we will discuss them separately.

%-----------------------------------------------------------------------------
\subsection{The odd dimensional case}\label{subsec:How_spinors_transform_d_odd}
%-----------------------------------------------------------------------------

First we look at the behavior under spin-base transformations $\mcS \in \mathrm{SL} (d_{\gamma} , \C)$.
To this end we remind ourselves that the Dirac matrices and their antisymmetric combinations form a complete basis in $\mathrm{Mat}(d_{\gamma} \times d_{\gamma}, \C)$, the $d_{\gamma} \times d_{\gamma}$ matrices \cite{Cornwell:1989bx}.
In the odd dimensional case we need only the antisymmetric combinations with an even number of Dirac matrices to decompose an arbitrary $M \in \mathrm{Mat}(d_{\gamma} \times d_{\gamma}, \C)$
\begin{align}\label{eq:gamma_matrices_basis}
 M = \sum\limits_{n = 0}^{\frac{d-1}{2}} m^{\mu_{1} \ldots \mu_{2n}} \gamma_{\mu_{1} \ldots \mu_{2n}}
\text{,}
\end{align}
with the ``coordinates'' $m^{\mu_{1} \ldots \mu_{2n}} \in \C$, whose indices are completely antisymmetrized.
The antisymmetric combinations of the Dirac matrices are given by
\begin{align}\label{eq:definition_antisymmetrized_gammas}
 \gamma_{\mu_{1} \ldots \mu_{n}} = \left\{
 \begin{matrix*}[l]
  \mrI &, n = 0\\
  \gamma_{[\mu_{1}} \ldots \gamma_{\mu_{n}]} &, n \geq 1
 \end{matrix*}
 \right.
 \text{,}
\end{align}
where we denote the normalized antisymmetrization with $[\ldots]$, e.g. $\gamma_{\mu \nu} = \gamma_{[\mu} \gamma_{\nu]} = \frac{1}{2} [\gamma_{\mu} , \gamma_{\nu} ]$.
Since in odd dimensions the basis elements are the $\gamma_{\mu_{1} \ldots \mu_{2n}}$, they transform homogeneously under spin-base transformations because the possible sign flip drops out.
In App. \ref{App:Special_relations_gamma_matrices} and \ref{App:Special_relations_gamma_matrices_II} we have collected some important properties of the Dirac matrices and the basis elements.

Now we look at the behavior under spin-base transformations $\mcS \in \mathrm{SL}(d_{\gamma} , \C)$ of $\psibar \gamma_{\mu} \psi$
\begin{align}
 \psibar \gamma_{\mu} \psi \to \psibar' \gamma'_{\mu} \psi' = \pm \psibar' \mcS \gamma_{\mu} \mcS^{-1} \psi' \stackrel{!}{=} \psibar \gamma_{\mu} \psi \text{,}
\end{align}
and demand invariance.
Without loss of generality we make the ansatz
\begin{align}
 \psi' &{}= \mcS \mcB \psi,
\\
 \spinmetric' &{}= (\mcS^{\dagger})^{-1} (\mcB^{\dagger})^{-1} \spinmetric \mcC \mcS^{-1},
\end{align}
where $\mcB, \mcC \in \mathrm{GL}(d_{\gamma} , \C)$ are arbitrary invertible matrices.
Note that the invertability of $\mcB$ and $\mcC$ is mandatory because otherwise we would violate the reversability of spin-base transformations and they would not form a group. Plugging in our ansatz we get
\begin{align}
 \pm \psibar \mcC \gamma_{\mu} \mcB \psi = \psibar \gamma_{\mu} \psi \text{.}
\end{align}
Because of the independence of $\psi$ and $\psi^{\dagger}$ we conclude
\begin{align}\label{eq:mcC_and_mcB_d_odd}
 \pm \mcC \gamma_{\mu} \mcB = \gamma_{\mu} \text{.}
\end{align}
By multiplying with $\pm \frac{1}{d} \mcB^{-1} \gamma^{\mu}$ from the right, we can read off
\begin{align}
 \mcC = \pm \frac{1}{d} \gamma_{\rho} \mcB^{-1} \gamma^{\rho} \text{.}
\end{align}
Inserting this back into \Eqref{eq:mcC_and_mcB_d_odd} we get
\begin{align}\label{eq:mcB_d_odd}
 \frac{1}{d} ( \gamma_{\rho} \mcB^{-1} \gamma^{\rho} ) \gamma_{\mu} \mcB = \gamma_{\mu} \text{.}
\end{align}
If we multiply with $\frac{1}{d} \gamma^{\mu}$ from the right we infer
\begin{align}
 \left( \frac{1}{d} (\gamma_{\rho} \mcB^{-1} \gamma^{\rho}) \right)^{-1} = \frac{1}{d} \gamma_{\lambda} \mcB \gamma^{\lambda} \text{.}
\end{align}
Therefore we can rewrite \Eqref{eq:mcB_d_odd} as
\begin{align}
 \gamma_{\mu} \mcB = \frac{1}{d} (\gamma_{\lambda} \mcB \gamma^{\lambda}) \gamma_{\mu} \text{.}
\end{align}
We finally multiply with $\frac{1}{d} \gamma^{\mu}$ from the left and find
\begin{align}\label{eq:mcB_solvable_d_odd}
 \mcB = \frac{1}{d} \gamma_{\mu} \left( \frac{1}{d} (\gamma_{\lambda} \mcB \gamma^{\lambda}) \right) \gamma^{\mu} \text{.}
\end{align}
Now we use that we can write $\mcB$ as
\begin{align}
 \mcB = \sum\limits_{n=0}^{\frac{d-1}{2}} b_{\rho_{1} \ldots \rho_{2n}} \gamma^{\rho_{1} \ldots \rho_{2n}}
\end{align}
and use the identity \Eqref{eq:App:gamma_contraction_base_elements} from App. \ref{App:Special_relations_gamma_matrices} to calculate
\begin{align}
 &\frac{1}{d} (\gamma_{\lambda} \mcB \gamma^{\lambda}) = \frac{1}{d} \sum\limits_{n=0}^{\frac{d-1}{2}} (d-4n) b_{\rho_{1} \ldots \rho_{2n}} \gamma^{\rho_{1} \ldots \rho_{2n}},
\\
 &\frac{1}{d} \gamma_{\mu} \left( \frac{1}{d} (\gamma_{\lambda} \mcB \gamma^{\lambda}) \right) \gamma^{\mu} = \frac{1}{d^{2}} \sum\limits_{n=0}^{\frac{d-1}{2}} (d-4n)^{2} b_{\rho_{1} \ldots \rho_{2n}} \gamma^{\rho_{1} \ldots \rho_{2n}} \text{.}
\end{align}
Together with \Eqref{eq:mcB_solvable_d_odd} and a comparison of coefficients we conclude
\begin{align}
 b_{\rho_{1} \ldots \rho_{2n}} = \left( 1 - 4 \frac{n}{d} \right)^{2} b_{\rho_{1} \ldots \rho_{2n}} , \quad n \in \left\{ 0 , \ldots, \frac{d-1}{2} \right\} \text{.}
\end{align}
These equations imply
\begin{align}
 &\mcB = b \cdot \mrI, \quad \mcC = \pm \frac{1}{b} \cdot \mrI, \quad b \in \C \backslash \{ 0 \} \text{.}
\end{align}
Since \Eqref{eq:spin_metric_det_1} has to be a spin-base independent statement also
\begin{align}
 \abs{\det \spinmetric'} = 1
\end{align}
has to hold. Therefore $b$ is restricted to a $\mathrm{U}(1)$ phase
\begin{align}
 b = \euler^{\cplx \varphi} \in \mathrm{U}(1) \text{.}
\end{align}
Summing up, we found $\psi \to \euler^{\cplx \varphi} \mcS \psi$.
The $\mathrm{SL}(d_{\gamma} , \C)$ part is the nontrivial spin-base transformation, whereas the $\mathrm{U}(1)$ phase is the aforementioned gauge symmetry which we are going to ignore.

% Observing that $\mcB \gamma_{\mu} \mcB^{-1} = \gamma_{\mu}$ we take $\mcS$ and $\mcB$ together and define the group of spin-base transformations $\mathrm{SB}(d_{\gamma})$ as%
% \footnote{
% Note the reminiscence to the group $\mathrm{Spin}^{c}(p,q) = \{ \euler^{\cplx \varphi} \mcS : \euler^{\cplx \varphi} \in \mathrm{U}(1) , \mcS \in \mathrm{Spin}(p,q) \}$.
% }
% \begin{align}
%  \mathrm{SB}(d_{\gamma}) = \{ \euler^{\cplx \varphi} \mcS : \euler^{\cplx \varphi} \in \mathrm{U}(1) , \mcS \in \mathrm{SL}(d_{\gamma} , \C) \}
% \end{align}
% with the usual matrix multiplication as the group product. This group must not be confused with $\mathrm{U}(1) \times \mathrm{SL}(d_{\gamma} , \C)$. E.g. the two elements $(\euler^{\cplx \frac{2 \pi}{d_{\gamma}}} , \mcS )$ and $(1 , \euler^{\cplx \frac{2 \pi}{d_{\gamma}}} \mcS )$ are different objects in $\mathrm{U}(1) \times \mathrm{SL}(d_{\gamma} , \C)$ whereas $\euler^{\cplx \frac{2 \pi}{d_{\gamma}}} \cdot \mcS$ and $1 \cdot \euler^{\cplx \frac{2 \pi}{d_{\gamma}}} \mcS$ are the same objects in $\mathrm{SB}(d_{\gamma})$.%
% \footnote{
% In fact the spin base group $\mathrm{SB}(d_{\gamma})$ turns out to be $U(1) \times \big( \mathrm{SL}(d_{\gamma} , \C) / \Z_{d_{\gamma}} \big)$.
% }

The transformation law for spin-base transformations %$\mcS_{\varphi} \in \mathrm{SB}(d_{\gamma})$
$\mcS \in \mathrm{SL}(d_{\gamma} , \C)$ in odd dimensions then reads
\begin{align}
% \begin{aligned}
%  \gamma_{\mu} &\to \pm \mcS_{\varphi} \gamma_{\mu} \mcS_{\varphi}^{-1} ,
% \\
%  \psi &\to \mcS_{\varphi} \psi,
% \\
%  \psibar &\to \pm \psibar \mcS_{\varphi}^{-1},
% \\
%  \spinmetric &\to \pm (\mcS_{\varphi}^{\dagger})^{-1} \spinmetric \mcS_{\varphi}^{-1} \text{.}
% \end{aligned}
\begin{aligned}
 \gamma_{\mu} &\to \pm \mcS \gamma_{\mu} \mcS^{-1} ,
\\
 \psi &\to \mcS \psi,
\\
 \psibar &\to \pm \psibar \mcS^{-1},
\\
 \spinmetric &\to \pm (\mcS^{\dagger})^{-1} \spinmetric \mcS^{-1} \text{.}
\end{aligned}
\end{align}
Note that the $\mathrm{U}(1)$ phase actually would drop out of the transformation law of $\gamma_{\mu}$ and $\spinmetric$ confirming that this symmetry is independent of the spin-base transformations.

Next we investigate the behavior under diffeomorphisms. Again we look at $\psibar \gamma_{\mu} \psi$ and demand that it behaves like a covariant vector
\begin{align}
 \psibar \gamma_{\mu} \psi \to \psibar' \gamma'_{\mu} \psi' = \psibar' \frac{\partial x^{\rho}}{\partial x'^{\mu}} \gamma_{\rho} \psi' \stackrel{!}{=} \frac{\partial x^{\rho}}{\partial x'^{\mu}} \psibar \gamma_{\rho} \psi \text{.}
\end{align}
Now we can go through the same steps as for the spin-base transformations and we find that the fermions have to transform like scalars under spacetime coordinate transformations again with an additional arbitrary phase transformation, which we neglect.
% But since we already have taken care of the phase in the spin-base transformations this additional phase transformation is redundant and we will neglect it for the spacetime coordinate transformations.
Therefore we find the transformation law under diffeomorphisms in odd dimensions as
\begin{align}
\begin{aligned}
 \gamma_{\mu} &\to \frac{\partial x^{\rho}}{\partial x'^{\mu}} \gamma_{\rho},
\\
 \psi &\to \psi,
\\
 \psibar &\to \psibar,
\\
 \spinmetric &\to \spinmetric \text{.}
\end{aligned}
\end{align}
An important remark is in order here. Since the Clifford algebra has two connected components in odd dimensions we had to introduce the sign flip for the spin-base transformations. This sign flip spoils full spin-base invariance of a mass term $\psibar \psi$, since this sign flip does not drop out as for $\psibar \gamma_{\mu} \psi$. This implies that $\psibar \psi$ transforms as a scalar under the continous part $\mcS_{\varphi}$, but as a pseudo scalar under the discrete sign flip.

%-----------------------------------------------------------------------------
\subsection{The even dimensional case}\label{subsec:How_spinors_transform_d_even}
%-----------------------------------------------------------------------------

To find the transformation behavior in even dimensions we proceed in a similar way as for the odd dimensional case.
First we introduce the complete basis \cite{Cornwell:1989bx} in $\mathrm{Mat}(d_{\gamma} \times d_{\gamma} , \C)$ in terms of the $\gamma_{\mu_{1} \ldots \mu_{n}}$. Such that we can rewrite an arbitrary $M \in \mathrm{Mat}(d_{\gamma} \times d_{\gamma} , \C)$ as
\begin{align}\label{eq:gamma_matrices_basis_even}
 M = \sum\limits_{n=0}^{d} m^{\mu_{1} \ldots \mu_{n}} \gamma_{\mu_{1} \ldots \mu_{n}},
\end{align}
where the $m^{\mu_{1} \ldots \mu_{n}}$ are the ``coordinates'' with respect to this basis, whose indices are completely antisymmetrized.

Additionally we introduce the matrix $\gamma_{\ast}$ defined in even dimensions as
\begin{align}
 \gamma_{\ast} \! = \frac{\cplx (- \cplx)^{d/2}}{d!} \tilde{\varepsilon}_{\mu_{1} \ldots \mu_{d}} \gamma^{\mu_{1}} \! \ldots \! \gamma^{\mu_{d}} \equiv \frac{\cplx (- \cplx)^{d/2}}{d!} \tilde{\varepsilon}_{\mu_{1} \ldots \mu_{d}} \gamma^{\mu_{1} \ldots \mu_{d}}.
\end{align}
Here $\tilde{\varepsilon}_{\mu_{1} \ldots \mu_{d}}$ denotes the totally antisymmetric Levi-Civita tensor $\tilde{\varepsilon}_{\mu_{1} \ldots \mu_{d}} = \sqrt{- \metric} \varepsilon_{\mu_{1} \ldots \mu_{d}}$ and $\varepsilon_{\mu_{1} \ldots \mu_{d}}$ is the totally antisymmetric Levi-Civita symbol $\varepsilon_{0 \ldots d-1} = 1$. The most important properties of $\gamma_{\ast}$ are
\begin{align}
 &\{ \gamma_{\ast} , \gamma_{\mu} \} = 0, \quad \tr \gamma_{\ast} = 0, \quad \gamma_{\ast}^{2} = \mrI \text{.}
\end{align}
Again we start with the spin-base transformations and analyze the behavior of $\psibar \gamma_{\mu} \psi$
\begin{align}
 \psibar \gamma_{\mu} \psi \to \psibar' \gamma_{\mu}' \psi' = \psibar' \mcS \gamma_{\mu} \mcS^{-1} \psi' \stackrel{!}{=} \psibar \gamma_{\mu} \psi
\end{align}
demanding that it behaves like a scalar. We employ again the general ansatz
\begin{align}
 &\psi' = \mcS \mcB \psi,
\\
 &\spinmetric' = (\mcS^{\dagger})^{-1} (\mcB^{\dagger})^{-1} \spinmetric  \mcC \mcS^{-1},
\end{align}
with $\mcB, \mcC \in \mathrm{GL}(d_{\gamma} , \C)$ arbitrary. Following the same route as before we find
\begin{align}
 \mcC \gamma_{\mu} \mcB = \gamma_{\mu}
\end{align}
and from there with similar manipulations
\begin{align}
 \mcC &= \frac{1}{d} \gamma_{\rho} \mcB^{-1} \gamma^{\rho},
\\
 \mcB &= \frac{1}{d} \gamma_{\mu} \left( \frac{1}{d} ( \gamma_{\lambda} \mcB \gamma^{\lambda} ) \right) \gamma^{\mu} \text{.}
\end{align}
Here we use the convenient basis $\gamma^{\mu_{1} \ldots \mu_{n}}$ for $\mcB$
\begin{align}
 \mcB = \sum\limits_{n=0}^{d} b_{\mu_{1} \ldots \mu_{n}} \gamma^{\mu_{1} \ldots \mu_{n}}
\end{align}
and calculate with the aid of the identity \Eqref{eq:App:gamma_contraction_base_elements} from App. \ref{App:Special_relations_gamma_matrices}
\begin{align}
 &\frac{1}{d} (\gamma_{\lambda} \mcB \gamma^{\lambda}) = \frac{1}{d} \sum\limits_{n=0}^{d} (-1)^{n} (d-2n) b_{\rho_{1} \ldots \rho_{n}} \gamma^{\rho_{1} \ldots \rho_{n}},
\\
 &\frac{1}{d} \gamma_{\mu} \left( \frac{1}{d} (\gamma_{\lambda} \mcB \gamma^{\lambda}) \right) \gamma^{\mu} = \frac{1}{d^{2}} \sum\limits_{n=0}^{d} (d-2n)^{2} b_{\rho_{1} \ldots \rho_{n}} \gamma^{\rho_{1} \ldots \rho_{n}} \text{.}
\end{align}
By comparison of the coefficients we can read off
\begin{align}
 b_{\rho_{1} \ldots \rho_{n}} = \left( 1 - 2 \frac{n}{d} \right)^{2} b_{\rho_{1} \ldots \rho_{n}} , \quad n \in \{ 0, \ldots , d\} \text{.}
\end{align}
This time the general solution is
\begin{align}
 &\mcB = b_{1} \euler^{b_{2} \gamma_{\ast}} = b_{1} ( \cosh b_{2} \cdot \mrI + \sinh b_{2} \cdot \gamma_{\ast}), \notag
\\
 &\mcC = \frac{1}{b_{1}} \, \euler^{b_{2} \gamma_{\ast}}, \quad b_{1} \in \C \backslash \{ 0 \} , \quad  b_{2} \in \C \text{.}
\end{align}
Since $\det \euler^{b_{2} \gamma_{\ast}} = 1$, the implementation of \Eqref{eq:spin_metric_det_1} restricts $b_{1}$ to a $\mathrm{U}(1)$ phase
\begin{align}
 b_{1} = \euler^{\cplx \varphi} \in \mathrm{U}(1) \text{.}
\end{align}
That means by solely demanding that the kinetic term is invariant under spin-base transformations we have another degree of freedom.
We can have not only a phase transformation $\euler^{\cplx \varphi}$ but also a non trivial chiral transformation $\euler^{b_{2} \gamma_{\ast}}$.

% But unlike for the odd dimensional case, we can demand additionally that the mass term $\psibar \psi$ behaves like a scalar under all spin-base transformations since the Clifford algebra has only one connected component in even dimensions.

As usual, the chiral symmetry can be broken explicitly by a mass term $\psibar \psi$.
We demand that it transforms as a scalar under all spin-base transformations since the Clifford algebra has only one connected component in even dimensions.

If we thus also demand that
\begin{align}
 \psibar \psi \to \psibar' \psi' = \psibar \mcC \mcB \psi \stackrel{!}{=} \psibar \psi,
\end{align}
we find that
\begin{align}
 \mcC \mcB = \euler^{ 2 b_{2} \gamma_{\ast}} = \cosh( 2 b_{2} ) \cdot \mrI + \sinh( 2 b_{2} ) \cdot \gamma_{\ast} \stackrel{!}{=} \mrI \text{.}
\end{align}
Remember that $b_{2} \in \C$. This equation leads to only two solutions for $\euler^{b_{2} \gamma_{\ast}}$
\begin{align}
 \euler^{b_{2} \gamma_{\ast}} = \pm \mrI \text{.}
\end{align}
The sign ambiguity can be compensated by a phase conversion,
\begin{align}
 \mcB = \pm \euler^{\cplx \varphi} \mrI = \euler^{\cplx \varphi'} \mrI, \quad \mcC = \pm \euler^{- \cplx \varphi} \mrI = \euler^{- \cplx \varphi'} \mrI \text{,}
\end{align}
with an appropriately chosen $\euler^{\cplx \varphi'} \in \mathrm{U}(1)$. Now we can apply the same arguments as before and ignore the phase again.
%insert the phase into the matrix $\mcS$ to get $\mcS_{\varphi'} = \euler^{\cplx \varphi'} \mcS \in \mathrm{SB}(d_{\gamma})$.

Therefore we conclude that spin-base transformations %$\mcS_{\varphi} \in \mathrm{SB}(d_{\gamma})$
$\mcS \in \mathrm{SL}(d_{\gamma} , \C)$
in even dimensions act as
\begin{align}\label{eq:spinbase_transformation_d_even}
% \begin{aligned}
%  \gamma_{\mu} &\to \mcS_{\varphi} \gamma_{\mu} \mcS_{\varphi}^{-1},
% \\
%  \psi &\to \mcS_{\varphi} \psi,
% \\
%  \psibar &\to \psibar \mcS_{\varphi}^{-1},
% \\
%  \spinmetric &\to (\mcS_{\varphi}^{\dagger})^{-1} \spinmetric \mcS_{\varphi}^{-1} \text{.}
% \end{aligned}
\begin{aligned}
 \gamma_{\mu} &\to \mcS \gamma_{\mu} \mcS^{-1},
\\
 \psi &\to \mcS \psi,
\\
 \psibar &\to \psibar \mcS^{-1},
\\
 \spinmetric &\to (\mcS^{\dagger})^{-1} \spinmetric \mcS^{-1} \text{.}
\end{aligned}
\end{align}

Finally, we investigate the diffeomorphisms by demanding that $\psibar \gamma_{\mu} \psi$ transforms as a covariant spacetime vector
\begin{align}
 \psibar \gamma_{\mu} \psi \to \psibar' \gamma_{\mu}' \psi' = \psibar' \frac{\partial x^{\rho}}{\partial x'^{\mu}} \gamma_{\rho} \psi' \stackrel{!}{=} \frac{\partial x^{\rho}}{\partial x'^{\mu}} \psibar \gamma_{\rho} \psi \text{.}
\end{align}
Once again we find the phase transformation $\euler^{\cplx \varphi}$ and the chiral transformation $\euler^{b_{2} \gamma_{\ast}}$. If we then proceed analogous to the spin-base transformations and demand that $\psibar \psi$ is a scalar
\begin{align}
 \psibar \psi \to \psibar' \psi' \stackrel{!}{=} \psibar \psi \text{,}
\end{align}
the chiral transformation turns out to be just a sign $\euler^{b_{2} \gamma_{\ast}} = \pm \mrI$.
This sign can be absorbed into the phase $\pm \euler^{\cplx \varphi} \mrI = \euler^{\cplx \varphi'} \mrI$, which we drop.
%. The phase for the spacetime coordinate transformations is redundant since there is already a phase inside the spin-base transformations.

We summarize the behavior under diffeomorphisms as
\begin{align}\label{eq:diffeomorphism_d_even}
\begin{aligned}
 \gamma_{\mu} &\to \frac{\partial x^{\rho}}{\partial x'^{\mu}} \gamma_{\rho},
\\
 \psi &\to \psi,
\\
 \psibar &\to \psibar,
\\
 \spinmetric &\to \spinmetric \text{.}
\end{aligned}
\end{align}
In even dimensions it is possible to demand that the kinetic term as well as the mass term is invariant under all types of coordinate transformations. If we do so, the behavior under spin-base transformations is given by \Eqref{eq:spinbase_transformation_d_even} and under spacetime coordinate transformations by \eqref{eq:diffeomorphism_d_even}.

%-----------------------------------------------------------------------------
\subsection{Relation to flat spacetime and vielbein formalism}\label{subsec:relation_to_vielbein_formalism}
%-----------------------------------------------------------------------------

To define fermions more formally one usually starts in flat space with the Lorentz group $\mathrm{SO}(d-1,1)$ and investigates its representations.
In four spacetime dimensions fermions are objects transforming under the $(\frac{1}{2},0) \oplus (0,\frac{1}{2})$ representation of $\mathrm{Spin}(3,1)$ which is the double cover of $\mathrm{SO}_{0}(3,1)$.
Here $\mathrm{SO}_{0}(3,1)$ is the connected component of the identity of $\mathrm{SO}(3,1)$.
Already on this stage it is apparent that a similar construction for the diffeomorphisms will be difficult.
This is because of two reasons, first the Lorentz transformations leave the metric invariant and thus the explicit form of the Clifford algebra.
Second the fermions are not representations of the Lorentz group $\mathrm{SO}(3,1)$ but of the double cover of the Lorentz group, which is the spin group $\mathrm{Spin}(3,1)$.
One may expect that something similar, probably more complicated holds for the diffeomorphisms.
In fact Ogievetsky and Polubarinov found a highly nonlinear way of assigning a diffeomorphism to transformations in spinor space \cite{Ogievetsky:1965ii,Pitts:2011jv}.
The standard way, however, to recover the Lorentz group is by introducing the vielbein, which then has the bein index carrying the Lorentz symmetry.
In order to make contact with the spin group the flat Clifford algebra
\begin{align}
 \{ \flatgamma_{a} , \flatgamma_{b} \} = 2 \eta_{a b} \mrI
\end{align}
is then introduced in tangential space at every point of the manifold.

We want to stress that spin-base invariance is in some sense already present in this construction.
It is now usually assumed, that the flat Dirac matrices $\flatgamma_{a}$ are chosen to be the same in every tangential space.
But of course there is no reason to do this, as every point of the manifold has its own tangential space, with its own base.%
\footnote{
In fact this is the reason why the $\mathrm{SO}(d-1,1)$ is local in the vielbein formalism.
}
If we allow the flat Dirac matrices to be different at the different tangential spaces, we find the $\mathrm{SL}(d_{\gamma} , \C)$ again as the corresponding transformation between the different choices of the bases.
We can now observe, that neither the vielbein $\vielbein_{\mu}^{\point a}$ nor the flat Dirac matrices $\flatgamma_{a}$ appear alone in the usual terms of the gravitational and matter action, it is exclusively the combination $\vielbein_{\mu}^{\point a} \flatgamma_{a}$, i.e. the full Dirac matrices $\gamma_{\mu}$.%
\footnote{
This becomes most apparent by comparing the later formulas for the spin connection $\hat{\Gamma}_{\mu}$, cf. \Eqref{eq:def_hatGamma}, and the spin metric $\spinmetric$, cf. \Eqref{eq:defining_spin_metric}, with their standard vierbein formalism analogs.
}
Therefore it seems rather artificial to decouple the Dirac matrices $\gamma_{\mu}$ into a vielbein $\vielbein_{\mu}^{\point a}$ and the flat Dirac matrices $\flatgamma_{a}$.

Finally we can explain what it means that spinors transform under Lorentz transformations as in \Eqref{eq:lorentz_transformation_psi}.
We have to read this transformation as a coordinate transformation composed of a spin-base transformation $\mcS = \mcS_{\text{Lor}}^{-1}$ and a diffeomorphism $\frac{\partial x'^{a}}{\partial x^{b}} = {\Lambda_{\mcS}}^{a}_{\point b}$ such that
\begin{align}
% \begin{aligned}
%  \flatgamma_{a} \!\! \rightarrow \! \frac{\partial x^{b}}{\partial x'^{a}} \mcS_{\varphi} \flatgamma_{b} \mcS_{\varphi}^{-1} \! = \! \Lambda_{a}^{\point b} \mcS_{\text{Lor}}^{-1}(\Lambda) \flatgamma_{b} \mcS_{\text{Lor}}(\Lambda) \! \equiv \! \flatgamma_{a}
% \end{aligned}
\begin{aligned}
 \flatgamma^{a} \!\! \rightarrow \! \frac{\partial x'^{a}}{\partial x^{b}} \mcS \flatgamma^{b} \mcS^{-1} \! = \! {\Lambda_{\mcS}}^{a}_{\point b} \mcS_{\text{Lor}}^{-1} \flatgamma^{b} \mcS_{\text{Lor}} \! \equiv \! \flatgamma^{a}
\end{aligned}
\end{align}
and
\begin{align}
%  \psi \rightarrow \mcS_{\varphi} \psi = \mcS_{\text{Lor}}^{-1} (\Lambda) \psi, \quad \psibar \rightarrow \psibar \mcS_{\varphi}^{-1} = \psibar \mcS_{\text{Lor}}(\Lambda) \text{.}
 \psi \rightarrow \mcS \psi = \mcS_{\text{Lor}}^{-1} \psi, \quad \psibar \rightarrow \psibar \mcS^{-1} = \psibar \mcS_{\text{Lor}} \text{.}
\end{align}
By contrast if we only perform a spacetime coordinate transformation the fermions do not change.
Strictly speaking there is no sense in saying that fermions change sign under a spatial rotation of $360^{\circ}$.
The standard sign change becomes only visible if also the spin base is transformed in a specific way. But of course the spin base can be rotated without the spacetime and vice versa.

The spin-base transformations and especially the invariance of the action with respect to these has an intuitive interpretation.
If we start with the Clifford algebra we have many different sets of Dirac matrices we can choose from for a given metric.
But all these different sets are connected to each other via a similarity transformation and in odd dimensions additionally via a sign flip.
With this in mind we can read the invariance under %$\mathrm{SB}(d_{\gamma})$
spin-base transformations
as an invariance of the choice of Dirac matrices, i.e.~ for any choice of compatible $\gamma_{\mu}$ we get the same physical answer.
And in order to satisfy this condition for all compatible representations of the Clifford algebra we really have to take the complete %$\mathrm{SB}(d_{\gamma})$.
$\mathrm{SL}(d_{\gamma} , \C)$ as shown in App. \ref{App:minimal_spin_base_group}.
% It seems at first that we could drop the $\mathrm{U}(1)$ part, since we would still cover all possible sets.
% But we would have to make a choice for the explicit representation of $\mathrm{SL}(d_{\gamma} , \C) / \Z_{d_{\gamma}}$ and need to explain how spinors transform under these transformations, as they always have this $\mathrm{U}(1)$ gauge freedom in their scalar product $\psibar \psi$.
% By including the $\mathrm{U}(1)$ we circumvent these problems.

This consideration also tells us that in odd dimensions physical results can depend on the choice of the connected component of the $\gamma_{\mu}$. We have an invariance with respect to %$\mathrm{SB}(d_{\gamma})$,
$\mathrm{SL}(d_{\gamma} , \C)$,
but if we e.g.~include a mass term we lose invariance under the sign flip.
And therefore the choice of the connected component can be an integral part of the theory.
This is, for instance, familiar from fermion-induced Chern-Simons terms \cite{Redlich:1983kn,Dunne:1998qy}

%-----------------------------------------------------------------------------
\section{General requirements}
\label{sec:general_requirements}
%-----------------------------------------------------------------------------

With the preparations of the previous chapter we now turn to the description of fermions in curved spacetimes.
Considering curved spacetimes and fermions we have to care about covariance with respect to coordinate transformations especially both kinds of them, spacetime coordinate transformations and spin-base transformations.
In order to describe spinors we need Dirac structure, defined via the Clifford algebra in irreducible representation
\begin{align}
 \{ \gamma_{\mu} , \gamma_{\nu} \} = 2 \metric_{\mu \nu} \mrI \text{,} \quad \gamma_{\mu} \in \mathrm{Mat}(d_{\gamma} \times d_{\gamma} , \C) \text{.}
\end{align}
Fermions $\psi$ are then complex Gra\ss{}mann valued fields transforming as ``vectors'' under the %above given matrix representation of the spin-base group $\mathrm{SB}(d_{\gamma})$ and are scalars under diffeomorphisms. The spin-base group is defined as
% \begin{align}
%  \mathrm{SB}(d_{\gamma}) = \{ \euler^{\cplx \varphi} \mcS : \euler^{\cplx \varphi} \in \mathrm{U}(1) , \mcS \in \mathrm{SL}(d_{\gamma} , \C) \}, \,\, d_{\gamma} = 2^{\lfloor \frac{d}{2} \rfloor},
% \end{align}
% with the standard matrix multiplication as the group product.
fundamental representation of the special linear group $\mathrm{SL}(d_{\gamma} , \C)$.
The dual vector $\psibar$ is related to the vector $\psi$ via the spin metric $\spinmetric$
\begin{align}
 \psibar = \psi^{\dagger} \spinmetric
\end{align}
whose determinant has to satisfy
\begin{align}
 \abs{\det \spinmetric} = 1 \text{,}
\end{align}
such that $\spinmetric$ does not introduce any scale between $\psi$ and $\psibar$.
The transformation law for fermions under a spin-base transformation %$\mcS_{\varphi} \in \mathrm{SB}(d_{\gamma})$
$\mcS \in \mathrm{SL}(d_{\gamma} , \C)$
reads
\begin{align}
%  \psi \to \psi' = \mcS_{\varphi} \psi, \quad \psibar \to \psibar' =  \psibar \mcS_{\varphi}^{-1}
 \psi \to \psi' = \mcS \psi, \quad \psibar \to \psibar' =  \psibar \mcS^{-1} 
\end{align}
and under diffeomorphisms
\begin{align}
 \psi \to \psi' = \psi, \quad \psibar \to \psibar' = \psibar \text{.}
\end{align}
Under spin-base transformations the spin metric changes as
\begin{align}
%  \spinmetric \to \spinmetric' = (\mcS_{\varphi}^{\dagger})^{-1} \spinmetric \mcS_{\varphi}^{-1}
 \spinmetric \to \spinmetric' = (\mcS^{\dagger})^{-1} \spinmetric \mcS^{-1} \text{,}
\end{align}
and under diffeomorphisms as a scalar
\begin{align}
 \spinmetric \to \spinmetric' = \spinmetric \text{.}
\end{align}
Of course also the Dirac matrices transform non trivially under spin-base transformations
\begin{align}
%  \gamma_{\mu} \to \mcS_{\varphi} \gamma_{\mu} \mcS_{\varphi}^{-1} \text{.}
 \gamma_{\mu} \to \mcS \gamma_{\mu} \mcS^{-1} \text{.}
\end{align}
In odd dimensions there are two connected components for the $\gamma_{\mu}$ such that there exists additionally the possibility of a change of the component via a sign flip
\begin{align}
 \gamma_{\mu} \to - \gamma_{\mu} \text{,} \quad d \text{ odd.}
\end{align}
Then the fermions and the spin metric transform like
\begin{align}
 \psi \to \psi, \quad \psibar \to - \psibar, \quad \spinmetric \to - \spinmetric \text{,} \quad d \text{ odd.}
\end{align}
In even dimensions there is only one connected component and therefore there is no such discrete transformation.
Under diffeomorphisms the $\gamma_{\mu}$ behave as covariant vectors
\begin{align}
 \gamma_{\mu} \to \frac{\partial x^{\rho}}{\partial x'^{\mu}} \gamma_{\rho} \text{.}
\end{align}
Since we aim at describing dynamics we also have to introduce a covariant derivative $\nabla_{\mu}$ with
\begin{align}\label{eq:axiom_cov_deriv}
 \begin{aligned}
  \rmi{} &\quad \text{linearity:}
\\
 {}& \quad \nabla_{\mu}( \psi_{1} + \psi_{2} ) = \nabla_{\mu} \psi_{1} + \nabla_{\mu} \psi_{2} \text{,}
\\
  \rmii{} &\quad \text{product rule:}
\\
 {}& \quad \nabla_{\mu}(\psi \bar{\psi}) = (\nabla_{\mu} \psi) \bar{\psi} + \psi (\nabla_{\mu} \bar{\psi}) \text{,}
\\
  \rmiii{} &\quad \text{metric compatibility:}
\\
 {}& \quad \nabla_{\mu} \bar{\psi} = \overline{\nabla_{\mu} \psi}\text{,}
\\
  \rmiv{} &\quad \text{covariance:}
\\
 {}& \quad \nabla_{\mu} (\psibar \gamma^{\nu} \psi) = D_{\mu} (\psibar \gamma^{\nu} \psi) \text{.}
 \end{aligned} \!\!
\end{align}
The first two properties are quite intuitive.
Demanding $\rmiii$ is the analog of metric compatibility
\begin{align}
 D_{\mu} T_{\nu} = \metric_{\nu \rho} D_{\mu} T^{\rho}
\end{align}
of the usual spacetime covariant derivative $D_{\mu}$ with
\begin{align}
 D_{\mu} T^{\nu} = \partial_{\mu} T^{\nu} + \Gamma_{\mu \rho}^{\nu} T^{\rho} \text{.}
\end{align}
Here $\Gamma_{\mu \rho}^{\nu}$ is the spacetime connection
\begin{align}
 \Gamma_{\mu \rho}^{\nu} = \christoffel{\nu \\ \mu \rho} + K^{\nu}_{\point \mu \rho}
\end{align}
composed of the mandatory Levi-Civita part $\christoffel{\nu \\ \mu \rho}$
\begin{align}
 \christoffel{\nu \\ \mu \rho} = \frac{1}{2} \metric^{\nu \lambda} ( \partial_{\mu} \metric_{\lambda \rho} + \partial_{\rho} \metric_{\lambda \mu} - \partial_{\lambda} \metric_{\mu \rho} )
\end{align}
and the possible contorsion tensor $K^{\nu}_{\point \mu \rho}$ which is related to the torsion tensor $C^{\nu}_{\point \mu \rho}$
\begin{align}
 C^{\nu}_{\point \mu \rho} &{}= 2 K^{\nu}_{\point[1] [\mu \rho]},
\\ \label{eq:torsion_contorsion_relation}
 K^{\nu}_{\point \mu \rho} &{}= \frac{1}{2} ( C^{\nu}_{\point \mu \rho} + C_{\rho \point \mu}^{\point \nu} - C_{\mu \rho}^{\point[2] \nu} ) \equiv - K_{\rho \mu}^{\point[2] \nu} \text{.}
\end{align}
The last condition connects the generalized covariant derivative $\nabla_{\mu}$ to the spacetime covariant derivative $D_{\mu}$ and provides the covariance with respect to diffeomorphisms.%
%
% \footnote{
% If we also want to incorporate the standard model and its symmetries, then $D_{\mu}$ has to be the covariant derivative which respects diffeomorphisms and the standard model symmetries. Especially $D_{\mu} \psibar \psi$ does not have to be just $\partial_{\mu} \psibar \psi$ but can also carry some gauge field dependence because of the unequal treatment of the right-handed and left-handed spinors.
% }

% Especially there could be a non abelian connection which carries group and Dirac indices.
% In the following we assume the covariant derivative $D_{\mu}$ to carry a connection $\cplx A_{\mu}$ for the spinor $\psi$
% \begin{align}
%  D_{\mu} \psi = \partial_{\mu} \psi + \cplx A_{\mu} \psi
% \end{align}
% and a connection $- \cplx A'_{\mu}$ for the conjugated spinor $\psibar$
% \begin{align}
%  D_{\mu} \psibar = \partial_{\mu} \psibar - \psibar \cplx A'_{\mu} \text{.}
% \end{align}
% The connection $\mcA_{\mu}$ carries Dirac indices as well as group indices.
% Therefore the covariant derivative $D_{\mu}$ acts on the Dirac matrices as
% \begin{align}
%  D_{\mu} \gamma^{\nu} = \partial_{\mu} \gamma^{\nu} + \christoffel{\nu \\ \mu \rho} \gamma^{\rho} + K^{\nu}_{\point \mu \rho} \gamma^{\rho} - \cplx A'_{\mu} \gamma^{\nu} + \gamma^{\nu} \cplx A_{\mu}
% \end{align}

We implicitly assume that the covariant derivative transforms a geometric object into the same geometric object of one spacetime tensorial rank higher.
For instance if the derivative acts on a spinor $\psi$, then $\nabla_{\mu}\psi$ is still a spinor with the same transformation law under spin-base transformations, but with the transformation law of a covariant vector under diffeomorphisms.

Finally the action of a unitary dynamical theory containing fermions should be real.
Therefore we demand that the kinetic and the mass term in their usual forms are real
\begin{align}\label{eq:kinetic_term_real}
 &{}\regint{x} (\psibar \slashed{\nabla} \psi)^{\ast} = \regint{x} \psibar \slashed{\nabla} \psi \text{,}
\\ \label{eq:mass_term_real}
 &{}\regint{x} (\psibar \psi)^{\ast} = \regint{x} \psibar \psi \text{.}
\end{align}
Here $\slashed{\nabla}$ denotes the Dirac operator
%
% \footnote{
% It is necessary to ensure that $\gamma_{\mu}$ only transforms under the diffeomorphisms and the spinbase transformations, but not under any other symmetry mediated by $D_{\mu}$. More explicitly we can assume the covariant derivative $D_{\mu}$ to have a connection $\cplx A_{\mu}$ for the spinor $\psi$, $D_{\mu} \psi = \partial_{\mu} \psi + \cplx A_{\mu} \psi$ and some other connection $- \cplx A'_{\mu}$ for the conjugated spinor $\psibar$, $D_{\mu} \psibar = \partial_{\mu} \psibar - \psibar \cplx A'_{\mu}$. Then we demand $\mcA'_{\mu} \gamma^{\nu} - \gamma^{\nu} \mcA_{\mu} = 0$ and ${A'_{\mu}}^{\dagger} \spinmetric - \spinmetric A_{\mu} = 0$ to hold. Note that this is satisfied for all gauge fields of the standard model.
%}
%
$\slashed{\nabla} = \gamma^{\mu} \nabla_{\mu}$ and $\regint{x}$ is a shorthand for the spacetime integral $\int \! \mrd^{d} x \sqrt{-g}$.
We tacitly assume that the considered manifolds and the fermionic fields allow us to freely integrate by parts under the integral without the occurrence of any boundary terms.

These basic requirements are the same as in \cite{Gies:2013noa}, where the spacetime dimension $d$ was fixed to $4$.
In the next section we construct the spin connection which will ensure the spin-base covariance for arbitrary integer dimensions $d \geq 2$.

%-----------------------------------------------------------------------------
\section{Spin metric and spin connection}
\label{sec:spin_connection}
%-----------------------------------------------------------------------------

Using our assumptions from the previous chapter let us analyze the properties of the necessary spin metric and spin connection.
Beginning with \Eqref{eq:mass_term_real} and the Gra\ss{}mann nature of fermions
\begin{align}
 (\psi^{\dagger} \spinmetric \psi)^{\ast} = \psi^{\mrT} \spinmetric^{\ast} \psi^{\ast} = - \psi^{\dagger} \spinmetric^{\dagger} \psi \text{,}
\end{align}
it turns out that the spin metric has to be antihermitean
\begin{align}
 \spinmetric^{\dagger} = - \spinmetric \text{.}
\end{align}
Additionally we define the Dirac conjugation of a matrix $M \in \mathrm{Mat}(d_{\gamma} \times d_{\gamma} , \C)$ analogous to the Dirac conjugation of a vector $\psi$ as
\begin{align}
 \bar{M} = \spinmetric^{-1} M^{\dagger} \spinmetric \text{.}
\end{align}
This Dirac conjugation is of particular interest for the complex conjugate of objects like
\begin{align}
 (\psibar M \psi)^{\ast} = \psibar \bar{M} \psi \text{.}
\end{align}
For the next step in our analysis we use the properties $\rmi$ - $\rmiv$ of \Eqref{eq:axiom_cov_deriv} to deduce
\begin{align}
 ( \partial_{\mu} \psibar) \psi + \psibar ( \partial_{\mu} \psi) \! = \! \partial_{\mu} \psibar \psi \! = \! \nabla_{\mu} \psibar \psi \! = \! (\nabla_{\mu} \psibar) \psi + \psibar (\nabla_{\mu} \psi) \text{.}
\end{align}
From here we conclude that the covariant derivative must carry a connection $\Gamma_{\mu}$
\begin{align}
 &\nabla_{\mu} \psi = \partial_{\mu} \psi + \Gamma_{\mu} \psi, \quad \nabla_{\mu} \psibar = \partial_{\mu} \psibar - \psibar \Gamma_{\mu} \text{.}
\end{align}
From the transformation laws under spin-base transformations and diffeomorphisms of spinors $\psi$ we find the transformation law of the connection $\Gamma_{\mu}$
\begin{align}
%  &\Gamma_{\mu} \to \mcS_{\varphi} \Gamma_{\mu} \mcS_{\varphi}^{-1} - (\partial_{\mu} \mcS_{\varphi}) \mcS_{\varphi}^{-1},
 &\Gamma_{\mu} \to \mcS \Gamma_{\mu} \mcS^{-1} - (\partial_{\mu} \mcS) \mcS^{-1},
\\
 &\Gamma_{\mu} \to \frac{\partial x^{\rho}}{\partial x'^{\mu}} \Gamma_{\rho} \text{.}
\end{align}
From $\rmiii$ we infer
%
% \footnote{
% We assume that the spin metric is invariant under all symmetries transmitted via the connection of $D_{\mu}$, such that $D_{\mu} \spinmetric = \partial_{\mu} \spinmetric$. Especially i.e.~the spin metric is a scalar with respect to spacetime coordinate transformations, but also with respect to e.g.~the standard model symmetries.
% }
%
\begin{align}
 \nabla_{\mu} \psibar = \overline{ \nabla_{\mu} \psi } = (\nabla_{\mu} \psi)^{\dagger} \spinmetric = \partial_{\mu} \psibar - \psibar \spinmetric^{-1} \partial_{\mu} \spinmetric + \psibar \bar{\Gamma}_{\mu} \text{,}
\end{align}
and deduce the metric compatibility equation
\begin{align}\label{eq:spin_metric_compatibility}
 \spinmetric^{-1} \partial_{\mu} \spinmetric = \Gamma_{\mu} + \bar{\Gamma}_{\mu} \text{.}
\end{align}
The following auxiliary matrix $\hat{\Gamma}_{\mu}$ turns out to be useful for our analysis. It is defined by
\begin{align}\label{eq:def_hatGamma}
 \DLC_{\mu} \gamma^{\nu} = \partial_{\mu} \gamma^{\nu} + \christoffel{\nu \\ \mu \rho} \gamma^{\rho} = - [ \hat{\Gamma}_{\mu} , \gamma^{\nu} ], \quad \tr \hat{\Gamma}_{\mu} = 0 \text{,}
\end{align}
where $\DLC_{\mu}$ is the (Levi-Civita) spacetime covariant derivative without torsion.
Such a matrix exists and is uniquely given by
\begin{align}\label{eq:explicit_solution_hat_gamma_I}
 &\hat{\Gamma}_{\mu} = \sum\limits_{n=1}^{d} \hat{m}_{\mu \rho_{1} \ldots \rho_{n}} \gamma^{\rho_{1} \ldots \rho_{n}}, \quad d \text{ even,}
\\\label{eq:explicit_solution_hat_gamma_II}
 &\hat{\Gamma}_{\mu} = \sum\limits_{n=1}^{\frac{d-1}{2}} \hat{m}_{\mu \rho_{1} \ldots \rho_{2n}} \gamma^{\rho_{1} \ldots \rho_{2n}}, \quad d \text{ odd,}
\\\label{eq:explicit_solution_hat_gamma_III}
 &\hat{m}_{\mu \rho_{1} \ldots \rho_{n}} = \frac{(-1)^{\frac{n ( n + 1 )}{2}} \tr\big( \gamma_{\rho_{1} \ldots \rho_{n}} [ ( \DLC_{\mu} \gamma^{\nu} ) , \gamma_{\nu} ] \big) }{2 \cdot n! \cdot \big( (1 - (-1)^{n}) d - 2n \big) \cdot d_{\gamma}} \text{.}
\end{align}
The proof is found in App. \ref{App:Existence_of_the_spin_connection}.
Note that $\hat{\Gamma}_{\mu}$ is completely determined in terms of the $\gamma_{\mu}$ and their first derivatives.
The matrix $\hat{\Gamma}_{\mu}$ transforms exactly like $\Gamma_{\mu}$ inhomogeneously under spin-base transformations
\begin{align}
 \hat{\Gamma}_{\mu} &\to \mcS \hat{\Gamma}_{\mu} \mcS^{-1} - ( \partial_{\mu} \mcS ) \mcS^{-1} \text{.}
\end{align}
To see this one considers the behavior of the defining equation of $\hat{\Gamma}_{\mu}$ \Eqref{eq:def_hatGamma} under spin-base transformations $\mcS \in \mathrm{SL}(d_{\gamma} , \C)$ and
\begin{align}
 \tr \big( (\partial_{\mu} \mcS) \mcS^{-1} \big) = \partial_{\mu} \tr \ln \mcS = \partial_{\mu} \euler^{\tr \ln \mcS} = \partial_{\mu} \det \mcS = 0 \text{.}
\end{align}
In order to investigate \Eqref{eq:kinetic_term_real} we calculate
\begin{align}
 \DLC_{\mu} \bar{\gamma}^{\nu} =& \DLC_{\mu} (\spinmetric^{-1} \gamma^{\nu \dagger} \spinmetric ) \notag
\\
 =& [ \bar{\gamma}^{\nu} , \spinmetric^{-1} (\partial_{\mu} \spinmetric) ] + \spinmetric^{-1} (\DLC \gamma^{\nu \dagger}) \spinmetric \notag
\\
 =& [ \bar{\gamma}^{\nu} , \Gamma_{\mu} + \bar{\Gamma}_{\mu} ] + \spinmetric^{-1} (\DLC \gamma^{\nu})^{\dagger} \spinmetric \notag
\\
 =& [ \bar{\gamma}^{\nu} , \Gamma_{\mu} + \bar{\Gamma}_{\mu} - \bar{\hat{\Gamma}}_{\mu}]
\end{align}
and recapitulate that
\begin{align}
 \partial_{\mu} \sqrt{- \metric} = \sqrt{- \metric} \christoffel{\rho \\ \rho \mu} \text{.}
\end{align}
With this in mind it is easy to evaluate
\begin{align}
 \regint{x} \psibar \slashed{\nabla} \psi =& \regint{x} (\psibar \slashed{\nabla} \psi)^{\ast} = \regint{x} (\overline{ \nabla_{\mu} \psi}) \bar{\gamma}^{\mu} \psi \notag = \regint{x} ( \nabla_{\mu} \psibar ) \bar{\gamma}^{\mu} \psi
\\
 =& - \regint{x} \psibar \big( (\DLC_{\mu} \bar{\gamma}^{\mu}) + \Gamma_{\mu} \bar{\gamma}^{\mu} + \bar{\gamma}^{\mu} \partial_{\mu}  \big) \psi \notag
\\
 =& \regint{x} \left[ \psibar (- \bar{\gamma}^{\mu}) \nabla_{\mu} \psi + \psibar [ \bar{\Gamma}_{\mu} - \bar{\hat{\Gamma}}_{\mu}  , \bar{\gamma}^{\mu} ] \psi \right] \!\! \text{.}
\end{align}
Since this statement has to be true for all spinors $\psi$ we identify
\begin{align}
 &\bar{\gamma}^{\mu} = - \gamma^{\mu},
\\ \label{eq:def:Delta_Gamma}
 &[ \Delta \Gamma_{\mu} , \gamma^{\mu} ] = 0 \text{.}
\end{align}
Here we have decomposed the spin connection $\Gamma_{\mu}$ without loss of generality into
\begin{align}
 \Gamma_{\mu} = \cplx \mcA_{\mu} \cdot \mrI + \hat{\Gamma}_{\mu} + \Delta \Gamma_{\mu} \text{.}
\end{align}
Apart from $\hat{\Gamma}_{\mu}$ defined above, we find a trace part $\mcA_{\mu}$
\begin{align}
 \mcA_{\mu} = - \frac{\cplx}{d_{\gamma}} \tr ( \Gamma_{\mu} ) \text{,}
\end{align}
and the spin torsion $\Delta \Gamma_{\mu}$ \cite{Gies:2013noa}
\begin{align}
 \Delta \Gamma_{\mu} = \Gamma_{\mu} - \hat{\Gamma}_{\mu} - \frac{1}{d_{\gamma}} \tr( \Gamma_{\mu} ) \cdot \mrI \text{.}
\end{align}
The transformation law under spin-base transformations for the components of the spin connection reads
\begin{align}
%  \Delta \Gamma_{\mu} &\to \mcS_{\varphi} \Delta \Gamma_{\mu} \mcS_{\varphi}^{-1}, 
% \\
%  \mcA_{\mu} &\to \mcA_{\mu} + \frac{\cplx}{d_{\gamma}} \tr \big( (\partial_{\mu} \mcS_{\varphi}) \mcS_{\varphi}^{-1} \big),
% \\
%  \hat{\Gamma}_{\mu} &\to \mcS_{\varphi} \hat{\Gamma}_{\mu} \mcS_{\varphi}^{-1} - ( \partial_{\mu} \mcS_{\varphi} ) \mcS_{\varphi}^{-1} + \frac{1}{d_{\gamma}} \tr \big( (\partial_{\mu} \mcS_{\varphi}) \mcS_{\varphi}^{-1} \big) \cdot \mrI \text{.}
%   \mcA_{\mu} \! &\to \! \mcA_{\mu} \text{,} \,\,\, \Delta \Gamma_{\mu} \! \to \! \mcS \Delta \Gamma_{\mu} \mcS^{-1} \text{,} \,\,\, \hat{\Gamma}_{\mu} \! \to \! \mcS \hat{\Gamma}_{\mu} \mcS^{-1} \! - \! ( \partial_{\mu} \mcS ) \mcS^{-1} \text{.}
  \mcA_{\mu} &\to \mcA_{\mu} \text{,} \quad \Delta \Gamma_{\mu} \to \mcS \Delta \Gamma_{\mu} \mcS^{-1} \text{.}
\end{align}
%
%
% Note that if we explicitly write $\mcS_{\varphi} = \euler^{\cplx \varphi} \mcS$, with $\euler^{\cplx \varphi} \in \mathrm{U}(1)$ and $\mcS \in \mathrm{SL}(d_{\gamma} , \C)$ then the transformation law reads%
% \footnote{
% Of course for a given $\mcS_{\varphi} \in \mathrm{SB}(d_{\gamma})$ there are several compatible $\euler^{\cplx \varphi} \in \mathrm{U}(1)$ and $\mcS \in \mathrm{SL}(d_{\gamma} , \C)$. But all of these choices give the exact same transformation law, because these different choices are related to each other via a discrete $\Z_{d_{\gamma}}$ transformation which has to be constant in order to be differentiable.
% }
% \begin{align}
%  \Delta \Gamma_{\mu} &\to \mcS \Delta \Gamma_{\mu} \mcS^{-1},
% \\
%  \mcA_{\mu} &\to \mcA_{\mu} - \partial_{\mu} \varphi,
% \\
%  \hat{\Gamma}_{\mu} &\to \mcS \hat{\Gamma}_{\mu} \mcS^{-1} - (\partial_{\mu} \mcS) \mcS^{-1} \text{.}
% \end{align}
%
We found the three important algebraic equations for the spin metric
\begin{align}\label{eq:defining_spin_metric}
 \gamma_{\mu}^{\dagger} = - \spinmetric \gamma_{\mu} \spinmetric^{-1}, \quad \spinmetric^{\dagger} = - \spinmetric, \quad \abs{\det \spinmetric} = 1 \text{.}
\end{align}
For a given set of Dirac matrices there is a unique spin metric (up to a sign) as proven in App. \ref{App:spin_metric}.

Next we use the Eqs. \eqref{eq:reality_DeltaGamma_1} and \eqref{eq:reality_DeltaGamma_2} from App. \ref{App:spin_metric} to infer
\begin{align}\label{eq:Delta_Gamma_antisymmetric}
 \overline{ \Delta \Gamma_{\mu} } = - \Delta \Gamma_{\mu}, \quad \Im \mcA_{\mu} = 0 \text{.}
\end{align}
If we compare the spin covariant derivative $\nabla_{\mu}$ with the spacetime covariant derivative $D_{\mu}$ we note a similar structure
\begin{align}
\begin{matrix}
 \nabla_{\mu} \psi & = & \partial_{\mu} \psi & + & \hat{\Gamma}_{\mu} \psi & + & \Delta \Gamma_{\mu} \psi & + & \cplx \mcA_{\mu} \psi ,
\\
 D_{\mu} T^{\nu} & = & \partial_{\mu} T^{\nu} & + & \christoffel{\nu \\ \mu \rho} T^{\rho} & + & K^{\nu}_{\point \mu \rho} T^{\rho} \text{.} &&
\end{matrix}
\end{align}
The first part is the ordinary partial derivative, the second part is the canonical (Levi-Civita) part, which is determined in terms of the Dirac matrices, respectively the metric.
The third part is a possible torsion term, whose dynamics is essentially independent of the Dirac matrices and has to be determined by other means, e.g.~an action principle.
For the fermionic fields there is another for the moment unrestricted contribution, $\mcA_{\mu}$, without analog in the spacetime covariant derivative.
This vector field is reminiscent to a $\mathrm{U}(1)$ gauge field from the standard model.
If we included a $\mathrm{U}(1)$ symmetry transformation for the fermions, then this field would behave exactly like a usual gauge field.
As discussed above, we ignore this gauge field in the following.

% On the other hand it is straightforward to enlarge the symmetry $\mathrm{SB}(d_{\gamma})$ with an additional unitary gauge group $\mcG$ to $\mcG \times \mathrm{SB}(d_{\gamma})$, in order to incorporate e.g.~the standard model, c.f.~\cite{Gies:2013noa}.
% If we admit the conjecture that all fundamental matter feels the same spacetime geometry together with our prerequisites from Sec. \ref{sec:general_requirements} it is inevitable to restrict any additional gauge group $\mcG$ to be unitary.
% Because then the spin metric has to be the same for all fermion types, i.e. the generalized spin metric has the form
% \begin{align}\label{eq:generalized_spin_metric}
%  \spinmetric_{\mcG \times \mathrm{SB}(d_{\gamma})} = \mrI_{\mcG} \otimes \spinmetric \text{,}
% \end{align}
% where $\mrI_{\mcG}$ is the unit element of $\mcG$.
% Then the transformation law under a general gauge transformation $\mfrg \otimes \mcS_{\varphi} \in \mcG \times \mathrm{SB}(d_{\gamma})$ reads
% \begin{align}
%  \spinmetric_{\mcG \times \mathrm{SB}(d_{\gamma})} \to{}& (\mfrg^{\dagger} \otimes \mcS_{\varphi}^{\dagger})^{-1} \spinmetric_{\mcG \times \mathrm{SB}(d_{\gamma})} (\mfrg \otimes \mcS_{\varphi})^{-1} \notag
% \\
%  &= (\mfrg \mfrg^{\dagger})^{-1} \otimes \big( (\mcS_{\varphi}^{\dagger})^{-1} \spinmetric \mcS_{\varphi}^{-1} \big) \text{.}
% \end{align}
% In order not to spoil the form \Eqref{eq:generalized_spin_metric} the elements of $\mcG$ have to satisfy
% \begin{align}
%  \mfrg \mfrg^{\dagger} = \mrI_{\mcG},
% \end{align}
% rendering the group $\mcG$ unitary.

Now we are in a very comfortable situation.
Given a set of Dirac matrices we can calculate everything we need to describe fermions in a curved spacetime.
There is a (up to a sign) unique spin metric $\spinmetric$ and a unique canonical (Levi-Civita) part of the connection $\hat{\Gamma}_{\mu}$.
%Additionally there is a ``gauge'' field $\mcA_{\mu}$, which takes care of the $\mathrm{U}(1)$ symmetry of the spinbase transformations.
Furthermore there is a rather undetermined object $\Delta \Gamma_{\mu}$, which we call spin torsion and whose dynamics we are going to investigate in the next section.

Let us first justify the name ``spin torsion'' by comparing it to spacetime torsion.
The spacetime torsion is the part of the spacetime connection, that even in local inertial coordinates at an arbitrary point is non vanishing, it cannot be transformed away with a spacetime coordinate transformation.
In order to be more precise, we need a notion of ``local inertial coordinates'' in our setup.
We want local inertial spacetime coordinates as well as local inertial spin bases.
There is a straightforward generalization for ``local inertial at a fixed spacetime point $z$''.
For the spacetime coordinates we demand that the spacetime metric aquires Minkowskian form and its first derivative vanishes
\begin{align}
 \metric_{\mu \nu}|_{z} = \eta_{\mu \nu}, \quad \partial_{\lambda} \metric_{\mu \nu}|_{z} = 0 \text{,}
\end{align}
i.e. the spacetime coordinate base is constant in a vicinity around $z$.
Since there is no preferred set of Dirac matrices compatible with the Clifford algebra, there is no ``Minkowskian'' form
of the $\gamma_{\mu}|_{z}$.%
\footnote{
In fact the Minkowskian form of $\metric_{\mu \nu}|_{z}$ is not important.
We could change the spacetime coordinates in a nontrivial, but constant way and would loose the Minkowskian form, but still the Christoffel symbols would vanish. The important point is the constant spacetime base.
}
Still, we can analogously demand that the spin base is adjusted in the same fashion around a vicinity of $z$
\begin{align}
 \partial_{\lambda} \gamma_{\mu}|_{z} = 0 \text{.}
\end{align}
These coordinates are by no means unique, e.g. for the spacetime coordinates we can always perform constant Lorentz transformations and for the spin bases we can perform constant similarity transformations.
However, the essential property of local inertial coordinates is that in these coordinates at the point $z$ the Christoffel symbol vanishes, but the contorsion tensor $K^{\nu}_{\point \mu \rho}|_{z}$ only vanishes if there is no torsion at this point.
We observe now the same behaviour for the spin connection.
The canonical (Levi-Civita) part $\hat{\Gamma}_{\mu}|_{z}$ vanishes, whereas the spin torsion $\Delta \Gamma_{\mu}|_{z}$ would only vanish if it was zero also before the coordinate transformation, i.e. if there was no spin torsion at all.

% Also the field $\mcA_{\mu}|_{z}$ can be transformed away with an appropriate phase transformation $\euler^{\cplx \varphi}$.
% We define $\varphi(x)$ only in an infinitesimal vicinity around $z$ and demand that it is a smoth spacetime scalar outside of this vicinity.
% This way we ensure $\varphi(x)$ to be a scalar everywhere.
% E.g. $\varphi(x) = (x - x_{z})^{\mu} \cdot \mcA_{\mu}(x)$ corresponds to $\partial_{\mu} \varphi|_{z} = \mcA_{\mu}|_{z}$,
% here $x_{z}^{\mu}$ are the spacetime coordinates for the point $z$.
% Then the transformed field $\mcA_{\mu}$ reads $\mcA_{\mu}|_{z} \to \mcA_{\mu}|_{z} - \partial_{\mu} \varphi|_{z} = 0$.

The dynamics of %the real-valued vector field $\mcA_{\mu}$ and the dynamics of
the spin torsion $\Delta \Gamma_{\mu}$ is still missing, as well as the actual degrees of freedom of $\Delta \Gamma_{\mu}$.
E.g.~for the spacetime covariant derivative the contorsion $K^{\nu}_{\point \mu \rho}$ is not an arbitrary tensor, but it has to be antisymmetric in the first and the last indices, c.f.~\Eqref{eq:torsion_contorsion_relation}, in order to satisfy the metric compatibility condition.
A similar statement holds for the spin torsion which has to be antisymmetric with respect to Dirac conjugation, c.f.~\Eqref{eq:Delta_Gamma_antisymmetric}, so that the spin-metric compatibility is satisfied.
Additionally we found the constraint \Eqref{eq:def:Delta_Gamma}, which ensures that the kinetic term is real.
A perfectly valid, but quite simple solution to this equation is $\Delta \Gamma_{\mu} \stackrel{!}{=} 0$.
But it is obvious that this is not the most general choice compatible with the constraints.

%-----------------------------------------------------------------------------
\section{Dynamics of spin torsion% and the gauge field
}
\label{sec:spin_torsion_and_the_gauge_field}
%-----------------------------------------------------------------------------

This section is devoted to the spin torsion and its degrees of freedom as well as the construction of a possible action governing the dynamics of %$\mcA_{\mu}$ and
$\Delta \Gamma_{\mu}$.
To find the most general form of the spin torsion we first decompose it into the basis of Dirac matrices
\begin{align}
 \Delta \Gamma_{\mu} &= \sum\limits_{n=1}^{d} \varrho_{\mu \rho_{1} \ldots \rho_{n}} \gamma^{\rho_{1} \ldots \rho_{n}}, \quad d \text{ even,}
\\
 \Delta \Gamma_{\mu} &= \sum\limits_{n=1}^{\frac{d-1}{2}} \varrho_{\mu \rho_{1} \ldots \rho_{2n}} \gamma^{\rho_{1} \ldots \rho_{2n}}, \quad d \text{ odd.}
\end{align}
Next we use the identities from App. \ref{App:Special_relations_gamma_matrices_II} to implement \Eqref{eq:def:Delta_Gamma} and \eqref{eq:Delta_Gamma_antisymmetric}.
The odd dimensional case is simpler, we employ \Eqref{eq:App:commutator_gamma_even_base_element} and find
\begin{align}
 0 =& [ \Delta \Gamma_{\mu} , \gamma^{\mu} ] = \sum\limits_{n=1}^{\frac{d-1}{2}} \varrho_{\mu \rho_{1} \ldots \rho_{2n}} [ \gamma^{\rho_{1} \ldots \rho_{2n}} , \gamma^{\mu} ] \notag
\\
 =& - 4 \sum\limits_{n=1}^{\frac{d-1}{2}} n \varrho_{\mu \rho_{1} \ldots \rho_{2n}} \metric^{\mu [ \rho_{1}} \gamma^{\rho_{2} \ldots \rho_{2n}]} \notag
\\
 =& - 4 \sum\limits_{n=1}^{\frac{d-1}{2}} n \varrho^{\rho_{1}}_{\point \rho_{1} \rho_{2} \ldots \rho_{2n}} \gamma^{\rho_{2} \ldots \rho_{2n}} \text{.}
\end{align}
From this we conclude
\begin{align}
 0 = \varrho^{\rho_{1}}_{\point \rho_{1} \rho_{2} \ldots \rho_{2n}} , \quad n \in \left\{ 1, \ldots , \frac{d-1}{2} \right\} \text{.}
\end{align}
In even dimensions we plug in our ansatz
\begin{align}
 0 = [ \Delta \Gamma_{\mu} , \gamma^{\mu} ] = \sum\limits_{n=1}^{d} \varrho_{\mu \rho_{1} \ldots \rho_{n}} [ \gamma^{\rho_{1} \ldots \rho_{n}} , \gamma^{\mu} ] \text{,}
\end{align}
and calculate for $k \in \{ 1, \ldots , d \}$
\begin{align}
 0 = \sum\limits_{n=1}^{d} \varrho_{\mu \rho_{1} \ldots \rho_{n}} \frac{1}{d_{\gamma}} \tr \big( [ \gamma^{\rho_{1} \ldots \rho_{n}} , \gamma^{\mu} ] \gamma_{\nu_{1} \ldots \nu_{k}}  \big) \text{.}
\end{align}
Since the trace of an odd number of Dirac matrices in even dimensions always vanishes
\begin{align}
 0 = \tr ( \gamma_{\mu_{1}} \ldots \gamma_{\mu_{2l+1}} ), \quad l \in \N_{0},
\end{align}
we have to distinguish two cases, $k$ even and $k$ odd.
Then we can neglect half of the sum for the respective choice of $k$.

For even $k$ we write $k = 2m$, $m \in \{ 1, \ldots, \frac{d}{2} \}$ and find with the identity \Eqref{eq:App:special_trace_gamma_commutator} from App. \ref{App:Special_relations_gamma_matrices_II}
\begin{align}
 0 =& \sum\limits_{l=1}^{\frac{d}{2}} \varrho^{\mu}_{\point \rho_{1} \ldots \rho_{2l-1}} \frac{1}{d_{\gamma}} \tr \big( [ \gamma^{\rho_{1} \ldots \rho_{2l-1}} , \gamma_{\mu} ] \gamma_{\nu_{1} \ldots \nu_{2m}} \big) \notag
\\
 =& \sum\limits_{l=1}^{\frac{d}{2}} \varrho^{\mu}_{\point \rho_{1} \ldots \rho_{2l-1}} \frac{1}{d_{\gamma}} \tr \big( [ \gamma_{\nu_{1} \ldots \nu_{2m}} , \gamma^{\rho_{1} \ldots \rho_{2l-1}} ] \gamma_{\mu} \big) \notag
\\
 =& \sum\limits_{l=1}^{\frac{d}{2}} \varrho^{\mu}_{\point \rho_{1} \ldots \rho_{2l-1}} (-1)^{l-1} \cdot 2 \cdot (2l)! \cdot \metric_{\mu [ \nu_{1}} \deltaA^{\rho_{1} \ldots \rho_{2l-1}}_{\nu_{2} \ldots \nu_{2l}} \cdot \delta^{l}_{m} \text{,}
\end{align}
where $\deltaA_{\mu_{1} \ldots \mu_{m}}^{\nu_{1} \ldots \nu_{m}}$ is the normalized and antisymmetrized Kronecker Delta.
Since $m \in \{ 1, \ldots, \frac{d}{2} \}$ is arbitrary we infer
\begin{align}
 0 = \varrho_{[\mu \rho_{1} \ldots \rho_{2m-1}]} \text{.}
\end{align}
Next we choose $k$ odd and write $k = 2m -1$ with $m \in \{ 1 , \ldots , \frac{d}{2} \}$.
Then the trace evaluates to
\begin{align}
  0 =& \sum\limits_{l=1}^{\frac{d}{2}} \varrho^{\mu}_{\point \rho_{1} \ldots \rho_{2l}} \frac{1}{d_{\gamma}} \tr \big( [ \gamma^{\rho_{1} \ldots \rho_{2l}} , \gamma_{\mu} ] \gamma_{\nu_{1} \ldots \nu_{2m-1}} \big) \notag
\\
 =& \sum\limits_{l=1}^{\frac{d}{2}} \varrho^{\mu}_{\point \rho_{1} \ldots \rho_{2l}} \frac{1}{d_{\gamma}} \tr \big( - 4 l \delta^{[\rho_{1}}_{\mu} \gamma^{\rho_{2} \ldots \rho_{2l}]} \gamma_{\nu_{1} \ldots \nu_{2m-1}} \big) \notag
\\
 =& 2 \sum\limits_{l=1}^{\frac{d}{2}} \varrho^{\mu}_{\point \rho_{1} \ldots \rho_{2l}} \cdot (2m)! \cdot  (-1)^{m} \delta_{\mu}^{[\rho_{1}} \deltaA^{\rho_{2} \ldots \rho_{2l}]}_{\nu_{1} \ldots \nu_{2m-1}} \cdot \delta^{l}_{m} \text{,}
\end{align}
where we have made use of Eqs. \eqref{eq:App:commutator_gamma_even_base_element} and \eqref{eq:App:Anm_d_even}.
Again since $m \in \{1, \ldots , \frac{d}{2} \}$ is arbitrary, we deduce
\begin{align}
 0 = \varrho^{\rho_{1}}_{\point \rho_{1} \rho_{2} \ldots \rho_{2m}} \text{.}
\end{align}
The second condition \Eqref{eq:Delta_Gamma_antisymmetric} reexpresses the metric compatibility and tells us whether the coefficients $\varrho_{\mu \rho_{1} \ldots \rho_{n}}$, respectively $\varrho_{\mu \rho_{1} \ldots \rho_{2n}}$ are purely real or purely imaginary.
We introduce the new variables
\begin{align}
 &\tilde{\varrho}_{\mu \rho_{1} \ldots \rho_{n}} = \cplx^{-\frac{n(n+1) + 2}{2}} \varrho_{\mu \rho_{1} \ldots \rho_{n}}, \,\, n \in \{ 1, \ldots, d \}, \,\, d \text{ even,}
\\
 &\tilde{\varrho}_{\mu \rho_{1} \ldots \rho_{2n}} = \cplx^{n - 1} \varrho_{\mu \rho_{1} \ldots \rho_{2n}}, \, n \! \in \! \left\{ 1, \ldots, \frac{d-1}{2} \right\} \!, \, d \text{ odd,}
\end{align}
and find that these have to be purely real employing the metric compatibility together with \Eqref{eq:App:gamma_bar_sign_identity} from App. \ref{App:Special_relations_gamma_matrices_II}:
\begin{align}
 &\tilde{\varrho}_{\mu \rho_{1} \ldots \rho_{n}} \in \R , \quad n \in \{ 1, \ldots, d \}, \quad d \text{ even,}
\\
 &\tilde{\varrho}_{\mu \rho_{1} \ldots \rho_{2n}} \in \R , \quad n \in \left\{ 1, \ldots , \frac{d-1}{2} \right\}, \quad d \text{ odd.}
\end{align}
Summing up, the spin torsion is given in even dimensions by
\begin{align}
 \Delta \Gamma_{\mu} = \sum\limits_{n=1}^{d} \tilde{\varrho}_{\mu \rho_{1} \ldots \rho_{n}} \cplx^{\frac{n ( n + 1 ) + 2}{2}} \gamma^{\rho_{1} \ldots \rho_{n}},
\end{align}
with the real coefficients
\begin{align}\label{eq:constrained_coefficients_Delta_Gamma}
 0 = \tilde{\varrho}^{\rho_{1}}_{\point \rho_{1} \rho_{2} \ldots \rho_{2m}}, \quad 0 = \tilde{\varrho}_{[\mu \rho_{1} \ldots \rho_{2m-1}]} , \quad m \in \left\{ 1, \ldots , \frac{d}{2} \right\}
\end{align}
and in odd dimensions
\begin{align}
 \Delta \Gamma_{\mu} = \sum\limits_{n=1}^{\frac{d-1}{2}} \tilde{\varrho}_{\mu \rho_{1} \ldots \rho_{2n}} \cplx^{- ( n - 1 )} \gamma^{\rho_{1} \ldots \rho_{2n}},
\end{align}
with the real coefficients
\begin{align}
 0 = \tilde{\varrho}^{\rho_{1}}_{\point \rho_{1} \rho_{2} \ldots \rho_{2m}}, \quad m \in \left\{ 1, \ldots , \frac{d}{2} \right\} \text{.}
\end{align}
Further we can count the degrees of freedom. In even dimensions, for each $\tilde{\varrho}_{\mu \rho_{1} \ldots \rho_{n}}$ we have $d \cdot \binom{d}{n}$ components. For even $n$ there are $\binom{d}{n - 1}$ constraints and for odd $n$ there are $\binom{d}{n + 1}$ constraints
\begin{align}
 d \! \cdot \! \sum\limits_{n = 1}^{d} \! \binom{d}{n} \! - \! \sum\limits_{n=1}^{\frac{d}{2}} \! \binom{d}{2n - 1} \! - \! \sum\limits_{n=1}^{\frac{d}{2}} \! \binom{d}{2n} = (d - 1) (d_{\gamma}^{2} - 1) \text{.}
\end{align}
Therefore we have in total $(d-1)(d_{\gamma}^{2} - 1)$ real degrees of freedom for spin torsion. In odd dimensions, for each $\tilde{\varrho}_{\mu \rho_{1} \ldots \rho_{2n}}$ we have $d \cdot \binom{d}{2n}$ components and $\binom{d}{2n-1}$ constraints
\begin{align}
 d \cdot \sum\limits _{n=1}^{\frac{d-1}{2}} \binom{d}{2n} - \sum\limits_{n=1}^{\frac{d-1}{2}} \binom{d}{2n-1} = (d-1)(d_{\gamma}^{2} - 1 )\text{.}
\end{align}
Hence, we also have $(d-1) (d_{\gamma}^{2} - 1)$ real degrees of freedom for spin torsion in odd dimensions.

For even dimensions this number decreases if we also demand chiral invariance ($\psi \to \gamma_{\ast} \psi$, $\psibar \to - \psibar \gamma_{\ast}$) of the kinetic operator
\begin{align}
 \psibar \slashed{\nabla} \psi \to - \psibar \gamma_{\ast} \slashed{\nabla} \gamma_{\ast} \psi \stackrel{!}{=} \psibar \slashed{\nabla} \psi \text{.}
\end{align}
This constraint leads to
\begin{align}
 0 =& \gamma^{\mu} (\nabla_{\mu} \gamma_{\ast}) = \gamma^{\mu} ( \partial_{\mu} \gamma_{\ast} + [ \hat{\Gamma}_{\mu} , \gamma_{\ast} ] ) + \gamma^{\mu} [ \Delta \Gamma_{\mu} , \gamma_{\ast} ] \notag
\\
 =& \gamma^{\mu} [ \Delta \Gamma_{\mu} , \gamma_{\ast} ] \text{.}
\end{align}
In order to implement this constraint we insert our series expansion for $\Delta \Gamma_{\mu}$ and use \Eqref{eq:def:Delta_Gamma}%
\footnote{
Note that if we decompose $[\gamma^{\mu} , \gamma^{\rho_{1} \ldots \rho_{2n-1}}]$ into our standard basis, we find $\tr( [\gamma^{\mu} , \gamma^{\rho_{1} \ldots \rho_{2n-1}}] \gamma_{\nu_{1} \ldots \nu_{2m-1}} ) = 0$ and $\tr ( [\gamma^{\mu} , \gamma^{\rho_{1} \ldots \rho_{2n-1}}] \gamma_{\nu_{1} \ldots \nu_{2m}} )
%= - \tr ( \gamma^{\rho_{1} \ldots \rho_{2n-1}} [\gamma^{\mu} , \gamma_{\lambda_{1} \ldots \lambda_{2m}} ] )
\sim \delta^{n}_{m} \deltaA^{\mu \rho_{1} \ldots \rho_{2n-1}}_{\nu_{1} \nu_{2} \ldots \nu_{2n}}$, c.f. \Eqref{eq:App:special_trace_gamma_commutator}.
}
\begin{align}
 0 =& \sum\limits_{n=1}^{d} \tilde{\varrho}_{\mu \rho_{1} \ldots \rho_{n}} \cplx^{\frac{n ( n + 1 ) + 2}{2}} \gamma^{\mu} [ \gamma^{\rho_{1} \ldots \rho_{n}} , \gamma_{\ast} ] \notag
\\
 =& - 2 \cplx \gamma_{\ast} \sum\limits_{n=1}^{\frac{d}{2}} \tilde{\varrho}_{\mu \rho_{1} \ldots \rho_{2n-1}} \cplx^{n} \gamma^{\mu} \gamma^{\rho_{1} \ldots \rho_{2n-1}} \notag
\\
 =& - \cplx \gamma_{\ast} \sum\limits_{n=1}^{\frac{d}{2}} \tilde{\varrho}_{\mu \rho_{1} \ldots \rho_{2n-1}} \cplx^{n} [ \gamma^{\mu} , \gamma^{\rho_{1} \ldots \rho_{2n-1}} ] \notag
\\
 & - \cplx \gamma_{\ast} \sum\limits_{n=1}^{\frac{d}{2}} \tilde{\varrho}_{\mu \rho_{1} \ldots \rho_{2n-1}} \cplx^{n} \{ \gamma^{\mu} , \gamma^{\rho_{1} \ldots \rho_{2n-1}} \} \notag
\\
 =& - \cplx \gamma_{\ast} \sum\limits_{n=1}^{\frac{d}{2}} \tilde{\varrho}_{\mu \rho_{1} \ldots \rho_{2n-1}} \cplx^{n} \{ \gamma^{\mu} , \gamma^{\rho_{1} \ldots \rho_{2n-1}} \} \text{.}
\end{align}
Next we employ the result \Eqref{eq:App:anticommutator_gamma_odd_base_element} from App.~\ref{App:Special_relations_gamma_matrices_II} and find
\begin{align}
 0 =& \sum\limits_{n=1}^{\frac{d}{2}} \tilde{\varrho}_{\mu \rho_{1} \ldots \rho_{2n-1}} \cplx^{n} \{ \gamma^{\mu} , \gamma^{\rho_{1} \ldots \rho_{2n-1}} \} \notag
\\
 =& 2 \sum\limits_{n=1}^{\frac{d}{2}} \tilde{\varrho}_{\mu \rho_{1} \ldots \rho_{2n-1}} (2n-1) \cplx^{n} \metric^{\mu [\rho_{1}} \gamma^{\rho_{2} \ldots \rho_{2n-1}]} \notag
\\
 =& 2 \sum\limits_{n=1}^{\frac{d}{2}} \tilde{\varrho}^{\rho_{1}}_{\point \rho_{1} \rho_{2} \ldots \rho_{2n-1}} (2n-1) \cplx^{n} \gamma^{\rho_{2} \ldots \rho_{2n-1}} \text{.}
\end{align}
Since the $\gamma^{\rho_{1} \ldots \rho_{n}}$ form a basis we can read off
\begin{align}
 0 = \tilde{\varrho}^{\rho_{1}}_{\point \rho_{1} \rho_{2} \ldots \rho_{2n-1}} , \quad n \in \left\{ 1, \ldots , \frac{d}{2} \right\} \text{.}
\end{align}
These additional constraints are independent from the first set \Eqref{eq:constrained_coefficients_Delta_Gamma}, again we can count the new constraints
\begin{align}
 \sum\limits_{n=1}^{\frac{d}{2}} \binom{d}{2n-2} = \frac{1}{2} d_{\gamma}^{2} - 1
\end{align}
leaving us with $( d - 2 ) ( d_{\gamma}^{2} - 1 ) + \frac{d_{\gamma}^{2}}{2}$ real degrees of freedom for chiral spin torsion in even dimensions.

Note that the coefficients $\tilde{\varrho}_{\mu \rho_{1} \ldots \rho_{n}}$, respectively the $\tilde{\varrho}_{\mu \rho_{1} \ldots \rho_{2n}}$ are spin-base independent: In accordance to the preceeding discussion we cannot transform away any of these coefficients with a spin-base transformation.

With the covariant derivative $\nabla_{\mu}$ at hand we can turn to a construction of an action similar to \cite{Gies:2013noa}.
The Einstein-Hilbert action is constructed from the field strength tensor, which in general relativity is the Riemann tensor $R_{\mu \nu \rho \lambda}$.
It is defined as
\begin{align}
 R_{\mu \nu \point \rho}^{\point[2] \lambda} T^{\rho} = [ D_{\mu} , D_{\nu} ] T^{\lambda} + C^{\sigma}_{\point \mu \nu} D_{\sigma} T^{\lambda}, \quad \forall T^{\rho} \text{ tensor.}
\end{align}
Following the same route we define the spin curvature $\Phi_{\mu \nu}$ to be
\begin{align}
 \Phi_{\mu \nu} \psi = [ \nabla_{\mu} , \nabla_{\nu} ] \psi + C^{\sigma}_{\point \mu \nu} \nabla_{\sigma} \psi \text{.}
\end{align}
More precisely $\Phi_{\mu \nu}$ reads
\begin{align}
 \Phi_{\mu \nu} =& \partial_{\mu} \Gamma_{\nu} - \partial_{\nu} \Gamma_{\mu} + [ \Gamma_{\mu} , \Gamma_{\nu} ]
\\
%  =& \cplx \mcF_{\mu \nu} \mrI \! + \! \hat{\Phi}_{\mu \nu} \! + \! 2 \partial_{[\mu} \Delta \Gamma_{\nu]} \! + \! 2 [ \hat{\Gamma}_{[\mu}, \Delta \Gamma_{\nu]}] \! + \! [ \Delta \Gamma_{\mu} , \Delta \Gamma_{\nu} ] ,
 =& \hat{\Phi}_{\mu \nu} + 2 \partial_{[\mu} \Delta \Gamma_{\nu]} + 2 [ \hat{\Gamma}_{[\mu}, \Delta \Gamma_{\nu]}] + [ \Delta \Gamma_{\mu} , \Delta \Gamma_{\nu} ] ,
\end{align}
%
% where $\mcF_{\mu \nu}$ is the field strength corresponding to $\mcA_{\mu}$
% \begin{align}
%  \mcF_{\mu \nu} = \partial_{\mu} \mcA_{\nu} - \partial_{\nu} \mcA_{\mu}
% \end{align}
% and 
where $\hat{\Phi}_{\mu \nu}$ is the curvature induced by $\hat{\Gamma}_{\mu}$, c.f. \cite{Gies:2013noa},
\begin{align}
 \hat{\Phi}_{\mu \nu} =& \partial_{\mu} \hat{\Gamma}_{\nu} - \partial_{\nu} \hat{\Gamma}_{\mu} + [ \hat{\Gamma}_{\mu} , \hat{\Gamma}_{\nu} ]
\\
 =& \frac{1}{8} {R_{(\mathrm{LC})}}_{\mu \nu \alpha \beta} [ \gamma^{\alpha} , \gamma^{\beta} ] \text{.}
\end{align}
By ${R_{(\mathrm{LC})}}_{\mu \nu \alpha \beta}$ we denote the Riemann curvature tensor induced by the Christoffel symbols
\begin{align}
 {R_{(\mathrm{LC})}}_{\mu \nu \point \rho}^{\point[2] \lambda} = \partial_{\mu} \christoffel{\lambda \\ \nu \rho} \! - \! \partial_{\nu} \christoffel{\lambda \\ \mu \rho} + \christoffel{\lambda \\ \mu \sigma} \christoffel{\sigma \\ \nu \rho} \! - \! \christoffel{\lambda \\ \nu \sigma} \christoffel{\sigma \\ \mu \rho} \text{.}
\end{align}
%
% Note the difference between $\mcF_{\mu \nu}$ here and in the previous work \cite{Gies:2013noa}.
% In the present case we restrict $\mcA_{\mu}$, to be the ``gauge'' field from the $\mathrm{U}(1)$ part of $\mathrm{SB}(d_{\gamma})$, where as in \cite{Gies:2013noa} it was the gauge field for the possible additional gauge group $\mcG$, which we neglect here.
% The inclusion of an additional unitary gauge group $\mcG$ is simple.
% One uses that the field strength of this gauge group together with the field strength of the $\mathrm{U}(1)$ part of $\mathrm{SB}(d_{\gamma})$ is the trace of $\Phi_{\mu \nu}$ with respect to the spinorial indices
% \begin{align}
%  \mcF_{\mu \nu} = \frac{- \cplx}{d_{\gamma}} \tr \Phi_{\mu \nu} \text{,}
% \end{align}
% i.e.~the gauge part $\mcG \times \mathrm{U}(1)$ decouples from the (pure) spinbase part $\mathrm{SL}(d_{\gamma} , \C) / \Z_{d_{\gamma}}$.

With this field strength we now construct an invariant (scalar) in order to give an action.
The simplest first order invariant without introduction of any new fields is
\begin{align}
%  \mfrL_{\Phi}^{(1)} = \frac{1}{d_{\gamma}} \tr ( \Phi_{\mu \nu} \gamma^{\nu \mu} ) \text{.}
 \mfrL_{\Phi} = \frac{1}{d_{\gamma}} \tr ( \Phi_{\mu \nu} \gamma^{\nu \mu} ) \text{.}
\end{align}
%
% There are several interesting second order invariants, but in order to keep it simple we are only interested in
% %
% \begin{align}
%  \mfrL_{\Phi}^{(2)} =& \frac{1}{d_{\gamma}^{2}} \tr ( \Phi_{\mu \nu} ) \tr ( \Phi^{\nu \mu} ) \equiv \mcF_{\mu \nu} \mcF^{\nu \mu} \text{.}
% %\\
% % \mfrL_{\Phi}^{(2,2)} =& \frac{1}{d_{\gamma}} \tr ( \Phi_{\mu \nu} \Phi^{\mu \nu} ),
% %\\
% % \mfrL_{\Phi}^{(2,3)} =& \frac{1}{d_{\gamma}} \tr ( \Phi_{\mu \nu} \Phi_{\rho \lambda} \gamma^{\mu \nu \rho \lambda} ),
% %\\
% % \mfrL_{\Phi}^{(2,4)} =& \big( \mfrL_{\Phi}^{(1,1)} \big)^{2} \text{.}
% \end{align}
%
With %these two invariants 
this invariant
we can construct an action reminiscent to the usual Einstein-Hilbert action %with a gauge field
\begin{align}
%  S_{\Phi} = \frac{1}{8 \pi G} \regint{x} \mfrL_{\Phi}^{(1)} + \frac{1}{4} \regint{x} \mfrL_{\Phi}^{(2)},
 S_{\Phi} = \frac{1}{8 \pi G} \regint{x} \mfrL_{\Phi},
\end{align}
where $G$ is some coupling constant.
For the explicit evaluation of %$\mfrL_{\Phi}^{(1)}$,
$\mfrL_{\Phi}$,
we first calculate in even dimensions
\begin{align}
 \DLC_{\mu} \Delta \Gamma_{\nu} ={}& \DLC_{\mu} \sum\limits_{n=1}^{d} \varrho_{\nu \rho_{1} \ldots \rho_{n}} \gamma^{\rho_{1} \ldots \rho_{n}} \notag
\\
 ={}& \sum\limits_{n=1}^{d} [ ( \DLC_{\mu} \varrho_{\nu \rho_{1} \ldots \rho_{n}} ) \gamma^{\rho_{1} \ldots \rho_{n}} \notag
\\
 & \qquad + \varrho_{\nu \rho_{1} \ldots \rho_{n}} ( \DLC_{\mu} \gamma^{\rho_{1} \ldots \rho_{n}} ) ] \notag
\\
 ={}& \sum\limits_{n=1}^{d} ( \DLC_{\mu} \varrho_{\nu \rho_{1} \ldots \rho_{n}} ) \gamma^{\rho_{1} \ldots \rho_{n}} - [ \hat{\Gamma}_{\mu} , \Delta \Gamma_{\nu} ] \text{.}
\end{align}
Analogously in odd dimensions
\begin{align}
 \DLC_{\mu} \Delta \Gamma_{\nu} = \sum\limits_{n=1}^{\frac{d-1}{2}} ( \DLC_{\mu} \varrho_{\nu \rho_{1} \ldots \rho_{2n}} ) \gamma^{\rho_{1} \ldots \rho_{2n}} - [ \hat{\Gamma}_{\mu} , \Delta \Gamma_{\nu} ]
\end{align}
holds.
Additionally it helps to rewrite the complete antisymmetrization
\begin{align}
 \tilde{\varrho}_{[\mu \rho_{1} \ldots \rho_{n}]} \! ={}& \! \frac{1}{n \! + \! 1} \! \left( \!\! \tilde{\varrho}_{\mu \rho_{1} \ldots \rho_{n}} \! + \! \sum\limits_{l=1}^{n} (-1)^{l} \tilde{\varrho}_{\rho_{l} \mu \rho_{1} \ldots \rho_{l - 1} \rho_{l + 1} \ldots \rho_{n}} \!\! \right)
\end{align}
of $\tilde{\varrho}_{\mu \rho_{1} \ldots \rho_{n}}$ as
\begin{align}
 \frac{1}{n} \sum\limits_{l=1}^{n} &(-1)^{l-1} \tilde{\varrho}_{\rho_{l} \mu \rho_{1} \ldots \rho_{l-1} \rho_{l+1} \ldots \rho_{n}} \notag
\\
 & = \frac{1}{n} \tilde{\varrho}_{\mu \rho_{1} \ldots \rho_{n}} - \frac{n + 1}{n} \tilde{\varrho}_{[\mu \rho_{1} \ldots \rho_{n}]} \text{.}
\end{align}
With the aid of the identities from App. \ref{App:Special_relations_gamma_matrices} and \ref{App:Special_relations_gamma_matrices_II} and the constraints for the $\varrho_{\mu \rho_{1} \ldots \rho_{n}}$ respectively $\varrho_{\mu \rho_{1} \ldots \rho_{2n}}$ it is then straightforward to calculate %$\mfrL_{\Phi}^{(1)}$
$\mfrL_{\Phi}$
in even dimensions
\begin{align}
%  \mfrL_{\Phi}^{(1)} ={}& \frac{1}{2} R_{(\mathrm{LC})} + 2 \sum\limits_{n=1}^{d} (-1)^{n} \cdot n! \cdot \Big[ \tilde{\varrho}^{\mu \rho_{1} \ldots \rho_{n}} \tilde{\varrho}_{\mu \rho_{1} \ldots \rho_{n}} \notag
% \\
%  & \! - \! ( n \! + \! 1 ) \tilde{\varrho}^{[\mu \rho_{1} \ldots \rho_{n}]} \tilde{\varrho}_{[\mu \rho_{1} \ldots \rho_{n}]} \! - \! n \tilde{\varrho}^{\mu \point \rho_{2} \ldots \rho_{n}}_{\point \mu} \tilde{\varrho}^{\nu}_{\point \nu \rho_{2} \ldots \rho_{n}} \Big]
 \mfrL_{\Phi} ={}& \frac{1}{2} R_{(\mathrm{LC})} + 2 \sum\limits_{n=1}^{d} (-1)^{n} \cdot n! \cdot \Big[ \tilde{\varrho}^{\mu \rho_{1} \ldots \rho_{n}} \tilde{\varrho}_{\mu \rho_{1} \ldots \rho_{n}} \notag
\\
 & \! - \! ( n \! + \! 1 ) \tilde{\varrho}^{[\mu \rho_{1} \ldots \rho_{n}]} \tilde{\varrho}_{[\mu \rho_{1} \ldots \rho_{n}]} \! - \! n \tilde{\varrho}^{\mu \point \rho_{2} \ldots \rho_{n}}_{\point \mu} \tilde{\varrho}^{\nu}_{\point \nu \rho_{2} \ldots \rho_{n}} \Big]
\end{align}
and in odd dimensions
\begin{align}
%  \mfrL_{\Phi}^{(1)} ={}& \frac{1}{2} R_{(\mathrm{LC})} + 2 \sum\limits_{n=1}^{\frac{d-1}{2}} (2n)! \cdot \Big[ \tilde{\varrho}^{\mu \rho_{1} \ldots \rho_{2n}} \tilde{\varrho}_{\mu \rho_{1} \ldots \rho_{2n}} \notag
% \\
%  & \hspace{2.5cm} - ( 2n + 1 ) \tilde{\varrho}^{[\mu \rho_{1} \ldots \rho_{2n}]} \tilde{\varrho}_{[\mu \rho_{1} \ldots \rho_{2n}]} \Big] \! \text{.}
 \mfrL_{\Phi} ={}& \frac{1}{2} R_{(\mathrm{LC})} + 2 \sum\limits_{n=1}^{\frac{d-1}{2}} (2n)! \cdot \Big[ \tilde{\varrho}^{\mu \rho_{1} \ldots \rho_{2n}} \tilde{\varrho}_{\mu \rho_{1} \ldots \rho_{2n}} \notag
\\
 & \hspace{2.5cm} - ( 2n + 1 ) \tilde{\varrho}^{[\mu \rho_{1} \ldots \rho_{2n}]} \tilde{\varrho}_{[\mu \rho_{1} \ldots \rho_{2n}]} \Big] \! \text{.}
\end{align}
While this is a compact form of the Lagrangian in terms of the $\tilde{\varrho}_{\mu \rho_{1} \ldots \rho_{n}}$ it is more convenient to rewrite it in a form which is respecting the constraints of the $\tilde{\varrho}_{\mu \rho_{1} \ldots \rho_{n}}$ explicitly.
We are dealing with tensors $T_{\mu \rho_{1} \ldots \rho_{n}}$ of the form $T_{\mu \rho_{1} \ldots \rho_{n}} = T_{\mu [\rho_{1} \ldots \rho_{n}]}$.
Hence, it is helpful to introduce the projectors onto the trace
\begin{align}
 (P_{\mathrm{Tr}}^{n})_{\mu \rho_{1} \ldots \rho_{n}}^{\point \alpha \lambda_{1} \ldots \lambda_{n}} = \frac{n (-1)^{n-1}}{d - (n-1)} \metric_{\mu [ \rho_{1}} \deltaA^{\lambda_{1} \ldots \lambda_{n}}_{\rho_{2} \ldots \rho_{n}] \nu} \metric^{\nu \alpha} \text{,}
\end{align}
as well as onto the totally antisymmetric part
\begin{align}
 (P_{\mathrm{A}}^{n})_{\mu \rho_{1} \ldots \rho_{n}}^{\point \alpha \lambda_{1} \ldots \lambda_{n}} = \deltaA_{\mu \rho_{1} \ldots \rho_{n}}^{\alpha \lambda_{1} \ldots \lambda_{n}} \text{,}
\end{align}
and onto the traceless part
\begin{align}
 (P_{\mathrm{TL}}^{n})_{\mu \rho_{1} \ldots \rho_{n}}^{\point \alpha \lambda_{1} \ldots \lambda_{n}} ={}& (\mathbb{1}^{n})_{\mu \rho_{1} \ldots \rho_{n}}^{\point \alpha \lambda_{1} \ldots \lambda_{n}} - (P_{\mathrm{Tr}}^{n})_{\mu \rho_{1} \ldots \rho_{n}}^{\point \alpha \lambda_{1} \ldots \lambda_{n}} \notag
\\
 & - (P_{\mathrm{A}}^{n})_{\mu \rho_{1} \ldots \rho_{n}}^{\point \alpha \lambda_{1} \ldots \lambda_{n}}
\end{align}
of such a tensor.
The identity projector reads
\begin{align}
 (\mathbb{1}^{n})_{\mu \rho_{1} \ldots \rho_{n}}^{\point \alpha \lambda_{1} \ldots \lambda_{n}} = \delta_{\mu}^{\alpha} \deltaA_{\rho_{1} \ldots \rho_{n}}^{\lambda_{1} \ldots \lambda_{n}} \text{.}
\end{align}
The projector properties such as idempotence%
\footnote{
The product $(P_{\mathrm{Tr}}^{n}) (P_{\mathrm{A}}^{n})$ of two projectors $(P_{\mathrm{Tr}}^{n})$ and $(P_{\mathrm{A}}^{n})$ is defined as $[(P_{\mathrm{Tr}}^{n}) (P_{\mathrm{A}}^{n})]_{\mu \rho_{1} \ldots \rho_{n}}^{\point \alpha \lambda_{1} \ldots \lambda_{n}} = (P_{\mathrm{Tr}}^{n})_{\mu \rho_{1} \ldots \rho_{n}}^{\point \beta \kappa_{1} \ldots \kappa_{n}}  (P_{\mathrm{A}}^{n})_{\beta \kappa_{1} \ldots \kappa_{n}}^{\point \alpha \lambda_{1} \ldots \lambda_{n}}$.
}
\begin{align}
 &(P_{\mathrm{Tr}}^{n})^{2} = (P_{\mathrm{Tr}}^{n}), \quad (P_{\mathrm{A}}^{n})^{2} = (P_{\mathrm{A}}^{n}), \quad (P_{\mathrm{TL}}^{n})^{2} = (P_{\mathrm{TL}}^{n}),
\end{align}
orthogonality
\begin{align}
\begin{aligned} 
&(P_{\mathrm{Tr}}^{n}) (P_{\mathrm{A}}^{n}) = (P_{\mathrm{A}}^{n}) (P_{\mathrm{Tr}}^{n}) = 0 \text{,}
\\
 &(P_{\mathrm{Tr}}^{n}) (P_{\mathrm{TL}}^{n}) = (P_{\mathrm{TL}}^{n}) (P_{\mathrm{Tr}}^{n}) = 0 \text{,}
\\
 &(P_{\mathrm{A}}^{n}) (P_{\mathrm{TL}}^{n}) = (P_{\mathrm{TL}}^{n}) (P_{\mathrm{A}}^{n}) = 0 \text{,}
\end{aligned}
\end{align}
and the partition of unity
\begin{align}
\begin{aligned}
(P_{\mathrm{Tr}}^{n}) + (P_{\mathrm{A}}^{n}) + (P_{\mathrm{TL}}^{n}) = (\mathbb{1}^{n}) \text{,}
\end{aligned}
\end{align}
are easily checked with the aid of
\begin{align}
 \deltaA^{\alpha_{1} \ldots \alpha_{n-1} \mu}_{\beta_{1} \ldots \beta_{n-1} \mu} = \frac{d - (n-1)}{n} \deltaA^{\alpha_{1} \ldots \alpha_{n-1}}_{\beta_{1} \ldots \beta_{n-1}} \text{.}
\end{align}
We denote the trace of $\tilde{\varrho}_{\mu \rho_{1} \ldots \rho_{n}}$ by $\tilde{\phi}_{\rho_{2} \ldots \rho_{n}}$,
\begin{align}
 \tilde{\phi}_{\rho_{2} \ldots \rho_{n}} = \tilde{\varrho}^{\mu}_{\point \mu \rho_{2} \ldots \rho_{n}} \text{,}
\end{align}
the antisymmetric part by $(\tilde{\varrho}_{\mathrm{A}})_{\mu \rho_{1} \ldots \rho_{n}}$,
\begin{align}
 (\tilde{\varrho}_{\mathrm{A}})_{\mu \rho_{1} \ldots \rho_{n}} = (P_{\mathrm{A}}^{n})_{\mu \rho_{1} \ldots \rho_{n}}^{\point \alpha \lambda_{1} \ldots \lambda_{n}} \tilde{\varrho}_{\alpha \lambda_{1} \ldots \lambda_{n}} \text{,}
\end{align}
and the traceless part by $(\tilde{\varrho}_{\mathrm{TL}})_{\mu \rho_{1} \ldots \rho_{n}}$,
\begin{align}
 (\tilde{\varrho}_{\mathrm{TL}})_{\mu \rho_{1} \ldots \rho_{n}} = (P_{\mathrm{TL}}^{n})_{\mu \rho_{1} \ldots \rho_{n}}^{\point \alpha \lambda_{1} \ldots \lambda_{n}} \tilde{\varrho}_{\alpha \lambda_{1} \ldots \lambda_{n}} \text{.}
\end{align}
Then we can decompose $\tilde{\varrho}_{\mu \rho_{1} \ldots \rho_{n}}$ as
\begin{align}
 \tilde{\varrho}_{\mu \rho_{1} \ldots \rho_{n}} \! ={}& \! (\tilde{\varrho}_{\mathrm{TL}})_{\mu \rho_{1} \ldots \rho_{n}} + (\tilde{\varrho}_{\mathrm{A}})_{\mu \rho_{1} \ldots \rho_{n}} \notag
\\
 & + \frac{n}{d - (n - 1)} \metric_{\mu [ \rho_{1}} \tilde{\phi}_{\rho_{2} \ldots \rho_{n}]} \text{.}
\end{align}
Especially the square of $\tilde{\varrho}_{\mu \rho_{1} \ldots \rho_{n}}$ then reads
\begin{align}
 \tilde{\varrho}^{\mu \rho_{1} \ldots \rho_{n}} \tilde{\varrho}_{\mu \rho_{1} \ldots \rho_{n}} \! =& ( \tilde{\varrho}_{\mathrm{TL}} )^{\mu \rho_{1} \ldots \rho_{n}} ( \tilde{\varrho}_{\mathrm{TL}} )_{\mu \rho_{1} \ldots \rho_{n}} \notag
\\
 &+ ( \tilde{\varrho}_{\mathrm{A}} )^{\mu \rho_{1} \ldots \rho_{n}} ( \tilde{\varrho}_{\mathrm{A}} )_{\mu \rho_{1} \ldots \rho_{n}} \notag
\\
 & + \frac{n}{d - (n - 1)} \tilde{\phi}^{\rho_{2} \ldots \rho_{n}} \tilde{\phi}_{\rho_{2} \ldots \rho_{n}} \text{.}
\end{align}
In other words, the trace, the antisymmetric part and the traceless part decouple from each other.

The Lagrangian in these variables in even dimensions then reads
\begin{align}
%  \mfrL_{\Phi}^{(1)} ={}& \frac{1}{2} R_{(\mathrm{LC})} + 2 \sum\limits_{n=1}^{d} (-1)^{n} n! \cdot (\tilde{\varrho}_{\mathrm{TL}})^{\mu \rho_{1} \ldots \rho_{n}} (\tilde{\varrho}_{\mathrm{TL}})_{\mu \rho_{1} \ldots \rho_{n}} \notag
% \\
%  & - 2 \sum\limits_{n=1}^{\frac{d}{2}} 2n \cdot (2n)! \cdot (\tilde{\varrho}_{\mathrm{A}})^{\mu \rho_{1} \ldots \rho_{2n}} (\tilde{\varrho}_{\mathrm{A}})_{\mu \rho_{1} \ldots \rho_{2n}} \notag
% \\
%  & + 2 \sum\limits_{n=1}^{\frac{d}{2}} (2n \! - \! 1) \! \cdot \! (2n \! - \! 1)! \! \cdot \! \frac{d \! - \! 2n \! + \! 1}{d \! - \! 2 n \! + \! 2 } \tilde{\phi}^{\rho_{2} \ldots \rho_{n}} \tilde{\phi}_{\rho_{2} \ldots \rho_{n}}
 \mfrL_{\Phi} ={}& \frac{1}{2} R_{(\mathrm{LC})} + 2 \sum\limits_{n=1}^{d} (-1)^{n} n! \cdot (\tilde{\varrho}_{\mathrm{TL}})^{\mu \rho_{1} \ldots \rho_{n}} (\tilde{\varrho}_{\mathrm{TL}})_{\mu \rho_{1} \ldots \rho_{n}} \notag
\\
 & - 2 \sum\limits_{n=1}^{\frac{d}{2}} 2n \cdot (2n)! \cdot (\tilde{\varrho}_{\mathrm{A}})^{\mu \rho_{1} \ldots \rho_{2n}} (\tilde{\varrho}_{\mathrm{A}})_{\mu \rho_{1} \ldots \rho_{2n}} \notag
\\
 & + 2 \sum\limits_{n=1}^{\frac{d}{2}} (2n \! - \! 1) \! \cdot \! (2n \! - \! 1)! \! \cdot \! \frac{d \! - \! 2n \! + \! 1}{d \! - \! 2 n \! + \! 2 } \tilde{\phi}^{\rho_{2} \ldots \rho_{n}} \tilde{\phi}_{\rho_{2} \ldots \rho_{n}} \text{,}
\end{align}
and in odd dimensions we have
\begin{align}
%  \mfrL_{\Phi}^{(1)} ={}& \frac{1}{2} R_{(\mathrm{LC})} + 2 \sum\limits_{n=1}^{\frac{d-1}{2}} (2n)! \cdot (\tilde{\varrho}_{\mathrm{TL}})^{\mu \rho_{1} \ldots \rho_{2n}} (\tilde{\varrho}_{\mathrm{TL}})_{\mu \rho_{1} \ldots \rho_{2n}} \notag
% \\
%  & - 2 \sum\limits_{n=1}^{\frac{d-1}{2}} 2n \cdot (2n)! \cdot (\tilde{\varrho}_{\mathrm{A}})^{\mu \rho_{1} \ldots \rho_{2n}} (\tilde{\varrho}_{\mathrm{A}})_{\mu \rho_{1} \ldots \rho_{2n}} \text{.}
 \mfrL_{\Phi} ={}& \frac{1}{2} R_{(\mathrm{LC})} + 2 \sum\limits_{n=1}^{\frac{d-1}{2}} (2n)! \cdot (\tilde{\varrho}_{\mathrm{TL}})^{\mu \rho_{1} \ldots \rho_{2n}} (\tilde{\varrho}_{\mathrm{TL}})_{\mu \rho_{1} \ldots \rho_{2n}} \notag
\\
 & - 2 \sum\limits_{n=1}^{\frac{d-1}{2}} 2n \cdot (2n)! \cdot (\tilde{\varrho}_{\mathrm{A}})^{\mu \rho_{1} \ldots \rho_{2n}} (\tilde{\varrho}_{\mathrm{A}})_{\mu \rho_{1} \ldots \rho_{2n}} \text{.}
\end{align}
In this form it is apparent that the resulting classical equations of motion after varying with respect to the spin torsion degrees of freedom are purely algebraic in the fields $(\tilde{\varrho}_{\mathrm{TL}})_{\mu \rho_{1} \ldots \rho_{n}}$, $(\tilde{\varrho}_{\mathrm{A}})_{\mu \rho_{1} \ldots \rho_{n}}$ and $\tilde{\phi}_{\rho_{2} \ldots \rho_{n}}$.
Therefore the spin torsion vanishes classically in the absence of, e.g.~spinorial sources.
The variation with respect to the metric %and the gauge fields $\mcA_{\mu}$
gives us the usual Einstein field equations.% in the presence of a $\mathrm{U}(1)$ gauge field.

\section{Lorentz symmetric gauge}
\label{sec:lorentz_symmetric_gauge}
%-----------------------------------------------------------------------------

In the usual vielbein setup one often needs the vielbein $\vielbein_{\mu}^{\point a}$ as a function of the metric $\metric_{\mu \nu}$ with respect to some fixed but arbitrary background metric $\bar{\metric}_{\mu \nu}$ and background vielbein $\bar{\vielbein}_{\mu}^{\point a}$.
Such relations define a gauge for the vielbein.
It turned out that the Lorentz-symmetric gauge is very useful and minimizes in practice the calculational effort \cite{Eichhorn:2011pc,Dona:2013qba,Dona:2014pla}.
In particular, corresponding $\mathrm{SO}(3,1)$ Faddeev-Popov ghosts do not contribute in perturbation theory \cite{Woodard:1984sj,vanNieuwenhuizen:1981uf}.
An interesting application of the generalized Weldon theorem \Eqref{eq:App:Weldon_Theorem} is the derivation of the analog of the Lorentz symmetric gauge for the Dirac matrices as the \textit{simplest}%
\footnote{
The notion of ``\textit{simplest}'' here will become apparent below. It means the least possible change of the Dirac structure while going from $\bar{\gamma}_{\mu}$ to $\gamma_{\mu}$.
}
possible choice (gauge) of the Dirac matrices $\gamma_{\mu} = \gamma_{\mu}(\metric)$.
We show how this is done in the following.
% Afterwards we give an outlook to a generalization of this gauge connecting two arbitrary metrics, which do not have to be in an infinitesimal neighborhood of each other.

With $\flucmet_{\mu \nu}$ we denote a metric fluctuation which parametrizes the full metric $\metric_{\mu \nu}$ with respect to an unspecified (arbitrary) background metric $\bar{\metric}_{\mu \nu}$,
\begin{align}
 \metric_{\mu \nu} = \bar{\metric}_{\mu \nu} + \flucmet_{\mu \nu} \text{.}
\end{align}
The background Dirac matrices are denoted with $\bar{\gamma}_{\mu}$.%
\footnote{During this section we will use the bar as a reference to the background. Especially the background Dirac matrices $\bar{\gamma}_{\mu} = \gamma_{\mu}(\metric = \bar{\metric})$ should not be confused with the Dirac conjugation of the Dirac matrices $\spinmetric^{-1} \gamma_{\mu}^{\dagger} \spinmetric$ which had the same bar notation in the previous chapters but are not present here.
}
We assume that we can expand%
\footnote{
We discuss the situation for a nonexpandable metric at the end of this section.
}
$\gamma_{\mu}(\metric)$ in powers of the fluctuation $\flucmet_{\mu \nu}$
\begin{align}\label{eq:gamma_matrix_flucmet_expansion}
 \gamma_{\mu}(\metric) = \sum\limits_{n=0}^{\infty} \!\! \left. \frac{\partial^{n} \gamma_{\mu}(\metric)}{n! \partial \metric_{\nu_{1} \nu_{2}} \ldots \partial \metric_{\nu_{2n-1} \nu_{2n}}} \right|_{\metric = \bar{\metric}} \!\! \flucmet_{\nu_{1} \nu_{2}} \ldots \flucmet_{\nu_{2n-1} \nu_{2n}} \text{.}
\end{align}
Using the generalized Weldon theorem \Eqref{eq:App:Weldon_Theorem} we can write
\begin{align}\label{eq:gamma_first_metric_derivative}
 \frac{\partial \gamma_{\mu}(\metric)}{\partial \metric_{\nu_{1} \nu_{2}}} = \frac{1}{2} \deltaS^{\nu_{1} \nu_{2}}_{\mu \rho} \gamma^{\rho}(\metric) + [ G^{\nu_{1} \nu_{2}}(\metric) , \gamma_{\mu}(\metric) ],
\end{align}
where $\deltaS^{\nu_{1} \ldots \nu_{m}}_{\mu_{1} \ldots \mu_{m}}$ is the normalized and symmetrized Kronecker delta and $G^{\nu_{1} \nu_{2}}$ is a Dirac valued function of the metric encoding the gauge choice.
I.e.~by fixing $G^{\nu_{1} \nu_{2}}(\metric)$ and $\bar{\gamma}_{\mu} = \gamma_{\mu}(\bar{\metric})$ we completely fix the function $\gamma_{\mu} = \gamma_{\mu}(\metric)$.
Since there is no preferred choice of Dirac matrices for a given metric, we can leave the background Dirac matrices $\bar{\gamma}_{\mu}$ arbitrary while compatible with the Clifford algebra.
We aim at optimizing the function $G^{\nu_{1} \nu_{2}}$ such that \Eqref{eq:gamma_matrix_flucmet_expansion} becomes as simple as possible.

In order to do so we expand $G^{\nu_{1} \nu_{2}}(\metric)$ in powers of the metric fluctuations
\begin{align}
 G^{\nu_{1} \nu_{2}} = \sum\limits_{n=0}^{\infty} \frac{1}{n!} \bar{G}^{\nu_{1} \nu_{2} \lambda_{1} \ldots \lambda_{2n}} \flucmet_{\lambda_{1} \lambda_{2}} \ldots \flucmet_{\lambda_{2n-1} \lambda_{2n}},
\end{align}
where the $\bar{G}^{\nu_{1} \nu_{2} \lambda_{1} \ldots \lambda_{2n}}$ are the expansion coefficients to be determined.
Since we aim at simplifying the function \Eqref{eq:gamma_matrix_flucmet_expansion} we have to simplify the derivatives of the Dirac matrices.
Looking at the first nontrivial term we find
\begin{align}
 \left. \frac{\partial \gamma_{\mu}(\metric)}{\partial \metric_{\nu_{1} \nu_{2}}} \right|_{\metric = \bar{\metric}} = \frac{1}{2} \deltaS^{\nu_{1} \nu_{2}}_{\mu \rho} \bar{\gamma}^{\rho} + [ \bar{G}^{\nu_{1} \nu_{2}} , \bar{\gamma}_{\mu} ] \text{.}
\end{align}
Taking into account that the symmetric part and the commutator part are completely independent, it is obvious that the best simplification we can find is $\bar{G}^{\nu_{1} \nu_{2}} = 0$.
With this we can go on to the second derivative
\begin{align}
 \left. \frac{\partial^{2} \gamma_{\mu}(\metric)}{\partial \metric_{\nu_{1} \nu_{2}} \partial \metric_{\nu_{3} \nu_{4}}} \right|_{\metric = \bar{\metric}} = - \omega^{\nu_{1} \ldots \nu_{4}}_{\point[4] \mu \rho} \bar{\gamma}^{\rho} + [ \bar{G}^{\nu_{1} \ldots \nu_{4}} , \bar{\gamma}_{\mu} ],
\end{align}
where $\omega^{\nu_{1} \ldots \nu_{4}}_{\point[4] \mu \rho} = \frac{1}{4} \deltaS^{\nu_{1} \nu_{2}}_{\mu \kappa} \bar{\metric}^{\kappa \sigma} \deltaS^{\nu_{3} \nu_{4}}_{\sigma \rho}$.
The tensor $\omega^{\nu_{1} \ldots \nu_{4}}_{\point[4] \mu \rho}$ has a symmetric and an antisymmetric part concerning the pair $(\mu, \rho)$.
Using that we can rewrite the antisymmetric part as a commutator.
We find
\begin{align}
 & \! \left. \frac{\partial^{2} \gamma_{\mu}(\metric)}{\partial \metric_{\nu_{1} \nu_{2}} \partial \metric_{\nu_{3} \nu_{4}}} \right|_{\metric = \bar{\metric}} \notag
\\
 &= \! - \omega^{\nu_{1} \ldots \nu_{4}}_{\point[4] (\mu \rho)} \bar{\gamma}^{\rho} \! + \! \left[ \bar{G}^{\nu_{1} \ldots \nu_{4}} \! + \! \frac{1}{8} \omega^{\nu_{1} \ldots \nu_{4}}_{\point[4] [\lambda_{1} \lambda_{2}]} [\bar{\gamma}^{\lambda_{1}} , \bar{\gamma}^{\lambda_{2}}] , \bar{\gamma}_{\mu} \right] \!\! ,
\end{align}
where $(\ldots)$ denotes the normalized symmetrization of indices.
Once again we have two independent terms and we find that the simplest choice is $\bar{G}^{\nu_{1} \ldots \nu_{4}} = - \frac{1}{8} \omega^{\nu_{1} \ldots \nu_{4}}_{\point[4] [\lambda_{1} \lambda_{2}]} [ \bar{\gamma}^{\lambda_{1}} , \bar{\gamma}^{\lambda_{2}} ]$.

We can iterate this process of identifying the symmetric part and the commutator part and eliminate the commutator part by appropriate choices of the $\bar{G}^{\nu_{1} \nu_{2} \lambda_{1} \ldots \lambda_{2n}}$.
By doing this we end up with an expansion of the Dirac matrices $\gamma_{\mu}(\metric)$ which is directly proportional to the background Dirac matrices $\bar{\gamma}_{\mu}$
\begin{align}\label{eq:gamma_expansion_linear_bargamma}
 \gamma_{\mu} = \sum\limits_{n=0}^{\infty} \frac{(-1)^{n-1}}{n!} (\omega_{n})^{\nu_{1} \ldots \nu_{2n}}_{\point[5] \mu \rho} \bar{\gamma}^{\rho} \flucmet_{\nu_{1} \nu_{2}} \ldots \flucmet_{\nu_{2n-1} \nu_{2n}} \text{.}
\end{align}
The $(\omega_{n})^{\nu_{1} \ldots \nu_{2n}}_{\point[5] \mu \rho} = (\omega_{n})^{\nu_{1} \ldots \nu_{2n}}_{\point[5] (\mu \rho)}$ can be calculated recursively applying the above given construction of the $\bar{G}^{\nu_{1} \nu_{2} \lambda_{1} \ldots \lambda_{2n}}$, where we already know the first three of them
\begin{align}\label{eq:omega_recursion_start_condition_0}
 &(\omega_{0})_{\mu \rho} = - \bar{\metric}_{\mu \rho},
\\\label{eq:omega_recursion_start_condition_1}
 &(\omega_{1})^{\nu_{1} \nu_{2}}_{\point[2] \mu \rho} = \frac{1}{2} \deltaS^{\rho_{1} \rho_{2}}_{\mu \rho},
\\\label{eq:omega_recursion_start_condition_2}
 &(\omega_{2})^{\nu_{1} \ldots \nu_{4}}_{\point[4] \mu \rho} = \omega^{\nu_{1} \ldots \nu_{4}}_{\point[4] (\mu \rho)} \text{.}
\end{align}
Unfortunately it is difficult to perform this iteration to all orders.
To circumvent this problem we remind ourselves that equation \Eqref{eq:gamma_first_metric_derivative} is a consequence of the Clifford algebra and insert the simplified ansatz \Eqref{eq:gamma_expansion_linear_bargamma} into the Clifford algebra
\begin{align}
 2 (\bar{\metric}_{\mu \nu} + \flucmet_{\mu \nu} ) \mrI ={}& \sum\limits_{n=0}^{\infty} \sum\limits_{m=0}^{\infty} \frac{(-1)^{n+m}}{n! \cdot m!} \{ \bar{\gamma}^{\alpha} , \bar{\gamma}^{\beta} \} \notag 
\\
 & \times (\omega_{n})^{\rho_{1} \ldots \rho_{2n}}_{\point[5] \mu \alpha} \flucmet_{\rho_{1} \rho_{2}} \ldots \flucmet_{\rho_{2n-1} \rho_{2n}} \notag
\\
 & \times  (\omega_{m})^{\lambda_{1} \ldots \lambda_{2m}}_{\point[5] \mu \beta} \flucmet_{\lambda_{1} \lambda_{2}} \ldots \flucmet_{\lambda_{2m-1} \lambda_{2m}} \text{.}
\end{align}
We can reorder the sums such that we sum over the powers of the fluctuations in increasing order.
For this we introduce the new summation variables $(s,l) = (n + m,m)$, where $l \in \{ 0, \ldots, s\}$ and $s \in \{ 0, \ldots , \infty\}$ and the shorthand
\begin{align}
 (\omega_{m} \flucmet^{m})^{\rho}_{\point \lambda} = \bar{\metric}^{\rho \alpha} (\omega_{m})^{\nu_{1} \ldots \nu_{2m}}_{\point[5] \alpha \lambda} \flucmet_{\nu_{1} \nu_{2}} \ldots \flucmet_{\nu_{2m-1} \nu_{2m}} \text{.}
\end{align}
Then we find
\begin{align}
 \bar{\metric}_{\mu \nu} \! + \! \flucmet_{\mu \nu} = \sum\limits_{s=0}^{\infty} \frac{(-1)^{s}}{s!} \sum\limits_{l=0}^{s} \binom{s}{l} \bar{\metric}_{\mu \kappa} (\omega_{s-l} \flucmet^{s-l})^{\kappa}_{\point \alpha} (\omega_{l} \flucmet^{l})^{\alpha}_{\point \nu} \text{.}
\end{align}
Since this equation has to be true for each power of $\flucmet_{\mu \nu}$ individually the equation splits into three parts.
First we have from $s=0$
\begin{align}
 \delta^{\mu}_{\nu} = (\omega_{0} \flucmet^{0})^{\mu}_{\point \alpha} (\omega_{0} \flucmet^{0})^{\alpha}_{\point \nu},
\end{align}
which is of course satisfied, cf.~ \Eqref{eq:omega_recursion_start_condition_0}.
For $s=1$ we find again just a trivially satisfied equation
\begin{align}
 \metric^{\mu \alpha} \flucmet_{\alpha \nu} = - (\omega_{1} \flucmet^{1})^{\mu}_{\point \alpha} (\omega_{0} \flucmet^{0})^{\alpha}_{\point \nu} - (\omega_{0} \flucmet^{0})^{\mu}_{\point \alpha} (\omega_{1} \flucmet^{1})^{\alpha}_{\point \nu},
\end{align}
cf.~\Eqref{eq:omega_recursion_start_condition_1}.
The last part is $s \geq 2$
\begin{align}
 0 = \sum_{l=0}^{s} \binom{s}{l} (\omega_{s-l} \flucmet^{s-l})^{\mu}_{\point \alpha} (\omega_{l} \flucmet^{l})^{\alpha}_{\point \nu},
\end{align}
which we can rewrite as
\begin{align}\label{eq:recursion_omega_n}
 (\omega_{s} \flucmet^{s})^{\mu}_{\point \nu} = \frac{1}{2} \sum\limits_{l=1}^{s-1} \binom{s}{l} (\omega_{s-l} \flucmet^{s-l})^{\mu}_{\point \alpha} (\omega_{l} \flucmet^{l})^{\alpha}_{\point \nu}
\end{align}
by splitting off the $l=0$ and $l=s$ parts and using \Eqref{eq:omega_recursion_start_condition_0}.
In other words we have a recursion relation with initial conditions \Eqref{eq:omega_recursion_start_condition_0} and \eqref{eq:omega_recursion_start_condition_1}.
This recursion obviously has a unique solution.
With the initial conditions we can show by induction that
\begin{align}
 (\omega_{s} \flucmet^{s})^{\mu}_{\point \nu} = c_{s} \cdot \flucmet^{\mu}_{\point \rho_{1}} \ldots \flucmet^{\rho_{s-1}}_{\point[3] \nu}, \quad s \geq 1 \text{,}
\end{align}
where $\flucmet^{\mu}_{\point \nu} = \bar{\metric}^{\mu \rho} \flucmet_{\rho \nu}$ and $c_{s}$ are just numbers to be determined.
Plugging our result into \Eqref{eq:recursion_omega_n} we get a recursion for the $c_{s}$
\begin{align}
 c_{s} = \frac{1}{2} \sum\limits_{l=1}^{s-1} \binom{s}{l} c_{s-l} c_{l} , \quad s \geq 2,
\end{align}
with initial condition $c_{1} = \frac{1}{2}$.
The explicit solution of this recursion reads
\begin{align}
 c_{s} = (-1)^{s-1} \frac{ \Gamma \! \left( \frac{3}{2} \right) }{ \Gamma \! \left( \frac{3}{2} - s \right) } ,
\end{align}
which can be shown by induction again.
As a result we have
\begin{align}
 \gamma_{\mu}(\metric) ={}& \sum\limits_{n=0}^{\infty} \frac{(-1)^{n-1}}{n!} \bar{\metric}_{\mu \kappa} (\omega_{n}\flucmet^{n})^{\kappa}_{\point \lambda} \bar{\gamma}^{\lambda} \notag
\\
 ={}& \bar{\metric}_{\mu \kappa} \sum\limits_{n=0}^{\infty} \frac{\Gamma \! \left( \frac{3}{2} \right)}{\Gamma \! \left( n + 1 \right) \Gamma \! \left( \frac{3}{2} - n \right)} (\flucmet^{n})^{\kappa}_{\point \lambda} \bar{\gamma}^{\lambda} \text{,}
\end{align}
where $(\flucmet^{n})^{\kappa}_{\point \lambda} = \flucmet^{\kappa}_{\point \rho_{1}} \flucmet^{\rho_{1}}_{\point \rho_{2}} \ldots \flucmet^{\rho_{n-2}}_{\point[3] \rho_{n-1}} \flucmet^{\rho_{n-1}}_{\point[3] \lambda}$.
This sum is exactly the series representation of the square root and we can write formally
\begin{align}
 \gamma_{\mu} ( \metric ) = \bar{\metric}_{\mu \kappa} [ \sqrt{\delta + \flucmet} ]^{\kappa}_{\point \lambda} \bar{\gamma}^{\lambda} \text{.}
\end{align}
This is precisely the representation given by Woodard for the vielbein \cite{Woodard:1984sj}
\begin{align}
 \vielbein_{\mu}^{\point a} = \bar{\metric}_{\mu \kappa} [ \sqrt{\delta + \flucmet} ]^{\kappa}_{\point \lambda} \bar{\vielbein}^{\lambda a} \text{.}
\end{align}
This calculation illustrates \textit{why} the Lorentz symmetric gauge proved so useful.

In view of contemporary nonperturbative quantum gravity calculations, an urgent question arises \cite{Pawlowski:2003sk,Manrique:2009uh,Codello:2013fpa,Becker:2014qya,Christiansen:2014raa,Deffayet:2012zc}.
Is there a way to fix the gauge without assuming that $\gamma_{\mu}(\metric)$ is expandable in the metric fluctuation $\flucmet_{\mu \nu}$ around the background metric $\bar{\metric}_{\mu \nu}$.
We will give a possibility here.

Let us assume two given metrics $\metric_{\mu \nu}$ and $\bar{\metric}_{\mu \nu}$,
%where $\metric_{\mu \nu}$ is supposed to be the full metric and $\bar{\metric}_{\mu \nu}$ the background metric as before.
% With $\gamma_{\mu}(\metric)$ we denote the Dirac matrix for $\metric_{\mu \nu}$ and with $\bar{\gamma}_{\mu}$ the Dirac matrix for $\bar{\metric}$.
with the same notation as before.
We have seen, that we can tune the spin-base such that the full Dirac matrices $\gamma_{\mu}(\metric)$ and the background Dirac matrices $\bar{\gamma}_{\mu}$ are related in a linear way $\gamma_{\mu} \sim \bar{\gamma}_{\nu}$.
Let us take such a form as an ansatz to find a nonperturbative gauge
\begin{align}
 \gamma_{\mu}(\metric) = \mcB_{\mu \nu}(\metric, \bar{\metric}) \bar{\gamma}^{\nu}, \quad \mcB_{\mu \nu}(\metric, \bar{\metric}) = \mcB_{\nu \mu}(\metric, \bar{\metric}) \text{,}
\end{align}
where we have to determine the complex functions $\mcB_{\mu \nu}(\metric, \bar{\metric})$.
The symmetry of $\mcB_{\mu \nu}(\metric, \bar{\metric} )$ is in the same spirit as our construction from above and is supposed to ensure the simplicity.
Plugging this ansatz into the Clifford algebra we find
\begin{align}\label{eq:mcB_in_clifford_in_coords}
 \metric_{\mu \nu} = \mcB_{\mu \rho}(\metric, \bar{\metric}) \bar{\metric}^{\rho \kappa} \mcB_{\nu \kappa}(\metric, \bar{\metric}) \text{.}
\end{align}
For clarity, we switch to an intuitive matrix formulation $\metric_{\mu \nu} \to \metric$, $\mcB_{\mu \nu}(\metric, \bar{\metric}) \to \mcB$ and additionally drop the arguments $(\metric, \bar{\metric})$ from now on.
By using the symmetry $\mcB  = \mcB^{\mrT}$ we can rewrite \Eqref{eq:mcB_in_clifford_in_coords}
\begin{align}
 \metric = \mcB \bar{\metric}^{-1} \mcB^{\mrT} = \mcB \bar{\metric}^{-1} \mcB  = \bar{\metric} (\bar{\metric}^{-1} \mcB)^{2} \text{.}
\end{align}
Therefore $\bar{\metric}^{-1} \mcB$ has to be a square root of $\bar{\metric}^{-1} \metric$, compatible with the symmetry condition.%
\footnote{
Note that for $\metric = \bar{\metric} + \flucmet$ we get the perturbative result from above $\mcB = \bar{\metric} \sqrt{\mrI + \bar{\metric}^{-1} \flucmet}$.
}
To simplify the structure we use that $\bar{\metric}$ is a real symmetric matrix and therefore has a (nonunique) symmetric square root $\chi$
\begin{align}
 \bar{\metric} = \chi^{2} , \quad \chi^{\mrT} = \chi \text{.}
\end{align}
Depending on the signature $\chi$ can be complex.
Then it follows that $\chi^{-1} \mcB \chi^{-1}$ is symmetric as long as $\mcB$ is symmetric.
Hence, we arrive at
\begin{align}
 (\chi^{-1} \mcB \chi^{-1})^{2} = \chi^{-1} \metric \chi^{-1},
\end{align}
where $\chi^{-1} \metric \chi^{-1}$ is obviously a symmetric matrix as well.
This is a quite comfortable situation, as we are looking for a symmetric square root of a symmetric matrix.
If we suppose there is a symmetric square root $\kappa$ of $\chi^{-1} \metric \chi^{-1}$
\begin{align}
 \chi^{-1} \metric \chi^{-1} = \kappa^{2} , \quad \kappa^{\mrT} = \kappa,
\end{align}
then we have a solution $\mcB$ with
\begin{align}\label{eq:solution_mcB}
 \mcB = \chi \kappa \chi
\end{align}
as can be checked easily.
Especially for the recently become prominent exponential parametrization \cite{Kawai:1993fq,Eichhorn:2013xr,Nink:2014yya,Percacci:2015wwa}
\begin{align}\label{eq:exponential_parametrization_metric}
 \metric = \bar{\metric} \euler^{\bar{\metric}^{-1} \flucmet} \text{,} \quad \flucmet^{\mrT} = \flucmet \text{,}
\end{align}
we find
\begin{align}\label{eq:exponential_solution_kappa}
 \kappa ={}& \euler^{\frac{1}{2} \chi^{-1} \flucmet \chi^{-1}}
\\ \label{eq:exponential_solution_mcB}
 \mcB ={}& \bar{\metric} \euler^{\frac{1}{2} \bar{\metric}^{-1} \flucmet} \text{.}
\end{align}
Unfortunately in general there is no guarantee that for a complex symmetric matrix a symmetric square root exists.
Still, any Euclidean metric corresponds to a symmetric, positive definite matrix.
Hence, there is a unique, symmetric, positive definite $\chi$.
Therefore we also have a unique, symmetric, positive definite $\kappa$, leading to a unique $\mcB$ given by \Eqref{eq:solution_mcB}.
As proven in App A of \cite{Nink:2014yya} one can uniquely parametrize any Euclidean metric $\metric$ by \Eqref{eq:exponential_parametrization_metric}, hence $\kappa$ and $\mcB$ are given by \Eqref{eq:exponential_solution_kappa} and \eqref{eq:exponential_solution_mcB}.

In general dimensions the situation for the Lorentzian signature is unclear so far.
The problem stems from the minus sign in the signature of the metric leading to a complex $\chi$.
The first nontrivial dimension is $d=2$.
One can show, however, that the only complex symmetric $2 \times 2$ matrices without symmetric square root are of the form $c \cdot \begin{pmatrix} \pm \cplx & 1 \\ 1 & \mp \cplx \end{pmatrix}$, with $c \in \C \backslash \{ 0 \}$.
Fortunately these matrices have vanishing determinant guaranteeing the existence of the symmetric square root of $\chi^{-1} \metric \chi^{-1}$ at least in two dimensions independent of the signature.
One can hope that this generalizes somehow to arbitrary integer dimensions $d \geq 2$, but this is beyond the scope of this paper.
% and signatures of the form $( - , + , \ldots , + )$.
% This could be possible e.g.~in an inductive way.
% Splitting the metric into the $2 \times 2$ block with signature $(-,+)$ and the rest of the metric which is strictly positive.
% For the strictly positive part everything is safe, and the problematic $2 \times 2$ block is protected by the nonvanishing determinant of the metric.

%-----------------------------------------------------------------------------
\section{Conclusion}
\label{sec:conc}
%-----------------------------------------------------------------------------

In this paper we have generalized the results of \cite{Gies:2013noa} to arbitrary integer dimensions $d \geq 2$.
It was demonstrated how the concept of spin-base invariance arises naturally from completely standard considerations.
We have pointed out the hidden spin-base invariance of the vielbein formalism and have shown how it artificially splits the full Dirac matrices into a vielbein and flat Dirac matrices.
Especially we have presented how the generalized Weldon theorem allows us to formulate a purely metric based description of fermions.

We have constructed all relevant quantities for the description of fermions in curved spacetimes from the Dirac matrices.
It is obvious that every manifold that admits a global vielbein also admits global Dirac matrices, but as shown in \cite{Gies:2015cka} the converse is not true.
The $2$-sphere serves as a simple example how our approach generalizes the usual vielbein formalism.
We stress that the vielbein formalism, if applicable, is always a special choice of Dirac matrices and therefore completely covered by our approach as long as there is no torsion.
Additionally to spacetime torsion the spin connection can carry spin torsion.
The name spin torsion is motivated by the fact, that this part of the spin connection transforms homogeneously under coordinate transformations and therefore cannot be transformed away locally by adjusting the coordinates.
Similarly to spacetime torsion we can impose conditions like metric compatibility for spin torsion leading to some algebraic constraints.
These constraints have been resolved completely such that we have been able to count the actual degrees of freedom of spin torsion.
Motivated through classical field theory we have given a possible action in terms of the field strength induced by the spin connection.
Using this action in vaccum, i.e.~no matter Lagrangian, we have found the standard vacuum Einstein field equations and identically vanishing spin torsion.

We further have found that the analog of the commonly used Lorentz symmetric gauge in terms of the Dirac matrices is in fact the simplest possible choice of Dirac matrices for a given background metric and metric fluctuation.
This explains why this gauge choice is so useful for explicit calculations.
Furthermore we have presented a possibility for an explicit gauge fixing of the Dirac matrices for two general metrics, which do not have to be linearly connected.

% Using the considerations from Sect.~VIII of \cite{Gies:2013noa} we possibly have a quantization scheme of gravity in terms of the metric even in the presence of fermions.

\acknowledgments 

The author thanks Martin Ammon, Holger Gies, Tobias Hellwig, Ren\'{e} Sondenheimer, Andreas Wipf, Luca Zambelli and Omar Zanusso for valuable
discussions, and Holger Gies for comments on the manuscript.
I acknowledge support by the DFG under grant GRK1523 and Gi 328/7-1.

%\newpage

\appendix

%-----------------------------------------------------------------------------
\section{Weldon theorem in arbitrary integer dimensions}
\label{App:Weldon_theorem}
%-----------------------------------------------------------------------------

An essential ingredient for our investigations is the Weldon theorem \cite{Weldon:2000fr}.
It states that the most general infinitesimal variation of the Dirac matrices compatible with the Clifford algebra can be written as
\begin{align}\label{eq:App:Weldon_Theorem}
 \delta \gamma_{\mu} = \frac{1}{2} (\delta \metric_{\mu \nu}) \gamma^{\nu} + [ \delta \mcS_{\gamma} , \gamma_{\mu} ], \quad \tr \delta \mcS_{\gamma} = 0 \text{,}
\end{align}
where $\delta \metric_{\mu \nu}$ is the infinitesimal variation of the metric and $\delta \mcS_{\gamma} \in \mathrm{Mat}(d_{\gamma} \times d_{\gamma} , \C)$ parametrizes an arbitrary infinitesimal similarity transformation.
With $\mathrm{Mat}(d_{\gamma} \times d_{\gamma} , \C)$ we denote the $d_{\gamma} \times d_{\gamma}$ matrices.
Especially there is a one-to-one mapping between $\delta \gamma_{\mu}$ on the one hand and $\delta \metric_{\mu \nu}$ and $\delta \mcS_{\gamma}$ on the other hand.
With this theorem we can proof the existence of $\hat{\Gamma}_{\mu}$ in App. \ref{App:Existence_of_the_spin_connection}, parametrize all possible Dirac matrices and perform derivatives of the $\gamma_{\mu}$ with respect to the metric.

Weldon has proven this theorem in $d = 4$ spacetime dimensions.
We give a general proof for arbitrary integer dimensions $d \geq 2$.
Starting with the Clifford algebra
\begin{align}
 \{ \gamma_{\mu} , \gamma_{\nu} \} = 2 \metric_{\mu \nu} \mrI
\end{align}
we perform an infinitesimal variation and arrive at
\begin{align}\label{eq:App:varied_Clifford_algebra}
 \{ \gamma_{\mu} + \delta \gamma_{\mu} , \gamma_{\nu} + \delta \gamma_{\nu} \} = 2 ( \metric_{\mu \nu} + \delta \metric_{\mu \nu} ) \mrI \text{.}
\end{align}
Now instead of solving this equation in general in one step, we start with one special solution namely
\begin{align}
 (\delta \gamma_{\mu})_{\text{special}} = \frac{1}{2} (\delta \metric_{\mu \nu}) \gamma^{\nu} \text{.}
\end{align}
This solution solves \Eqref{eq:App:varied_Clifford_algebra} not exactly but only to the first order in $\delta \metric_{\mu \nu}$, which is of course sufficient since we are only interested in infinitesimal variations.
Now we employ the well known theorem that every solution to the Clifford algebra to a given metric is connected to each other via a similarity transformation and in odd dimensions via a sign flip (if necessary) \cite{Cornwell:1989bx}.
Since we only deal with infinitesimal variations, we cannot leave the connected component.
This excludes the sign flip also in odd dimensions.
Therefore the most general solution $\delta \gamma_{\mu}$ must be connected to $(\delta \gamma_{\mu})_{\text{special}}$ via a similarity transformation and actually this transformation has to be an infinitesimal one $\euler^{\delta \mcS_{\gamma}} \simeq \mrI + \delta \mcS_{\gamma}$
\begin{align}
 \gamma_{\mu} + \delta \gamma_{\mu} \stackrel{!}{=} \euler^{\delta \mcS_{\gamma}} \big( \gamma_{\mu} + (\delta \gamma_{\mu})_{\text{special}} \big) \euler^{ - \delta \mcS_{\gamma}} \text{.}
\end{align}
By expanding this equation we can read off
\begin{align}
\begin{aligned}
 \delta \gamma_{\mu} &{}= ( \delta \gamma_{\mu} )_{\text{special}} + [ \delta \mcS_{\gamma} , \gamma_{\mu} ] = \frac{1}{2} (\delta \metric_{\mu \nu}) \gamma^{\nu} + [ \delta \mcS_{\gamma} , \gamma_{\mu} ] \text{.}
\end{aligned}
\end{align}
Since the trace part completely drops out of the commutator it is sufficient to restrict $\delta \mcS_{\gamma}$ to be traceless.
The last relation proves that we can decompose every Dirac matrix fluctuation compatible with the Clifford algebra as in \Eqref{eq:App:Weldon_Theorem}.

We still have to proof the uniqueness of $\delta \metric_{\mu \nu}$ and $\delta \mcS_{\gamma}$ for a given $\delta \gamma_{\mu}$, where we impose that any metric fluctuation has to be symmetric $\delta \metric_{\mu \nu} = \delta \metric_{\nu \mu}$ and that any spin-base fluctuation has to be traceless $\tr \delta \mcS_{\gamma} = 0$.
Now let us suppose we have two sets of compatible metric fluctuations and spin-base fluctuations $\delta \metric_{\mu \nu}, \delta \mcS_{\gamma}$ (unprimed decomposition) and $\delta \metric'_{\mu \nu}, \delta \mcS'_{\gamma}$ (primed decomposition) for a given $\delta \gamma_{\mu}$,
\begin{align}
 \delta \gamma_{\mu} = \frac{1}{2} (\delta \metric_{\mu \nu}) \gamma^{\nu} + [ \delta \mcS_{\gamma} , \gamma_{\mu} ] = \frac{1}{2} (\delta \metric'_{\mu \nu}) \gamma^{\nu} + [ \delta \mcS'_{\gamma} , \gamma_{\mu} ] \text{.}
\end{align}
By calculating the trace of $\gamma_{\mu} \delta \gamma_{\nu} + \gamma_{\nu} \delta \gamma_{\mu}$ first in the unprimed decomposition
\begin{align}
 \frac{1}{d_{\gamma}} \tr ( \gamma_{\mu} \delta \gamma_{\nu} + \gamma_{\nu} \delta \gamma_{\mu} ) = \delta \metric_{\mu \nu}
\end{align}
and then again in the primed decomposition
\begin{align}
 \frac{1}{d_{\gamma}} \tr ( \gamma_{\mu} \delta \gamma_{\nu} + \gamma_{\nu} \delta \gamma_{\mu} ) = \delta \metric'_{\mu \nu}
\end{align}
we find that the two metric fluctuations have to be equal $\delta \metric_{\mu \nu} = \delta \metric'_{\mu \nu}$.
From here it is obvious that $[ \delta \mcS_{\gamma} , \gamma_{\mu} ] = [ \delta \mcS'_{\gamma} , \gamma_{\mu} ]$, implying that
\begin{align}
 \big[ [ \delta \mcS_{\gamma} , \gamma_{\mu} ] , \gamma^{\mu} \big] = \big[ [ \delta \mcS'_{\gamma} , \gamma_{\mu} ] , \gamma^{\mu} \big] \text{.}
\end{align}
Now we can use that the $\gamma^{\mu_{1} \ldots \mu_{n}}$  in even dimensions or respectively the $\gamma^{\mu_{1} \ldots \mu_{2n}}$ in odd dimensions form a basis in $\mathrm{Mat}(d_{\gamma} \times d_{\gamma} , \C)$ \cite{Cornwell:1989bx}.%
\footnote{
The $\gamma_{\mu_{1} \ldots \mu_{n}}$ are the normalized and antisymmetrized combinations of the Dirac matrices, cf. \Eqref{eq:definition_antisymmetrized_gammas}.
}
Next we observe that the contracted commutator of Dirac matrices $\big[[\cdot,\gamma_{\mu}],\gamma^{\mu}]$ does not mix the base elements, and only eliminates the part proportional to the identity, cf. \Eqref{eq:App:gamma_contraction_commutator} from App. \ref{App:Special_relations_gamma_matrices}.
Hence the two matrices $\delta \mcS_{\gamma}$ and $\delta \mcS'_{\gamma}$ are equal up to a trace term.
Since we know that they are traceless they have to be equal $\delta \mcS_{\gamma} = \delta \mcS'_{\gamma}$.
This proves the uniqueness of $\delta \metric_{\mu \nu}$ and $\delta \mcS_{\gamma}$.

%-----------------------------------------------------------------------------
\section{Dirac matrices in terms of the vielbein}
\label{App:impossible_vielbein}
%-----------------------------------------------------------------------------

In the following we investigate what happens if one tries to give the Dirac matrix path integral a meaning using the vielbein formalism.
In other words we aim at decomposing an arbitrary Dirac matrix fluctuation $\delta \gamma_{\mu}$ (compatible with the Clifford algebra) uniquely into a vielbein fluctuation $\delta \vielbein_{\mu}^{\point a}$ and a fluctuation of some other quantity.

We begin by assuming the existence of a vielbein degree of freedom $\vielbein_{\mu}^{\point a}$ in the Dirac matrix formalism.
Then we can define a set of spacetime dependent flat Dirac matrices $\flatgamma_{a}$ by
\begin{align}\label{eq:App:Dirac_vielbein_decomposition}
 \flatgamma_{a} = \vielbein^{\mu}_{\point a} \gamma_{\mu}
\end{align}
satisfying a Clifford algebra for the flat metric $\eta_{a b}$.
Hence, we can express the Dirac matrices as $\gamma_{\mu} = \vielbein_{\mu}^{\point a} \flatgamma_{a}$.
A general vielbein fluctuation can be decomposed uniquely into a metric fluctuation $\delta \metric_{\mu \nu}$ and a Lorentz fluctuation $\delta \Lambda^{a}_{\point b}$
\begin{align}\label{eq:App:vielbein_metric_decompostion}
 \delta \vielbein_{\mu}^{\point a} = \frac{1}{2} (\delta \metric_{\mu \nu}) \vielbein^{\nu a} + \vielbein_{\mu}^{\point b} \delta \Lambda^{a}_{\point b},
\end{align}
where
\begin{align}
 \delta \metric_{\mu \nu} ={}& (\delta \vielbein_{\mu}^{\point a} ) \vielbein_{\nu a} + (\delta \vielbein_{\nu}^{\point a} ) \vielbein_{\mu a},
\\
 \delta \Lambda^{a}_{\point b} ={}& \frac{1}{2} (\delta \vielbein_{\rho}^{\point a}) \vielbein^{\rho}_{\point b} - \frac{1}{2} (\delta \vielbein_{\rho}^{\point c}) \, \eta_{c b} \, \vielbein^{\rho}_{\point d} \, \eta^{d a} \text{.}
\end{align}
From a given Dirac matrix fluctuation $\delta \gamma_{\mu}$ we can read off the corresponding metric fluctuation $\delta \metric_{\mu \nu}$ and the corresponding spin-base fluctuation $\delta \mcS_{\gamma}$ from the Weldon theorem
\begin{align}
 \delta \gamma_{\mu} = \frac{1}{2} (\delta \metric_{\mu \nu}) \gamma^{\nu} + [ \delta \mcS_{\gamma} , \gamma_{\mu} ] \text{,}
\end{align}
cf. App. \ref{App:Weldon_theorem}.
On the other hand we can calculate the fluctuations of the decomposition \Eqref{eq:App:Dirac_vielbein_decomposition}
\begin{align}
 \delta \gamma_{\mu} = (\delta \vielbein_{\mu}^{\point a}) \flatgamma_{a} + \vielbein_{\mu}^{\point a} (\delta \flatgamma_{a}) \text{,}
\end{align}
where $\delta \vielbein_{\mu}^{\point a}$ is the vielbein fluctuation and $\delta \flatgamma_{a}$ is a fluctuation of a flat Dirac matrix.
Here we can decompose the vielbein fluctuation like in \Eqref{eq:App:vielbein_metric_decompostion}.
Additionally we know that the $\flatgamma_{a}$ have to satisfy the flat Clifford algebra.
Therefore the fluctuation $\delta \flatgamma_{a}$ has to be a pure spin-base fluctuation $\delta \mcS_{(\text{f})}$
\begin{align}
 \delta \flatgamma_{a} = [ \delta \mcS_{(\text{f})} , \flatgamma_{a}] \text{.}
\end{align}
Then we arrive at
\begin{align}
 \delta \gamma_{\mu} = \frac{1}{2} (\delta \metric_{\mu \nu}) \gamma^{\nu} + \left[ \delta \mcS_{(\text{f})} + \frac{1}{8} (\delta \Lambda^{b}_{\point c}) [ \flatgamma_{b} , \flatgamma^{c} ] , \gamma_{\mu} \right] \! \text{,}
\end{align}
where we have used the identity $\big[ [ \flatgamma_{b} , \flatgamma^{c} ] , \flatgamma_{a} \big] = 4 \delta^{c}_{a} \flatgamma_{b} - 4 \eta_{a b} \flatgamma^{c}$.
Here we see that $\delta \gamma_{\mu}$ fixes the metric fluctuation part $\delta \metric_{\mu \nu}$ of the vielbein fluctuation $\delta \vielbein_{\mu}^{\point a}$.
Besides it follows that
\begin{align}
 \delta \mcS_{\gamma} = \delta \mcS_{(\text{f})} + \frac{1}{8} (\delta \Lambda^{b}_{\point c}) [ \flatgamma_{b} , \flatgamma^{c} ] \text{.}
\end{align}
This implies that there are in principle infinitely many possible Lorentz fluctuations $\delta \Lambda^{a}_{\point b}$ for a given spin-base fluctuation $\delta \mcS_{\gamma}$.
In order to cure this ambiguity, we have to find a way to extract a unique Lorentz fluctuation from $\delta \mcS_{\gamma}$.
The obvious way is to restrict $\delta \mcS_{(\text{f})}$ to the traceless matrices, without the set $\{ \omega_{\mu \nu} [\gamma^{\mu} , \gamma^{\nu} ]: \omega_{\mu \nu} \in \R \}$, where $[\gamma^{\mu}, \gamma^{\nu}]$ is one part of the Dirac matrix base for $\mathrm{Mat}(d_{\gamma} \times d_{\gamma} , \C)$.
%Then we can extract $\delta \Lambda^{a}_{\point b}$ unambiguously from $\delta \mcS_{\gamma}$.
Unfortunately the remaining degree of freedom $\delta \mcS_{(\text{f})}$ then corresponds to a symmetry whose representation becomes spacetime dependent and dependent on the position in configuration space (within a path integral).
Even worse, as $\delta \mcS_{\gamma}$ is an element of the complex algebra $\mathfrak{sl}(d_{\gamma} , \C)$, the matrix $\delta \mcS_{(\text{f})}$ has then to be an element of $\mathfrak{sl}(d_{\gamma} , \C) \backslash \{ \omega_{\mu \nu} [ \gamma^{\mu} , \gamma^{\nu} ] : \omega_{\mu \nu} \in \R \}$.
This set obviously does not form an algebra any more (except for $d = 3$, $d_{\gamma} = 2$), and hence the construction of a meaningful integral for $\delta \mcS_{(\text{f})}$ is an open problem.

Summing up we found, that if one insists on integrating over Dirac matrices in terms of a vielbein, it will be difficult to define a meaningful remaining quantity, necessary to cover all possible Dirac matrices.
Either the remaining quantity, will be a spacetime dependent representation of an object which most likely will not form a group, leading to a complicated construction of the corresponding path integral.
Or one already needs a revised and presumably inconvenient way of assigning a vielbein fluctuation to the Dirac matrix fluctuation.

%-----------------------------------------------------------------------------
\section{Minimal spin-base group}
\label{App:minimal_spin_base_group}
%-----------------------------------------------------------------------------

%In this section we investigate answer the question which group 
In this section we will show, that $\mathrm{SL}(d_{\gamma}, \C)$ is the unique group%
\footnote{
In fact we are dealing with the fundamental representation of $\mathrm{SL}(d_{\gamma} , \C)$ and not the group itself. But we will keep this terminology in the following for simplicity. By fundamental representation we mean the defining matrix representation of $\mathrm{SL}(d_{\gamma} , \C)$, which is $\{ \mcS \in \mathrm{Mat}(d_{\gamma} \times d_{\gamma} , \C) : \det \mcS = 1 \}$ together with the matrix multiplication as the group law.
}
$\minsb \leq \mathrm{GL}(d_{\gamma} , \C)$ satisfying
\begin{align}
&\begin{aligned}
 \rmi{} & \,\, \forall \gamma_{\mu} , \gamma'_{\mu} \text{ compatible with the Clifford algebra}
\\
 & \,\, \exists \mcS \in \minsb : \gamma'_{\mu} = \left\{ \begin{matrix} \mcS \gamma_{\mu} \mcS^{-1} & \text{, $d$ even} \\ \pm \mcS \gamma_{\mu} \mcS^{-1} & \text{, $d$ odd} \end{matrix} \right. \! \text{,}
\end{aligned}
\\
&\begin{aligned}
 \rmii{} & \,\, \forall \gamma_{\mu} \text{ compatible with the Clifford algebra, it holds}
\\
 & \abs{\{ \mcS \in \minsb : \mcS \gamma_{\mu} \mcS^{-1} = \gamma_{\mu} \}}
\\
 & \quad
 = \min\limits_{\ttm{\begin{matrix} \mathrm{SB_{\text{test}}} \leq \mathrm{GL}(d_{\gamma} , \C) \\ \text{compatible with } \rmi \end{matrix}}} \abs{\{ \mcS \in \mathrm{SB_{\text{test}}} : \mcS \gamma_{\mu} \mcS^{-1} = \gamma_{\mu}  \}} \text{,}
\end{aligned}
\end{align}
where we denote the cardinality of a set $\mfrS$ with $\abs{\mfrS}$.

The existence of a group satisfying $\rmi$ is guaranteed by the Clifford algebra and is independent of the metric \cite{Cornwell:1989bx}.
In addition condition $\rmii$ is independent of the actual choice of the Dirac matrices, i.e. if it is satisfied for a specific set $\gamma_{\mu}$ compatible with the Clifford algebra, then it is satisfied for any.
This follows from Schur's lemma%
\footnote{
Schur's lemma basically says that a matrix $M \in \mathrm{Mat}(d_{\gamma} \times d_{\gamma} , \C)$ which commutes with every base element is proportional to the identity matrix. Since we can construct a basis in $\mathrm{Mat}(d_{\gamma} \times d_{\gamma}, \C)$ from the $\gamma_{\mu_{1} \ldots \mu_{n}}$ it suffices if $M$ commutes with the $\gamma_{\mu}$, as it then obviously also commutes with the $\gamma_{\mu_{1} \ldots \mu_{n}}$.
}
and $\mcS \gamma_{\mu} \mcS^{-1} = \gamma_{\mu} \Leftrightarrow \mcS = \frac{1}{d_{\gamma}} (\tr \mcS) \cdot \mrI$ for $\mcS \in \mathrm{GL}(d_{\gamma} , \C)$.

Now let us construct the group $\minsb$.
We start by observing that every element of $\mathrm{GL}(d_{\gamma} , \C)$ can be written as $\euler^{M}$ for some $M \in \mathrm{Mat}(d_{\gamma} \times d_{\gamma} , \C)$.
Next we can split $M$ into its trace part $\frac{1}{d_{\gamma}} (\tr M) \cdot \mrI$ and the traceless part $\hat{M} = M - \frac{1}{d_{\gamma}} (\tr M) \cdot \mrI$.
Since the trace part is proportional to the identity matrix it commutes with every element of $\mathrm{Mat}(d_{\gamma} \times d_{\gamma} , \C)$.
Therefore the trace part is trivial for the similarity transformations.
By the use of Jacobi's formula we find $\det \euler^{\hat{M}} = 1$, leading us to $\minsb \leq \mathrm{SL}(d_{\gamma} , \C)$.

If we calculate the set of trivial elements (condition $\rmii$) for $\mathrm{SL}(d_{\gamma} , \C)$, we find
\begin{align}
\begin{aligned}
 &{}\{ \mcS \in \mathrm{SL}(d_{\gamma} , \C) : \mcS \gamma_{\mu} \mcS^{-1} = \gamma_{\mu} \} = \cen \big( \mathrm{SL}(d_{\gamma} , \C) \big) \text{,}
\end{aligned}
\end{align}
where $\cen \big( \mathrm{SL}(d_{\gamma} , \C) \big) = \big\{ \euler^{\cplx \frac{2 \pi}{d_{\gamma}} n} \cdot \mrI : n \in \{ 0 , \ldots , d_{\gamma} - 1 \} \big\}$ is the center of $\mathrm{SL}(d_{\gamma} , \C)$ and has finite cardinality $\abs{\cen \big( \mathrm{SL}(d_{\gamma} , \C) \big)} = d_{\gamma}$.

In order to determine which elements of $\mathrm{SL}(d_{\gamma} , \C)$ we definitely need, we use condition $\rmi$.
Let us consider two different transformations $\mcS_{1}, \mcS_{2} \in \mathrm{SL}(d_{\gamma} , \C)$ connecting a given pair $\gamma_{\mu}, \gamma'_{\mu}$ compatible with the Clifford algebra.
It turns out that they have to be related by a center element
\begin{align}
 \mcS_{1} \gamma_{\mu} \mcS_{1}^{-1} = \gamma'_{\mu} = \mcS_{2} \gamma_{\mu} \mcS_{2}^{-1} \Rightarrow [ \mcS_{2}^{-1} \mcS_{1} , \gamma_{\mu} ] = 0 \text{.}
\end{align}
With this observation we can define an equivalence relation $\sim$, $\mcS_{1} \sim \mcS_{2}$ iff $\exists \mcC \in \cen \big( \mathrm{SL}(d_{\gamma} , \C) \big)$ so that $\mcS_{1} = \mcC \mcS_{2}$.
For a given $\gamma_{\mu}$ every equivalence class generates a different $\gamma'_{\mu}$, and we already know that all $\gamma'_{\mu}$ are generated in this way.

In the next step we will show that the center gets generated by a specific equivalence class.
Since we need at least one representative of each equivalence class the whole $\mathrm{SL}(d_{\gamma} , \C)$ gets generated as well by applying the generated center elements to the representatives of the equivalence classes.

Let us define the matrix $M$ as
\begin{align}
 M = \cplx \frac{2 \pi}{d_{\gamma}} \mrI - \cplx 2 \pi A \text{,}
\end{align}
where $A$ can be any matrix satisfying
\begin{align}
 A^{2} = A, \quad \tr A = 1 \text{.}
\end{align}
One such $A$ is the matrix with a $1$ in the upper left corner and $0$ everywhere else.
The matrix $M$ is by construction traceless and satisfies
\begin{align}
 \euler^{M} = \euler^{\cplx \frac{2 \pi}{d_{\gamma}}} \cdot \mrI \text{,}
\end{align}
i.e. it generates the center of $\mathrm{SL}(d_{\gamma} , \C)$ and belongs to the equivalence class of the identity element.
This relation can be verified by observing
\begin{align}
 \euler^{a \cdot A} = \mrI + \sum\limits_{n=1}^{\infty} \frac{a^n}{n!} A = \mrI + (\euler^{a} - 1) A , \quad a \in \C \text{.}
\end{align}
Next we calculate
\begin{align}
 \euler^{\frac{1}{d_{\gamma}} M} = \euler^{\cplx \frac{2 \pi}{d_{\gamma}^2}} \mrI + \euler^{\cplx \frac{2 \pi}{d_{\gamma}^{2}}}(\euler^{- \cplx \frac{2 \pi}{d_{\gamma}}} - 1) A.
\end{align}
The determinant of this matrix is equal to 1 and additionally it is not proportional to the identity matrix and therefore is a nontrivial spin-base transformation, i.e. it belongs to a non-trivial equivalence class.
Hence, there has to be at least one $n \in \{0 , \ldots , d_{\gamma} - 1 \}$, so that $\euler^{\cplx \frac{2 \pi}{d_{\gamma}} n} \euler^{\frac{1}{d_{\gamma}} M} \in \minsb$.
Because $\minsb$ is supposed to be a group, it has to be closed under the group law.
This implies that also $(\euler^{\cplx \frac{2 \pi}{d_{\gamma}} n} \euler^{\frac{1}{d_{\gamma}} M})^{d_{\gamma}}$ has to be an element of $\minsb$, as $d_{\gamma}$ is an integer.
By calculating
\begin{align}
 (\euler^{\cplx \frac{2 \pi}{d_{\gamma}} n} \euler^{\frac{1}{d_{\gamma}} M})^{d_{\gamma}} = \euler^{M} = \euler^{\cplx \frac{2 \pi}{d_{\gamma}}} \cdot \mrI
\end{align}
we see that $\euler^{\cplx \frac{2 \pi}{d_{\gamma}}} \cdot \mrI \in \minsb$, implying that the whole center gets generated.
Therefore we have $\mathrm{SL}(d_{\gamma} , \C) \leq \minsb$.
With this finding we conclude that $\minsb = \mathrm{SL}(d_{\gamma} , \C)$.

%-----------------------------------------------------------------------------
\section{Special relations for the Dirac matrices -- Part I}
\label{App:Special_relations_gamma_matrices}
%-----------------------------------------------------------------------------

In order to proof the uniqueness of $\hat{\Gamma}_{\mu}$ and give the explicit expressions \Eqref{eq:explicit_solution_hat_gamma_I}, \eqref{eq:explicit_solution_hat_gamma_II} and \eqref{eq:explicit_solution_hat_gamma_III} we will need some identities for the Dirac matrices.
Let us introduce the shorthands
\begin{align}\label{eq:App:definition_Anm}
 (A^{n}_{m})^{\mu_{1} \ldots \mu_{n}}_{\nu_{1} \ldots \nu_{m}} = \frac{1}{d_{\gamma}} \tr ( \gamma^{\mu_{1} \ldots \mu_{n}} \gamma_{\nu_{1} \ldots \nu_{m}} ), 
\\
 (A^{n,k}_{m,r})^{\mu_{1} \ldots \mu_{n}}_{\nu_{1} \ldots \nu_{m}} = (A^{n-1}_{m-1})^{\mu_{1} \ldots \mu_{k-1} \mu_{k+1} \ldots \mu_{n}}_{\nu_{1} \ldots \nu_{r-1} \nu_{r+1} \ldots \nu_{m}} \text{.}
\end{align}
Some of these $(A^{n}_{m})$ are easily calculated directly from their definition \Eqref{eq:App:definition_Anm}
\begin{align}
 &(A^{0}_{0}) = 1,
\\
 &(A^{n}_{0})^{\mu_{1} \ldots \mu_{n}} = 0, \quad n > 0 , \quad d \text{ even,}
\\
 &(A^{2n}_{0})^{\mu_{1} \ldots \mu_{2n}} = 0, \quad n > 0 , \quad d \text{ odd,}
\\
 &(A^{1}_{1})^{\mu}_{\nu} = \delta^{\mu}_{\nu}, \label{eq:App:starting_condition_Anm_d_general}
\\
 &(A^{n}_{1})^{\mu_{1} \ldots \mu_{n}}_{\nu} = 0, \quad n > 1, \quad d \text{ even,} \label{eq:App:starting_condition_Anm_d_even}
\\
 &(A^{2n}_{2})^{\mu_{1} \ldots \mu_{2n}}_{\nu_{1} \nu_{2} } = 0, \quad n > 1, \quad d \text{ odd,} \label{eq:App:starting_condition_Anm_d_odd}
\\
 &(A^{2n-1}_{1})^{\mu_{1} \ldots \mu_{2n-1}}_{\nu} = 0, \quad n > 1, \quad d \text { odd.} \label{eq:App:starting_condition_Anm_d_odd_2}
\end{align}
The identities to be proven are
\begin{align}\label{eq:App:gamma_contraction_commutator}
 \big[ [ \gamma^{\mu_{1} \ldots \mu_{n}} , \gamma^{\nu} ] , \gamma_{\nu} \big] = 2 \big( ( 1 - (-1)^{n} ) d + (-1)^{n} 2 n \big) \gamma^{\mu_{1} \ldots \mu_{n}} \text{,}
\end{align}
and the traces of the base elements for even dimensions with $n,m \in \{1, \ldots, d \}$
\begin{align}
 \label{eq:App:Anm_d_even}
 &(A^{n}_{m})^{\mu_{1} \ldots \mu_{n}}_{\nu_{1} \ldots \nu_{m}} = (-1)^{\frac{n(n-1)}{2}} \,\, n! \,\, \delta^{n}_{m} \deltaA^{\mu_{1} \ldots \mu_{n}}_{\nu_{1} \ldots \nu_{n}}
\end{align}
as well as the traces of the base elements for odd dimensions with $n,m \in \{0, \ldots, \frac{d-1}{2} \}$
\begin{align}
 \label{eq:App:Anm_d_odd}
 &(A^{2n}_{2m})^{\mu_{1} \ldots \mu_{2n}}_{\nu_{1} \ldots \nu_{2m}} = (-1)^{n} \,\, (2n)! \,\, \delta^{n}_{m} \deltaA^{\mu_{1} \ldots \mu_{2n}}_{\nu_{1} \ldots \nu_{2n}},
\\
 \label{eq:App:Anm_d_odd_2}
 &(A^{2n+1}_{2m+1})^{\mu_{1} \ldots \mu_{2n+1}}_{\nu_{1} \ldots \nu_{2m+1}} = (-1)^{n} \,\, (2n+1)! \,\, \delta^{n}_{m} \deltaA^{\mu_{1} \ldots \mu_{2n+1}}_{\nu_{1} \ldots \nu_{2n+1}} \text{.}
\end{align}
Here $\deltaA_{\mu_{1} \ldots \mu_{m}}^{\nu_{1} \ldots \nu_{m}}$ denotes the normalized and antisymmetrized Kronecker delta.
As a first step we rewrite the Clifford algebra as
\begin{align}
 \gamma_{\nu} \gamma^{\mu} = - \gamma^{\mu} \gamma_{\nu} + 2 \delta^{\mu}_{\nu} \mrI
\end{align}
to find that for $n \in \N^{\ast}$%
\footnote{
We denote the natural numbers including zero with $\N_{0}$ and the natural numbers excluding zero with $\N^{\ast} = \N_{0} \backslash \{ 0 \}$.
}
\begin{align}
 \gamma_{\nu} \gamma^{\mu_{1}} \! \ldots \! \gamma^{\mu_{n}} \gamma^{\nu} = - \gamma^{\mu_{1}} \gamma_{\nu} \gamma^{\mu_{2}} \! \ldots \! \gamma^{\mu_{n}} \gamma^{\nu} + 2 \gamma^{\mu_{2}} \! \ldots \! \gamma^{\mu_{n}} \gamma^{\mu_{1}} \text{.}
\end{align}
Now we can iterate this process $n$ times to get
\begin{align}
 \gamma_{\nu} \gamma^{\mu_{1}} \! &\ldots \! \gamma^{\mu_{n}} \gamma^{\nu} \notag
\\
 =& (-1)^{n} \gamma^{\mu_{1}} \! \ldots \! \gamma^{\mu_{n}} \gamma_{\nu} \gamma^{\nu} \notag
\\
 & - 2 \sum\limits_{l = 1}^{n} (-1)^{l} \gamma^{\mu_{1}} \! \ldots \! \gamma^{\mu_{l-1}} \gamma^{\mu_{l+1}} \! \ldots \! \gamma^{\mu_{n}} \gamma^{\mu_{l}} \notag
\\
 =& (-1)^{n} \, d \, \gamma^{\mu_{1}} \! \ldots \! \gamma^{\mu_{n}} \notag 
\\
 &- 2 \sum\limits_{l = 1}^{n} (-1)^{l} \gamma^{\mu_{1}} \! \ldots \! \gamma^{\mu_{l-1}} \gamma^{\mu_{l+1}} \! \ldots \! \gamma^{\mu_{n}} \gamma^{\mu_{l}} \text{.}
\end{align}
With this equation we can infer
\begin{align}\label{eq:App:gamma_contraction_base_elements}
 \gamma_{\nu} \gamma^{\mu_{1} \ldots \mu_{n}} \gamma^{\nu} \! &= \! (-1)^n d \gamma^{\mu_{1} \ldots \mu_{n}} \! - 2 \! \sum\limits_{l=1}^{n} (-1)^{l} \gamma^{\mu_{1} \ldots \mu_{n}} (-1)^{n-l} \notag 
\\
 &= (-1)^{n} ( d - 2 n ) \gamma^{\mu_{1} \ldots \mu_{n}} \text{,}
\end{align}
and from there we deduce the first of two necessary results to give an explicit expression of $\hat{\Gamma}_{\mu}$
\begin{align}
 \big[ [ \gamma^{\mu_{1} \ldots \mu_{n}} , \gamma^{\nu} ] , \gamma_{\nu} \big] = 2 \big( (1 - (-1)^{n} ) d + (-1)^n 2 n \big) \gamma^{\mu_{1} \ldots \mu_{n}} \text{.}
\end{align}
Note that we did not assume $d$ to be even, this result holds in any integer dimension $d \geq 2$.

The second result is concerning the trace of two basis elements $\gamma^{\mu_{1} \ldots \mu_{n}}$ in even dimensions and $\gamma^{\mu_{1} \ldots \mu_{2n}}$ or $\gamma^{\mu_{1} \ldots \mu_{2n+1}}$ in odd dimensions.

At first we leave $d$ without restrictions and look at $n, m \in \N^{\ast}$
\begin{align}\label{eq:App:trace_gamma_basis_preparation}
 \frac{1}{d_{\gamma}} &\tr( \gamma^{\mu_{1}} \ldots \gamma^{\mu_{n}} \gamma_{\nu_{1}} \ldots \gamma_{\nu_{m}} ) \notag
\\
 =& - \frac{1}{d_{\gamma}} \tr ( \gamma^{\mu_{1}} \ldots \gamma^{\mu_{n-1}} \gamma_{\nu_{1}} \gamma^{\mu_{n}} \gamma_{\nu_{2}} \ldots \gamma_{\nu_{m}} ) \notag
\\
 & + \frac{2}{d_{\gamma}} \delta^{\mu_{n}}_{\nu_{1}} \tr ( \gamma^{\mu_{1}} \ldots \gamma^{\mu_{n-1}} \gamma_{\nu_{2}} \ldots \gamma_{\nu_{m}} ) \text{.}
\end{align}
This time it is a little more difficult to iterate and antisymmetrize the indices in \Eqref{eq:App:trace_gamma_basis_preparation}.
For the first term we get after iterating
\begin{align}
 \frac{(-1)^{m}}{d_{\gamma}} \tr ( \gamma^{\mu_{n}} \gamma^{\mu_{1}} \ldots \gamma^{\mu_{n-1}} \gamma_{\nu_{1}} \ldots \gamma_{\nu_{m}} ) \text{,}
\end{align}
and after antisymmetrization
\begin{align}
 - (-1)^{n + m} (A^{n}_{m})^{\mu_{1} \ldots \mu_{n}}_{\nu_{1} \ldots \nu_{m}} \text{.}
\end{align}
The iterated second term becomes
\begin{align}
\begin{aligned}
 - \frac{2}{d_{\gamma}} \sum\limits_{l=1}^{m} (-1)^{l} \delta^{\mu_{n}}_{\nu_{l}} \tr( &\gamma^{\mu_{1}} \! \ldots \! \gamma^{\mu_{n-1}} 
\\
 &\times \gamma_{\nu_{1}} \! \ldots \! \gamma_{\nu_{l-1}} \gamma_{\nu_{l+1}} \! \ldots \! \gamma_{\nu_{m}} ) \text{.}
\end{aligned}
\end{align}
If we now perform the antisymmetrization we can split it into the antisymmetrization of the indices inside the trace and the indices outside the trace to reach
\begin{align}
 - \frac{2}{n m} \sum\limits_{l=1}^{m} \sum\limits_{k=1}^{n} \sum\limits_{r = 1}^{m} (-1)^{n+k+r} \delta^{\mu_{k}}_{\nu_{r}} (A^{n,k}_{m,r})^{\mu_{1} \ldots \mu_{n}}_{\nu_{1} \ldots \nu_{m}} \text{.}
\end{align}
Plugging this into \Eqref{eq:App:trace_gamma_basis_preparation} we find
\begin{align}\label{eq:App:recursion_Anm_preparation}
 (A^{n}_{m})&^{\mu_{1} \ldots \mu_{n}}_{\nu_{1} \ldots \nu_{m}} \notag
\\
 =& - (-1)^{n+m} (A^{n}_{m})^{\mu_{1} \ldots \mu_{n}}_{\nu_{1} \ldots \nu_{m}} \notag
\\
  & - \sum\limits_{k=1}^{n} \sum\limits_{r=1}^{m} \frac{ 2 (-1)^{n + k + r}}{n} \delta^{\mu_{k}}_{\nu_{r}} (A^{n,k}_{m,r})^{\mu_{1} \ldots \mu_{n}}_{\nu_{1} \ldots \nu_{m}} \text{.}
\end{align}
Because the even and the odd dimensional case are conceptually a little different we will discuss them separately now starting with the even dimensional one.

It is obvious that
\begin{align}
 (A^{n}_{m})^{\mu_{1} \ldots \mu_{n}}_{\nu_{1} \ldots \nu_{m}} = 0 , \quad (n + m) \text{ odd,}
\end{align}
since the trace then contains an odd number of Dirac matrices and hence always vanishes in even dimensions. Therefore we can restrict ourselves to the case where $(n+m)$ is even. In this case we conclude from \Eqref{eq:App:recursion_Anm_preparation}
\begin{align}
 (A^{n}_{m})^{\mu_{1} \ldots \mu_{n}}_{\nu_{1} \ldots \nu_{m}} = - \sum\limits_{k=1}^{n} \sum\limits_{r=1}^{m} \frac{(-1)^{n+k+r}}{n} \delta^{\mu_{k}}_{\nu_{r}} (A^{n,k}_{m,r})^{\mu_{1} \ldots \mu_{n}}_{\nu_{1} \ldots \nu_{m}} \text{.}
\end{align}
Therefore the $(A^{n}_{m})$ are directly proportional to the $(A^{n-1}_{m-1})$. Via iteration and the conditions \Eqref{eq:App:starting_condition_Anm_d_even} we find that
\begin{align}\label{eq:App:Anm_n_neq_m_d_even}
 (A^{n}_{m}) = 0 , \quad n \neq m \text{.}
\end{align}
For $n = m$  we get the recursion relation
\begin{align}
 (A^{n}_{n})^{\mu_{1} \ldots \mu_{n}}_{\nu_{1} \ldots \nu_{n}} = - n (-1)^{n} \delta^{[\mu_{1}}_{[\nu_{1}} (A^{n-1}_{n-1})^{\mu_{2} \ldots \mu_{n}]}_{\nu_{2} \ldots \nu_{n}]}
\end{align}
with the initial condition
\begin{align}
 (A^{1}_{1})^{\mu}_{\nu} = \delta^{\mu}_{\nu} \text{.}
\end{align}
This relation can easily be solved explicitly and we find
\begin{align}\label{eq:App:Ann_solution}
 (A^{n}_{n}) = (-1)^{\frac{n(n-1)}{2}} \,\, n! \,\, \deltaA^{\mu_{1} \ldots \mu_{n}}_{\nu_{1} \ldots \nu_{n}},
\end{align}
where $\deltaA^{\mu_{1} \ldots \mu_{n}}_{\nu_{1} \ldots \nu_{n}}$ is the normalized and antisymmetrized Kronecker delta.
Together with \Eqref{eq:App:Anm_n_neq_m_d_even} this proves \Eqref{eq:App:Anm_d_even}
\begin{align}
 (A^{n}_{m})^{\mu_{1} \ldots \mu_{n}}_{\nu_{1} \ldots \nu_{m}} = (-1)^{\frac{n(n-1)}{2}} \,\, n! \,\, \delta^{n}_{m} \deltaA^{\mu_{1} \ldots \mu_{n}}_{\nu_{1} \ldots \nu_{n}}\text{.}
\end{align}
To show the last relation we restrict ourselves to odd dimensions. Therefore we can shift $n \to 2n$ and $m \to 2m$ in \Eqref{eq:App:recursion_Anm_preparation} and find
\begin{align}
 (A^{2n}_{2m})^{\mu_{1} \ldots \mu_{2n}}_{\nu_{1} \ldots \nu_{2m}} = - \sum\limits_{k=1}^{2n}\sum\limits_{r=1}^{2m} \frac{(-1)^{k+r}}{2 n} \delta^{\mu_{k}}_{\nu_{r}} (A^{2n,k}_{2m,r})^{\mu_{1} \ldots \mu_{2n}}_{\nu_{1} \ldots \nu_{2m}} \text{.}
\end{align}
Again we find a directly proportional relation from $(A^{2n}_{2m})$ to $(A^{2n-1}_{2m-1})$.
Note that $(A^{2n-1}_{2m-1})$ are not the traces we are looking for since they have an odd number of upper and an odd number of lower indices.
But we can further relate the $(A^{2n-1}_{2m-1})$ directly proportional to $(A^{2(n-1)}_{2(m-1)})$ because \Eqref{eq:App:recursion_Anm_preparation} is true for all $n,m \in \N^{\ast}$ and $(2n-1 + 2m - 1)$ is an even number.
Therefore we deduce a direct proportionality between $(A^{2n}_{2m})$ and $(A^{2(n-1)}_{2(m-1)})$ and with the iteration of that and the conditions \Eqref{eq:App:starting_condition_Anm_d_odd} and \eqref{eq:App:starting_condition_Anm_d_odd_2} we get
\begin{align}
 (A^{2n}_{2m}) = (A^{2n+1}_{2m+1}) = 0, \quad n \neq m \text{.}
\end{align}
Here we note that \Eqref{eq:App:Ann_solution} uses only $n = m$ in \Eqref{eq:App:recursion_Anm_preparation} and the initial condition \Eqref{eq:App:starting_condition_Anm_d_general}, with both of them valid in even and odd dimensions.
Therefore \Eqref{eq:App:Ann_solution} is also valid odd dimensions.
Hence, we easily conclude
\begin{align}
 &(A^{2n}_{2n})^{\mu_{1} \ldots \mu_{2n}}_{\nu_{1} \ldots \nu_{2n}} = (-1)^{n} \,\, (2n)! \,\, \deltaA^{\mu_{1} \ldots \mu_{2n}}_{\nu_{1} \ldots \nu_{2n}},
\\
 &(A^{2n+1}_{2n+1})^{\mu_{1} \ldots \mu_{2n+1}}_{\nu_{1} \ldots \nu_{2n+1}} = (-1)^{n} \,\, (2n+1)! \,\, \deltaA^{\mu_{1} \ldots \mu_{2n+1}}_{\nu_{1} \ldots \nu_{2n+1}} \text{.}
\end{align}
The last two relations prove \Eqref{eq:App:Anm_d_odd} and \eqref{eq:App:Anm_d_odd_2}
\begin{align}
 (A^{2n}_{2m})^{\mu_{1} \ldots \mu_{2n}}_{\nu_{1} \ldots \nu_{2m}} = (-1)^{n} \,\, (2n)! \,\, \delta^{n}_{m} \deltaA^{\mu_{1} \ldots \mu_{2n}}_{\nu_{1} \ldots \nu_{2n}},
\\
 (A^{2n+1}_{2m+1})^{\mu_{1} \ldots \mu_{2n+1}}_{\nu_{1} \ldots \nu_{2m+1}} = (-1)^{n} \,\, (2n+1)! \,\, \delta^{n}_{m} \deltaA^{\mu_{1} \ldots \mu_{2n+1}}_{\nu_{1} \ldots \nu_{2n+1}} \text{.}
\end{align}
%

%-----------------------------------------------------------------------------
\section{Special relations for the Dirac matrices -- Part II}
\label{App:Special_relations_gamma_matrices_II}
%-----------------------------------------------------------------------------

For the explicit implementation of some conditions concerning the spin torsion $\Delta \Gamma_{\mu}$ we need more identities for the $\gamma_{\mu}$.
We have to prove
\begin{align}\label{eq:App:gamma_bar_sign_identity}
 &\bar{\gamma}_{\mu_{1} \ldots \mu_{n}} = (-1)^{\frac{n(n+1)}{2}} \gamma_{\mu_{1} \ldots \mu_{n}},
\\
\label{eq:App:commutator_gamma_even_base_element}
 &[ \gamma_{\mu} , \gamma^{\rho_{1} \ldots \rho_{2 n}} ] = 4 n \delta_{\mu}^{[\rho_{1}} \gamma^{\rho_{2} \ldots \rho_{2n}]},
\\
\label{eq:App:anticommutator_gamma_odd_base_element}
 &\{ \gamma_{\mu} , \gamma^{\rho_{1} \ldots \rho_{2n + 1}} \} = 2 (2 n + 1) \delta_{\mu}^{[\rho_{1}} \gamma^{\rho_{2} \ldots \rho_{2n + 1}]},
\end{align}
where $n \in \N_{0}$ is arbitrary.
Additionally we need
\begin{align}\label{eq:App:special_trace_gamma_commutator}
& \frac{1}{d_{\gamma}} \tr ( [ \gamma_{\nu_{1} \ldots \nu_{2m}} , \gamma^{\rho_{1} \ldots \rho_{2n-1}} ] \gamma_{\mu} ) \notag
\\
 &\quad  = (-1)^{n-1} \cdot 2 \cdot (2n)! \cdot \metric_{\mu [ \nu_{1}} \deltaA_{\nu_{2} \ldots \nu_{2n}]}^{\rho_{1} \ldots \rho_{2n-1}} \cdot \delta^{n}_{m},
\end{align}
where $n,m \in \N^{\ast}$.
The last important identity reads
\begin{align}\label{eq:App:special_trace_commutator}
 \frac{1}{d_{\gamma}} &\tr ( [ \gamma^{\rho_{1} \ldots \rho_{n}} , \gamma_{\lambda_{1} \ldots \lambda_{m}} ] \gamma_{\mu \nu} ) \notag
\\
 &= - 4 \cdot n \cdot n! \cdot (-1)^{\frac{n(n-1)}{2}} \cdot \delta_{[\mu}^{[\rho_{1}} \metric_{\nu][\lambda_{1}} \deltaA^{\rho_{2} \ldots \rho_{n}]}_{\lambda_{2} \ldots \lambda_{n}]} \cdot \delta^{n}_{m},
\end{align}
where $(n+m)$ has to be even and $n,m \in \N^{\ast}$, i.e.~either both have to be even or both have to be odd.

The proof of the first three identities is rather simple.
First we use that $\bar{\gamma}_{\mu} = \spinmetric^{-1} \gamma_{\mu}^{\dagger} \spinmetric = - \gamma_{\mu}$ to show
\begin{align}
 \spinmetric^{-1} (\gamma_{\mu_{1}} \! \ldots \gamma_{\mu_{n}})^{\dagger} \spinmetric = \spinmetric^{-1} \gamma_{\mu_{n}}^{\dagger} \! \ldots \gamma_{\mu_{1}}^{\dagger} \spinmetric = (-1)^{n} \gamma_{\mu_{n}} \! \ldots \gamma_{\mu_{1}} \text{.}
\end{align}
Next we antisymmetrize the indices on both sides to prove the first identity
\begin{align}
 \bar{\gamma}_{\mu_{1} \ldots \mu_{n}} = (-1)^{n} (-1)^{\sum\limits_{l=1}^{n-1} l} \gamma_{\mu_{1} \ldots \mu_{n}} = (-1)^{\frac{n(n+1)}{2}} \gamma_{\mu_{1} \ldots \mu_{n}} \text{.}
\end{align}
The second proof follows a similar track, we start with
\begin{align}
 \gamma_{\mu} \gamma^{\rho_{1}} \ldots \gamma^{\rho_{2n}} = 2 \delta_{\mu}^{\rho_{1}} \gamma^{\rho_{2}} \ldots \gamma^{\rho_{2n}} - \gamma^{\rho_{1}} \gamma_{\mu} \gamma^{\rho_{2}} \ldots \gamma^{\rho_{2n}} \text{.}
\end{align}
Again we iterate $2n$ times
\begin{align}
 \gamma_{\mu} \gamma^{\rho_{1}} \ldots \gamma^{\rho_{2n}} =& 2 \sum\limits_{l = 1}^{2n} (-1)^{l-1} \delta_{\mu}^{\rho_{l}} \gamma^{\rho_{1}} \ldots \gamma^{\rho_{l-1}} \gamma^{\rho_{l+1}} \ldots \gamma^{\rho_{2n}} \notag
\\
 & + \gamma^{\rho_{1}} \ldots \gamma^{\rho_{2n}} \gamma_{\mu} \text{.}
\end{align}
If we now also antisymmetrize the indices we can read off
\begin{align}
 \gamma_{\mu} \gamma^{\rho_{1} \ldots \rho_{2n}} =& 2 \sum\limits_{l=1}^{2n} \delta_{\mu}^{[\rho_{1}} \gamma^{\rho_{2} \ldots \rho_{2n}]} + \gamma^{\rho_{1} \ldots \rho_{2n}} \gamma_{\mu}
\\
 =& 4 n \delta_{\mu}^{[\rho_{1}} \gamma^{\rho_{2} \ldots \rho_{2n}]} + \gamma^{\rho_{1} \ldots \rho_{2n}} \gamma_{\mu} \text{.}
\end{align}
The last relation proves the identity. In order to show the third statement we perform analogous steps
\begin{align}
 \gamma_{\mu} \gamma^{\rho_{1} \ldots \rho_{2n + 1}} =& 2 \sum\limits_{l=1}^{2n + 1} \delta_{\mu}^{[\rho_{1}} \gamma^{\rho_{2} \ldots \rho_{2n + 1}]} - \gamma^{\rho_{1} \ldots \rho_{2n + 1}} \gamma_{\mu}
\\
 =& 2 ( 2n + 1) \delta_{\mu}^{[\rho_{1}} \gamma^{\rho_{2} \ldots \rho_{2n + 1}]} - \gamma^{\rho_{1} \ldots \rho_{2n + 1}} \gamma_{\mu} \text{.}
\end{align}
With the identities Eqs. \eqref{eq:App:Anm_d_even} and \eqref{eq:App:Anm_d_odd_2} from App. \ref{App:Special_relations_gamma_matrices} it is straightforward to calculate
\begin{align}
 \frac{1}{d_{\gamma}} &\tr ( [ \gamma_{\nu_{1} \ldots \nu_{2m}} , \gamma^{\rho_{1} \ldots \rho_{2n-1}} ] \gamma_{\mu} ) \notag
\\
 &= \frac{1}{d_{\gamma}} \tr ( [ \gamma_{\mu} , \gamma_{\nu_{1} \ldots \nu_{2m}} ] \gamma^{\rho_{1} \ldots \rho_{2n-1}} ) \notag
\\
 &= \frac{4m}{d_{\gamma}} \metric_{\mu [ \nu_{1}} \tr ( \gamma_{\nu_{2} \ldots \nu_{2m}]} \gamma^{\rho_{1} \ldots \rho_{2n-1}} ) \notag
\\
 &= (-1)^{n-1} \cdot 2 \cdot (2n)! \cdot \metric_{\mu [ \nu_{1}} \deltaA^{\rho_{1} \ldots \rho_{2n-1}}_{\nu_{2} \ldots \nu_{2n}]} \cdot \delta^{n}_{m},
\end{align}
which proves the fourth identity.

Now we are left with the proof of the last identity.
Which is only true for $(n+m)$ even and $n,m \in \N^{\ast}$.
Employing our usual trick we get
\begin{align}
 \frac{1}{d_{\gamma}} &\tr ( \gamma^{\rho_{1} \ldots \rho_{n}} \gamma_{\lambda_{1} \ldots \lambda_{m}} \gamma_{\mu \nu} ) \notag
\\
 =& n \delta^{[\rho_{1}}_{\nu} \frac{1}{d_{\gamma}} \tr ( \gamma^{\rho_{2} \ldots \rho_{n}]} \gamma_{\lambda_{1} \ldots \lambda_{m}} \gamma_{\mu} ) \notag
\\
  &+ m (-1)^{n} \metric_{\nu [ \lambda_{1}} \frac{1}{d_{\gamma}} \tr ( \gamma_{\lambda_{2} \ldots \lambda_{m}]} \gamma_{\mu} \gamma^{\rho_{1} \ldots \rho_{n}} )
\end{align}
and
\begin{align}
 \frac{1}{d_{\gamma}} &\tr ( \gamma_{\lambda_{1} \ldots \lambda_{m}} \gamma^{\rho_{1} \ldots \rho_{n}} \gamma_{\mu \nu} ) \notag
\\
 =& m \metric_{\nu [ \lambda_{1}} \frac{1}{d_{\gamma}} \tr ( \gamma_{\lambda_{2} \ldots \lambda_{m}]} \gamma^{\rho_{1} \ldots \rho_{n}} \gamma_{\mu} ) \notag
\\
  &+ n (-1)^{m} \delta^{[\rho_{1}}_{\nu} \frac{1}{d_{\gamma}} \tr ( \gamma^{\rho_{2} \ldots \rho_{n}]} \gamma_{\mu} \gamma_{\lambda_{1} \ldots \lambda_{m}} ) \text{.}
\end{align}
There are two distinct cases, $n,m$ even and $n,m$ odd.
Starting with $n,m$ even we shift $n \to 2l$ and $m \to 2k$ and find
\begin{align}
 \frac{1}{d_{\gamma}} &\tr ( [ \gamma^{\rho_{1} \ldots \rho_{2l}} , \gamma_{\lambda_{1} \ldots \lambda_{2k}} ] \gamma_{\mu \nu} ) \notag
\\
 =& 2 l \delta^{[\rho_{1}}_{\nu} \frac{1}{d_{\gamma}} \tr ( [ \gamma^{\rho_{2} \ldots \rho_{2l}]} , \gamma_{\lambda_{1} \ldots \lambda_{2k}} ] \gamma_{\mu} ) \notag
\\
 &- 2 k \metric_{\nu [ \lambda_{1}} \frac{1}{d_{\gamma}} \tr ( [ \gamma_{\lambda_{2} \ldots \lambda_{2k}]} , \gamma^{\rho_{1} \ldots \rho_{2l}} ] \gamma_{\mu} ) \notag
\\
 =& - 4 \cdot 2l \cdot (2l)! \cdot (-1)^{\frac{2l (2l-1)}{2}} \! \cdot \! \delta^{[\rho_{1}}_{[\mu} \metric_{\nu] [\lambda_{1}} \deltaA^{\rho_{2} \ldots \rho_{2l}]}_{\lambda_{2} \ldots \lambda_{2l}]} \cdot \delta^{l}_{k} \text{.}
\end{align}
This gives us the relation \Eqref{eq:App:special_trace_commutator} for $n,m$ even.
On the other hand we now can take $n,m$ odd and therefore shift $n \to 2l - 1$ and $m \to 2k - 1$.
Now the commutator reads
\begin{align}
 \frac{1}{d_{\gamma}} &\tr ( [ \gamma^{\rho_{1} \ldots \rho_{2l-1}} , \gamma_{\lambda_{1} \ldots \lambda_{2k-1}} ] \gamma_{\mu \nu} ) \notag
\\
 =& (2 l - 1) \delta^{[\rho_{1}}_{\nu} \frac{1}{d_{\gamma}} \tr ( \gamma^{\rho_{2} \ldots \rho_{2l-1}]} \{ \gamma_{\lambda_{1} \ldots \lambda_{2k - 1}} , \gamma_{\mu} \} ) \notag
\\
 &- (2 k - 1) \metric_{\nu [ \lambda_{1}} \frac{1}{d_{\gamma}} \tr ( \gamma_{\lambda_{2} \ldots \lambda_{2k - 1}]} \{ \gamma^{\rho_{1} \ldots \rho_{2l - 1}} , \gamma_{\mu} \} ) \notag
\\
 =& - 4 \cdot (2l - 1) \cdot (2l - 1)! \cdot (-1)^{\frac{(2l - 1)( (2l - 1) - 1 )}{2}} \! \notag
\\
 & \times \delta^{[\rho_{1}}_{[\mu} \metric_{\nu] [\lambda_{1}} \deltaA^{\rho_{2} \ldots \rho_{2l - 1}]}_{\lambda_{2} \ldots \lambda_{2l - 1}]} \cdot \delta^{l}_{k} \text{,}
\end{align}
proving the last identity for $n,m$ odd.

%-----------------------------------------------------------------------------
\section{Existence and uniqueness of the spin connection}
\label{App:Existence_of_the_spin_connection}
%-----------------------------------------------------------------------------

In this appendix we prove the existence and the uniqueness of the spin connection $\hat{\Gamma}_{\mu}$ implicitly defined as
\begin{align}\label{eq:App:definition_hatGamma}
 \partial_{\mu} \gamma^{\nu} + \christoffel{\nu \\ \mu \rho} \gamma^{\rho} = - [ \hat{\Gamma}_{\mu} , \gamma^{\nu} ],\quad \tr \hat{\Gamma}_{\mu} = 0 \text{.}
\end{align}
We follow the idea of Weldon in \cite{Weldon:2000fr} to prove the existence.
First we expand the $\gamma_{\mu}$ and the metric around some arbitrary spacetime point $x$
\begin{align}
 \gamma^{\nu}(x + \mrd x) &{}\simeq \gamma^{\mu}(x) + \mrd x^{\mu} \partial_{\mu} \gamma^{\nu}(x),\\
 \metric^{\nu \lambda}(x + \mrd x) &{}\simeq \metric^{\nu \lambda}(x) + \mrd x^{\mu} \partial_{\mu} \metric^{\nu \lambda}(x) \text{.}
\end{align}
Next we plug the variations of the metric and the Dirac matrices into the Weldon theorem to get
\begin{align}
 \mrd x^{\mu} \partial_{\mu} \gamma^{\nu} = \frac{1}{2} \mrd x^{\mu} ( \partial_{\mu} \metric^{\nu \lambda} ) \gamma_{\lambda} + [  \delta S_{\gamma} , \gamma^{\nu} ] \text{.}
\end{align}
Since this equation has to be fulfilled for all infinitesimal changes of the coordinates $\mrd x^{\mu}$, we can also expand $\delta \mcS_{\gamma} = \mrd x^{\mu} (\mcS_{\gamma})_{\mu}$, where the $(\mcS_{\gamma})_{\mu}$ are specified by the explicit choice $\gamma_{\mu}(x)$ (as a function of spacetime) and will therefore transform non homogeneously under coordinate transformations, i.e.~spacetime coordinate as well as spin-base transformations.

Additionally we employ the metric compatibility of the Christoffel symbol
\begin{align}
 \partial_{\mu} \metric^{\nu \lambda} = - \christoffel{\nu \\ \mu \rho} \metric^{\rho \lambda} - \christoffel{\lambda \\ \mu \rho} \metric^{\rho \nu}
\end{align}
to conclude
\begin{align}
 \partial_{\mu} \gamma^{\nu} + \christoffel{\nu \\ \mu \rho} \gamma^{\rho} = - \left[ \frac{1}{8} \christoffel{\alpha \\ \mu \rho} \metric^{\rho \beta} [\gamma_{\alpha} , \gamma_{\beta}] - (\mcS_{\gamma})_{\mu} , \gamma^{\nu} \right] \text{,}
\end{align}
where we took advantage of the identity
\begin{align}
 \big[ [ \gamma_{\alpha} , \gamma_{\beta} ] , \gamma^{\nu} \big] = 4 \delta^{\nu}_{\beta} \gamma_{\alpha} - 4 \delta^{\nu}_{\alpha} \gamma_{\beta} \text{.}
\end{align}
This means that $\partial_{\mu} \gamma^{\nu} + \christoffel{\nu \\ \mu \rho} \gamma^{\rho}$ can be written as a commutator.
Furthermore at least one $\hat{\Gamma}_{\mu}$ fulfilling \Eqref{eq:App:definition_hatGamma} exists.

Since we know that there exists a solution to \Eqref{eq:App:definition_hatGamma}, we can expand $\hat{\Gamma}_{\mu}$ with the basis elements from \Eqref{eq:gamma_matrices_basis} and \eqref{eq:gamma_matrices_basis_even}
\begin{align}
 &\hat{\Gamma}_{\mu} = \sum\limits_{n=1}^{d} \hat{m}_{\mu \rho_{1} \ldots \rho_{n}} \gamma^{\rho_{1} \ldots \rho_{n}} , \quad d \text{ even,}
\\
 &\hat{\Gamma}_{\mu} = \sum\limits_{n=1}^{\frac{d-1}{2}} \hat{m}_{\mu \rho_{1} \ldots \rho_{2n}} \gamma^{\rho_{1} \ldots \rho_{2n}} , \quad d \text{ odd.}
\end{align}
From \Eqref{eq:App:definition_hatGamma} we infer by calculating the commutator with $\gamma_{\nu}$
\begin{align}
 \big[ [ \hat{\Gamma}_{\mu} , \gamma^{\nu} ] , \gamma_{\nu} \big] = - [ (\DLC_{\mu} \gamma^{\nu}) , \gamma_{\nu} ] \text{.}
\end{align}
Plugging in our ansatz for $\hat{\Gamma}_{\mu}$ and using the identity \Eqref{eq:App:gamma_contraction_commutator} from chapter \ref{App:Special_relations_gamma_matrices} we get for the left hand side in the even dimensional case
\begin{align}
 \big[ [& \hat{\Gamma}_{\mu} , \gamma^{\nu} ] , \gamma_{\nu} \big] \notag
\\
 &= \sum\limits_{n=1}^{d} 2 \big( (1 - (-1)^{n}) d + (-1)^{n} 2 n \big) \hat{m}_{\mu \rho_{1} \ldots \rho_{n}} \gamma^{\rho_{1} \ldots \rho_{n}}
\end{align}
and in the odd dimensional case
\begin{align}
 \big[ [& \hat{\Gamma}_{\mu} , \gamma^{\nu} ] , \gamma_{\nu} \big] = \sum\limits_{n=1}^{\frac{d-1}{2}} 8 n \,\, \hat{m}_{\mu \rho_{1} \ldots \rho_{2n}} \gamma^{\rho_{1} \ldots \rho_{2n}} \text{.}
\end{align}
The right hand side can be expanded into the basis from \Eqref{eq:gamma_matrices_basis} as well
\begin{align}
 - [ (\DLC_{\mu} \gamma^{\nu}) , \gamma_{\nu} ] = \sum\limits_{n=1}^{d} \hat{a}_{\mu \rho_{1} \ldots \rho_{n}} \gamma^{\rho_{1} \ldots \rho_{n}}, d \text{ even,}
\\
 - [ (\DLC_{\mu} \gamma^{\nu}) , \gamma_{\nu} ] = \sum\limits_{n=1}^{\frac{d-1}{2}} \hat{a}_{\mu \rho_{1} \ldots \rho_{2n}} \gamma^{\rho_{1} \ldots \rho_{2n}}, d \text{ odd.}
\end{align}
The coefficients $\hat{a}_{\mu \rho_{1} \ldots \rho_{n}}$ or respectively $\hat{a}_{\mu \rho_{1} \ldots \rho_{2n}}$ can be calculated employing the orthogonality of the trace \Eqref{eq:App:Anm_d_even} and \eqref{eq:App:Anm_d_odd}
\begin{align}
 \hat{a}_{\mu \rho_{1} \ldots \rho_{n}} &= - \frac{(-1)^{\frac{n(n-1)}{2}}}{n! \, d_{\gamma}} \tr\big( \gamma_{\rho_{1} \ldots \rho_{n}} [ \DLC_{\mu} \gamma^{\nu} , \gamma_{\nu} ] \big),
\\
 \hat{a}_{\mu \rho_{1} \ldots \rho_{2n}} &= - \frac{(-1)^{n}}{(2n)! \, d_{\gamma}} \tr\big( \gamma_{\rho_{1} \ldots \rho_{2n}} [ \DLC_{\mu} \gamma^{\nu} , \gamma_{\nu} ] \big) \text{.}
\end{align}
Since the $\gamma^{\rho_{1} \ldots \rho_{n}}$ or respectively the $\gamma^{\rho_{1} \ldots \rho_{2n}}$ form a basis we are allowed to compare the coefficients and find
\begin{align}
 \hat{m}_{\mu \rho_{1} \ldots \rho_{n}} &= \frac{\hat{a}_{\mu \rho_{1} \ldots \rho_{n}}}{2 \big( ( 1 - (-1)^{n} ) d + (-1)^{n} 2 n \big)}, \quad d \text{ even,}
\\
 \hat{m}_{\mu \rho_{1} \ldots \rho_{2n}} &= \frac{\hat{a}_{\mu \rho_{1} \ldots \rho_{2n}}}{8  n}, \quad d \text{ odd.}
\end{align}
With the last equations we have shown the uniqueness and have given an explicit expression for $\hat{\Gamma}_{\mu}$ in terms of the $\gamma_{\mu}$ and their first derivatives.

%-----------------------------------------------------------------------------
\section{Spin metric}
\label{App:spin_metric}
%-----------------------------------------------------------------------------

The spin metric is an important quantity in our investigations. We found that it is restricted to satisfy
\begin{align}
\begin{aligned}
 \rmi {}& \quad \gamma_{\mu}^{\dagger} = - \spinmetric \gamma_{\mu} \spinmetric^{-1} \text{,}\\
 \rmii {}& \quad \abs{\det \spinmetric} = 1 \text{,}\\
 \rmiii {}& \quad \spinmetric^{\dagger} = - \spinmetric\text{.}
\end{aligned}
\end{align}
In \cite{Gies:2013noa} it was already shown that these conditions are sufficient to determine (up to a sign) the spin metric $\spinmetric$ in terms of the Dirac matrices $\gamma_{\mu}$ in $4$ spacetime dimensions.
The way this proof was done is actually true for all even dimensions.
And we will see that with minor modifications this proof also generalizes to all integer dimensions $d \geq 2$.

In order to be self consistent we give the full proof again.
As a first step we show the uniqueness (up to a sign) of the spin metric.
Let us assume that there is at least one spin metric $h_{1}$,
which satisfies all three conditions. Then we know, if there is
another spin metric $\spinmetric_{2}$, they must be related via
\begin{align}
 [ \spinmetric_{2}^{-1} \spinmetric_{1} , \gamma_{\mu} ] = 0 \text{,}
\end{align}
because both spin metrics have to fulfill
\begin{align}
 \spinmetric_{2} \gamma_{\mu} \spinmetric_{2}^{-1} = - \gamma_{\mu}^{\dagger} = \spinmetric_{1} \gamma_{\mu} \spinmetric_{1}^{-1} \text{.}
\end{align}
Therefore, using Schur's Lemma,
\begin{align}
 \spinmetric_{2} = z \spinmetric_{1} , \quad z \in \C \text{,}
\end{align}
has to hold. With $\rmii$, it follows that
\begin{align}
 \abs{z} = 1 \text{.}
\end{align}
But if both spin metrics satisfy the condition $\rmiii$, then
\begin{align}
 z^{\ast} \spinmetric_{1} = - z^{\ast} \spinmetric_{1}^{\dagger}  = - \spinmetric_{2}^{\dagger} = \spinmetric_{2} = z \spinmetric_{1}
\end{align}
has to hold. Therefore both spin metrics have to be identical up to a sign,
\begin{align}
 \spinmetric_{2} = \pm \spinmetric_{1} \text{.}
\end{align}
This demonstrates the uniqueness (up to a sign) of the spin
metric.

Now we only need to prove the existence of one such spin metric
$h$. For this, we first introduce the Matrix $\hat{M}$ satisfying
\begin{align}
 \gamma_{\mu}^{\dagger} = - \euler^{\hat{M}} \gamma_{\mu} \euler^{- \hat{M}} , \quad \tr \hat{M} = 0 \label{eq:def_M} \text{.}
\end{align}
The existence of such a matrix in every dimension is guaranteed by the Clifford algebra and our sign conventions.
In even dimensions the existence is obvious since $\gamma_{\mu}^{\dagger}$ and $- \gamma_{\mu}$ satisfy the Clifford algebra and therefore there must exist a connecting similarity transformation.
For odd dimensions we use that the hermitean conjugation can change the connected component of the representation of the Clifford algebra depending on the signature of the metric.
According to \cite{Cornwell:1989bx} the number of ``$+$'' signs in the signature tells us whether the connected component is changed or not.
For an even number of plus signs the connected component is changed, whereas for an odd number it is not.
In our case we have $d-1$ plus signs in the signature, i.e.~for $d$ odd we have an even number leading to a change of the connected component and therefore we need the minus sign in \Eqref{eq:def_M}.
The trace of $\hat{M}$ can always be set to zero,
because the trace part commutes with all matrices and therefore drops out of \Eqref{eq:def_M}.

The hermitean conjugate of \Eqref{eq:def_M} is
\begin{align}
 \gamma_{\mu} = - \euler^{- \hat{M}^{\dagger}} \gamma_{\mu}^{\dagger} \euler^{\hat{M}^{\dagger}}.
\end{align}
Therefore, also 
\begin{align}
 \euler^{\hat{M}} \gamma_{\mu} \euler^{-\hat{M}} = - \gamma_{\mu}^{\dagger} = \euler^{\hat{M}^{\dagger}} \gamma_{\mu} \euler^{- \hat{M}^{\dagger}}
\end{align}
has to hold. Schur's Lemma again implies there exists a $\varphi$ such that
\begin{align}
 \euler^{\hat{M}^{\dagger}} = \euler^{\cplx \varphi} \euler^{\hat{M}} , \quad \varphi \in \R \text{.}
\end{align}
This equation fixes $\euler^{\cplx \varphi}$ once we have chosen a
specific $\hat{M}$. Now we also know, that $\det \euler^{\hat{M}} = 1$
and therefore the same has to hold for $\det
\euler^{\hat{M}^{\dagger}} = 1$. From this, we conclude that $\varphi$ is
limited to
\begin{align}
 \varphi \in \left\{ n \frac{2 \pi}{d_{\gamma}} : n \in \{ 0 , \ldots , d_{\gamma} - 1 \} \right\} \text{.}
\end{align}
The desired spin metric $h$ is then given by
\begin{align}
 h = \cplx \euler^{\cplx \frac{\varphi}{2}} \euler^{\hat{M}} \text{.}
\end{align}
It is straightforward to show, that this metric satisfies $\rmi$ - $\rmiii$.

It is helpful to note that the determinant of the spin metric is also fixed and even independent of the set of Dirac matrices.
To show this we just use that $d_{\gamma}$ is even for $d \geq 2$ and therefore the sign ambiguity of the spin metric is not important for the determinant and additionally the %for the spin metric non trivial part $\mcS \in \mathrm{SL}(d_{\gamma} , \C)$ of a spin-base transformation $\mcS_{\varphi} \in \mathrm{SB}(d_{\gamma})$ has determinant one.
determinant of a spin-base transformation $\mcS \in \mathrm{SL}(d_{\gamma} , \C)$ is equal to one.
With our previous investigations there are only two possibilities, namely
\begin{align}
 \det \spinmetric = \pm 1 \text{.}
\end{align}
Since we always can choose local inertial coordinates in an arbitrary point $x$ of the manifold it is sufficient to calculate the determinant in this frame with a special chosen set of Dirac matrices compatible with the Clifford algebra.
Hence, we can take a representation which fulfills%
\footnote{
An explicit example is $\gamma_{0}(x) = \cplx \sigma_{1} (\otimes \sigma_{0})^{\lfloor \frac{d}{2} \rfloor - 1 }$, $\gamma_{1}(x) = \sigma_{2} ( \otimes \sigma_{0} )^{\lfloor \frac{d}{2} \rfloor - 1}$, $\gamma_{2j}(x) = (\sigma_{3} \otimes)^{j} \sigma_{1} ( \otimes \sigma_{0} )^{\lfloor \frac{d}{2} \rfloor - 1 - j}$, $\gamma_{2j+1}(x) = (\sigma_{3} \otimes)^{j} \sigma_{2} ( \otimes \sigma_{0} )^{\lfloor \frac{d}{2} \rfloor - 1 - j}$, $j \in \{ 1, \ldots , \lfloor \frac{d}{2} \rfloor - 1 \}$ and for odd dimensions we additionally need $\gamma_{d-1}(x) = \sigma_{3} ( \otimes \sigma_{3} )^{\lfloor \frac{d}{2} \rfloor - 1}$.
}
\begin{align}
 \gamma_{0}(x)^{\dagger} = - \gamma_{0}(x), \quad \gamma_{i}(x)^{\dagger} = \gamma_{i}(x), \quad i \in \{ 1, \ldots , d-1 \} \text{.}
\end{align}
In such a representation the spin metric at the spacetime point $x$ is given by $\pm \gamma_{0}(x)$ since
\begin{align}
 &\gamma_{0}^{\dagger}(x) \! = - \gamma_{0}(x) = - \gamma_{0}(x) \gamma_{0}(x) \big( \gamma_{0}(x) \big)^{-1},
\\
 &\gamma_{i}^{\dagger}(x) \! = \gamma_{i}(x) = - \gamma_{0}(x) \gamma_{i}(x) \big( \gamma_{0}(x) \big)^{-1}, i \! \in \! \{ 1 , \ldots , d-1 \},
\\
 &\det \gamma_{0}(x) \! = \det \euler^{\frac{\pi}{2} \gamma_{0}(x)} = \euler^{\tr \frac{\pi}{2} \gamma_{0}(x)} = 1 \text{.}
\end{align}
With this special choice we prove the general relation
\begin{align}
 \det \spinmetric = 1 \text{.}
\end{align}
%
% Note that we have given a different proof for the existence of the spin metric with the preceding construction.
% 
We continue with implementing the spin metric compatibility
as expressed in \Eqref{eq:spin_metric_compatibility}.
This tells us that
\begin{align}
 \Gamma_{\mu} + \bar{\Gamma}_{\mu} = \spinmetric^{-1} \partial_{\mu} \spinmetric
\end{align}
has to hold.
Taking into account that (cf. \Eqref{eq:def_hatGamma})
\begin{align}
 - {D_\ttm{(\mathrm{LC})}}_{\mu} \spinmetric \gamma^{\nu} \spinmetric^{-1} \! = \! {D_\ttm{(\mathrm{LC})}}_{\mu} {\gamma^{\nu}}^{\dagger} \! = \! ({D_\ttm{(\mathrm{LC})}}_{\mu} \gamma^{\nu})^{\dagger} \! = - [ \hat{\Gamma}_{\mu} , \gamma^{\nu} ]^{\dagger},
\end{align}
we arrive at
\begin{align}
 \big[ \spinmetric^{-1} ( \partial_{\mu} \spinmetric ) - \hat{\Gamma}_{\mu} - \bar{\hat{\Gamma}}_{\mu} , \gamma^{\nu} \big] = 0 \text{.}
\end{align}
Because $\tr \hat{\Gamma}_{\mu} = 0$, this implies
\begin{align}
 \hat{\Gamma}_{\mu} + \bar{\hat{\Gamma}}_{\mu} = \spinmetric^{-1} \partial_{\mu} \spinmetric - \frac{1}{d_{\gamma}} \tr ( \spinmetric^{-1} \partial_{\mu} \spinmetric ) \, \mrI \text{.}
\end{align}
Now we use $\det \spinmetric = 1$ to deduce
\begin{align}
 0 = \partial_{\mu} \det \spinmetric = \partial_{\mu} \euler^{\tr \ln \spinmetric} = \partial_{\mu} \tr ( \ln \spinmetric ) = \tr ( \spinmetric^{-1} \partial_{\mu} \spinmetric ) \text{.}
\end{align}
This leaves us with
\begin{align}\label{eq:reality_DeltaGamma_1}
 \Gamma_{\mu} + \bar{\Gamma}_{\mu} = h^{-1} \partial_{\mu} h = \hat{\Gamma}_{\mu} + \bar{\hat{\Gamma}}_{\mu} \text{,}
\end{align}
which implies that
\begin{align}\label{eq:reality_DeltaGamma_2}
 \Re \tr \Gamma_{\mu} = 0 \text{.}
\end{align}
These two identities are used in Sect.~\ref{sec:spin_connection} to constrain spin torsion.


\begin{thebibliography}{89}


%\cite{Ashtekar:2014kba}
\bibitem{Ashtekar:2014kba} 
  A.~Ashtekar, M.~Reuter and C.~Rovelli,
  %``From General Relativity to Quantum Gravity,''
  arXiv:1408.4336 [gr-qc].
  %%CITATION = ARXIV:1408.4336;%%
  %5 citations counted in INSPIRE as of 05 Feb 2015


\bibitem{Ashtekar:2004eh} 
  A.~Ashtekar and J.~Lewandowski,
  %``Background independent quantum gravity: A Status report,''
  Class.\ Quant.\ Grav.\  {\bf 21}, R53 (2004)
  [gr-qc/0404018]; %.
  %%CITATION = GR-QC/0404018;%%
  %791 citations counted in INSPIRE as of 17 Jul 2013


%\cite{Weinberg:1980gg} %gravity from asymptotic safety
\bibitem{Weinberg:1980gg}
  S.~Weinberg,
  %``Ultraviolet Divergences In Quantum Theories Of Gravitation,''
%\href{http://www.slac.stanford.edu/spires/find/hep/www?irn=784877}{SPIRES entry}
{\it  In *Hawking, S.W., Israel, W.: General Relativity*, 790-831}.


%\cite{Reuter:1996cp} %asymptotic safety in metric degrees of freedom
\bibitem{Reuter:1996cp}
  M.~Reuter,
  %``Nonperturbative Evolution Equation for Quantum Gravity,''
  Phys.\ Rev.\  D {\bf 57}, 971 (1998)
  [arXiv:hep-th/9605030]; %.
  %%CITATION = PHRVA,D57,971;%%
%\cite{Niedermaier:2006wt}
%\bibitem{Niedermaier:2006wt}
  M.~Niedermaier and M.~Reuter,
  %``The Asymptotic Safety Scenario in Quantum Gravity,''
  Living Rev.\ Rel.\  {\bf 9}, 5 (2006); %.
  %%CITATION = 00222,9,5;%%
%\cite{Percacci:2007sz}
%\bibitem{Percacci:2007sz} 
  R.~Percacci,
  %``Asymptotic Safety,''
  In *Oriti, D. (ed.): Approaches to quantum gravity* 111-128
  [arXiv:0709.3851 [hep-th]]; %.
  %%CITATION = ARXIV:0709.3851;%%
  %106 citations counted in INSPIRE as of 04 Mar 2013
%\cite{Reuter:2012id}
%\bibitem{Reuter:2012id} 
  M.~Reuter and F.~Saueressig,
  %``Quantum Einstein Gravity,''
  New J.\ Phys.\  {\bf 14}, 055022 (2012)
  [arXiv:1202.2274 [hep-th]].
  %%CITATION = ARXIV:1202.2274;%%
  %32 citations counted in INSPIRE as of 04 Mar 2013


%\cite{Harst:2012ni}
\bibitem{Harst:2012ni} %asymptotic safety in vielbein degrees of freedom
  U.~Harst and M.~Reuter,
  %``The 'Tetrad only' theory space: Nonperturbative renormalization flow and Asymptotic Safety,''
  JHEP {\bf 1205}, 005 (2012)
  [arXiv:1203.2158 [hep-th]]; %.
  %%CITATION = ARXIV:1203.2158;%%
  %15 citations counted in INSPIRE as of 27 May 2014
%
%\cite{Dona:2012am}
%\bibitem{Dona:2012am} 
  P.~Don{\`a} and R.~Percacci,
  %``Functional renormalization with fermions and tetrads,''
  Phys.\ Rev.\ D {\bf 87}, 045002 (2013)
  [arXiv:1209.3649 [hep-th]]; %.
  %%CITATION = ARXIV:1209.3649;%%
  %8 citations counted in INSPIRE as of 17 Jul 2013

%\cite{Plebanski:1977zz}
\bibitem{Plebanski:1977zz} %gravity as connection
  J.~F.~Plebanski,
  %``On the separation of Einsteinian substructures,''
  J.\ Math.\ Phys.\  {\bf 18}, 2511 (1977); %.
  %%CITATION = JMAPA,18,2511;%%
  %215 citations counted in INSPIRE as of 17 Jul 2013
%
%\cite{Capovilla:1989ac}
%\bibitem{Capovilla:1989ac} 
  R.~Capovilla, T.~Jacobson and J.~Dell,
  %``General Relativity Without the Metric,''
  Phys.\ Rev.\ Lett.\  {\bf 63}, 2325 (1989); %.
  %%CITATION = PRLTA,63,2325;%%
  %142 citations counted in INSPIRE as of 17 Jul 2013
%\cite{Capovilla:1991kx}
%\bibitem{Capovilla:1991kx} 
%  R.~Capovilla, T.~Jacobson and J.~Dell,
  %``A Pure spin connection formulation of gravity,''
  Class.\ Quant.\ Grav.\  {\bf 8}, 59 (1991); %.
  %%CITATION = CQGRD,8,59;%%
  %104 citations counted in INSPIRE as of 17 Jul 2013
%
%\cite{Krasnov:2011up}
%\bibitem{Krasnov:2011up} 
  K.~Krasnov,
  %``Gravity as a diffeomorphism invariant gauge theory,''
  Phys.\ Rev.\ D {\bf 84}, 024034 (2011)
  [arXiv:1101.4788 [hep-th]]; %.
  %%CITATION = ARXIV:1101.4788;%%
  %12 citations counted in INSPIRE as of 17 Jul 2013
%\cite{Krasnov:2011pp}
%\bibitem{Krasnov:2011pp} 
%  K.~Krasnov,
  %``Pure Connection Action Principle for General Relativity,''
  Phys.\ Rev.\ Lett.\  {\bf 106}, 251103 (2011)
  [arXiv:1103.4498 [gr-qc]].
  %%CITATION = ARXIV:1103.4498;%%
  %13 citations counted in INSPIRE as of 17 Jul 2013


%\cite{Perez:2004hj}
\bibitem{Perez:2004hj} %LQG and spin foams
  A.~Perez,
  %``Introduction to loop quantum gravity and spin foams,''
  gr-qc/0409061.
  %%CITATION = GR-QC/0409061;%%
  %153 citations counted in INSPIRE as of 18 Feb 2015
%
%\cite{Alexandrov:2010un}
%\bibitem{Alexandrov:2010un} 
  S.~Alexandrov and P.~Roche,
  %``Critical Overview of Loops and Foams,''
  Phys.\ Rept.\  {\bf 506}, 41 (2011)
  [arXiv:1009.4475 [gr-qc]]; %.
  %%CITATION = ARXIV:1009.4475;%%
  %26 citations counted in INSPIRE as of 18 Feb 2015
%
%\cite{Dittrich:2014mxa}
%\bibitem{Dittrich:2014mxa} 
  B.~Dittrich, S.~Mizera and S.~Steinhaus,
  %``Decorated tensor network renormalization for lattice gauge theories and spin foam models,''
  arXiv:1409.2407 [gr-qc].
  %%CITATION = ARXIV:1409.2407;%%
  %5 citations counted in INSPIRE as of 18 Feb 2015


%\cite{Bombelli:1987aa}
\bibitem{Bombelli:1987aa}  %gravity from Causal Sets
  L.~Bombelli, J.~Lee, D.~Meyer and R.~Sorkin,
  %``Space-Time as a Causal Set,''
  Phys.\ Rev.\ Lett.\  {\bf 59}, 521 (1987); %.
  %%CITATION = PRLTA,59,521;%%
  %359 citations counted in INSPIRE as of 18 Feb 2015
%
%\cite{Dowker:aza}
%\bibitem{Dowker:aza}
  F.~Dowker,
  %``Introduction to causal sets and their phenomenology,''
  Gen.\ Rel.\ Grav.\  {\bf 45}, no. 9, 1651 (2013).
  %%CITATION = GRGVA,45,1651;%%
  %2 citations counted in INSPIRE as of 18 Feb 2015


%\cite{Ambjorn:1998xu}
\bibitem{Ambjorn:1998xu} %gravity in CDT
  J.~Ambj{\o}rn and R.~Loll,
  %``Nonperturbative Lorentzian quantum gravity, causality and topology change,''
  Nucl.\ Phys.\ B {\bf 536}, 407 (1998)
  [hep-th/9805108]; %.
  %%CITATION = HEP-TH/9805108;%%
  %209 citations counted in INSPIRE as of 18 Feb 2015
%
%\cite{Ambjorn:2014gsa}
%\bibitem{Ambjorn:2014gsa} 
  J.~Ambj{\o}rn, A.~G{\"o}rlich, J.~Jurkiewicz, A.~Kreienbuehl and R.~Loll,
  %``Renormalization Group Flow in CDT,''
  Class.\ Quant.\ Grav.\  {\bf 31}, 165003 (2014)
  [arXiv:1405.4585 [hep-th]].
  %%CITATION = ARXIV:1405.4585;%%
  %12 citations counted in INSPIRE as of 18 Feb 2015


%\cite{Horava:2009uw}
\bibitem{Horava:2009uw}  %Horava Lifshitz Gravity
  P.~Ho{\v r}ava,
  %``Quantum Gravity at a Lifshitz Point,''
  Phys.\ Rev.\ D {\bf 79}, 084008 (2009)
  [arXiv:0901.3775 [hep-th]]; %.
  %%CITATION = ARXIV:0901.3775;%%
  %1069 citations counted in INSPIRE as of 18 Feb 2015
%
%\cite{Benedetti:2013pya}
%\bibitem{Benedetti:2013pya} 
  D.~Benedetti and F.~Guarnieri,
  %``One-loop renormalization in a toy model of Hořava-Lifshitz gravity,''
  JHEP {\bf 1403}, 078 (2014)
  [arXiv:1311.6253 [hep-th]]; %.
  %%CITATION = ARXIV:1311.6253;%%
  %6 citations counted in INSPIRE as of 18 Feb 2015
%
%\cite{Rechenberger:2012dt}
%\bibitem{Rechenberger:2012dt} 
  S.~Rechenberger and F.~Saueressig,
  %``A functional renormalization group equation for foliated spacetimes,''
  JHEP {\bf 1303}, 010 (2013)
  [arXiv:1212.5114]; %.
  %%CITATION = ARXIV:1212.5114;%%
  %12 citations counted in INSPIRE as of 18 Feb 2015
%
%\cite{D'Odorico:2014iha}
%\bibitem{D'Odorico:2014iha} 
  G.~D'Odorico, F.~Saueressig and M.~Schutten,
  %``Asymptotic Freedom in Hořava-Lifshitz Gravity,''
  Phys.\ Rev.\ Lett.\  {\bf 113}, no. 17, 171101 (2014)
  [arXiv:1406.4366 [gr-qc]].
  %%CITATION = ARXIV:1406.4366;%%
  %3 citations counted in INSPIRE as of 18 Feb 2015


%\cite{Ambjorn:2010hu}
\bibitem{Ambjorn:2010hu} %CDT and Horava-Lifshitz
  J.~Ambj{\o}rn, A.~G{\"o}rlich, S.~Jordan, J.~Jurkiewicz and R.~Loll,
  %``CDT meets Horava-Lifshitz gravity,''
  Phys.\ Lett.\ B {\bf 690}, 413 (2010)
  [arXiv:1002.3298 [hep-th]]; %.
  %%CITATION = ARXIV:1002.3298;%%
  %56 citations counted in INSPIRE as of 18 Feb 2015
%
%\cite{Anderson:2011bj}
%\bibitem{Anderson:2011bj} 
  C.~Anderson, S.~J.~Carlip, J.~H.~Cooperman, P.~Ho{\v r}ava, R.~K.~Kommu and P.~R.~Zulkowski,
  %``Quantizing Horava-Lifshitz Gravity via Causal Dynamical Triangulations,''
  Phys.\ Rev.\ D {\bf 85}, 044027 (2012)
  [arXiv:1111.6634 [hep-th]].
  %%CITATION = ARXIV:1111.6634;%%
  %42 citations counted in INSPIRE as of 18 Feb 2015


%\cite{Percacci:2002ie}
\bibitem{Percacci:2002ie} 
  R.~Percacci and D.~Perini,
  %``Constraints on matter from asymptotic safety,''
  Phys.\ Rev.\ D {\bf 67}, 081503 (2003)
  [hep-th/0207033].
  %%CITATION = HEP-TH/0207033;%%
  %139 citations counted in INSPIRE as of 05 Feb 2015


%\cite{Percacci:2003jz}
\bibitem{Percacci:2003jz} 
  R.~Percacci and D.~Perini,
  %``Asymptotic safety of gravity coupled to matter,''
  Phys.\ Rev.\ D {\bf 68}, 044018 (2003)
  [hep-th/0304222].
  %%CITATION = HEP-TH/0304222;%%
  %159 citations counted in INSPIRE as of 05 Feb 2015


%\cite{Eichhorn:2011pc}
\bibitem{Eichhorn:2011pc} 
  A.~Eichhorn and H.~Gies,
  %``Light fermions in quantum gravity,''
  New J.\ Phys.\  {\bf 13}, 125012 (2011)
  [arXiv:1104.5366 [hep-th]].
  %%CITATION = ARXIV:1104.5366;%%
  %19 citations counted in INSPIRE as of 25 Sep 2013


%\cite{Dona:2013qba}
\bibitem{Dona:2013qba} 
  P.~Donà, A.~Eichhorn and R.~Percacci,
  %``Matter matters in asymptotically safe quantum gravity,''
  Phys.\ Rev.\ D {\bf 89}, no. 8, 084035 (2014)
  [arXiv:1311.2898 [hep-th]].
  %%CITATION = ARXIV:1311.2898;%%
  %23 citations counted in INSPIRE as of 05 Feb 2015


%\cite{Dona:2014pla}
\bibitem{Dona:2014pla} 
  P.~Donà, A.~Eichhorn and R.~Percacci,
  %``Consistency of matter models with asymptotically safe quantum gravity,''
  arXiv:1410.4411 [gr-qc].
  %%CITATION = ARXIV:1410.4411;%%
  %3 citations counted in INSPIRE as of 05 Feb 2015


%\cite{Weyl:1929} 
\bibitem{Weyl:1929}
  H.~Weyl, 
  %``Elektron und Gravitation I'', 
  Z.\ Phys.\ {\bf 56}, 330 (1929).


%\cite{Fock:1929}
\bibitem{Fock:1929}
  V.~Fock and D.~Ivanenko,
  %``G\'eometrie quantique lin\'eaire et d\'eplacement parall\'ele,''
  Compt.\ Rend.\ Acad.\ Sci.\ Paris {\bf 188}, 1470, (1929).


%\cite{DeWitt:1965jb}
\bibitem{DeWitt:1965jb} 
  B.~S.~DeWitt,
  ``Dynamical theory of groups and fields,''
  Gordon \& Breach, New York, 1965.


%\cite{Buchbinder:1992rb}
\bibitem{Buchbinder:1992rb} 
  I.~L.~Buchbinder, S.~D.~Odintsov and I.~L.~Shapiro,
  ``Effective action in quantum gravity,''
  Bristol, UK: IOP (1992).


%\cite{Schroedinger:1932}
\bibitem{Schroedinger:1932}
  E.~Schr\"odinger, 
  % ``Diracsches Elektron im Schwerefeld I%
  Sitz.ber.~Preuss.~Akad.~Wiss.~(Berlin), Phys.-math.~Kl., 105 (1932).


%\cite{Bargmann:1932}
\bibitem{Bargmann:1932}
  V.~Bargmann, 
  % ``Bemerkungen zur allgemein-relativistischen Fassung der Quantentheorie''
  Sitz.ber.~Preuss.~Akad.~Wiss.~(Berlin), Phys.-math.~Kl., 346 (1932).


%\cite{Finster:1997gn}
\bibitem{Finster:1997gn} 
  F.~Finster,
  %``Local U(2,2) symmetry in relativistic quantum mechanics,''
  J.\ Math.\ Phys.\  {\bf 39}, 6276 (1998)
  [hep-th/9703083].
  %%CITATION = HEP-TH/9703083;%%
  %20 citations counted in INSPIRE as of 17 Jul 2013


%\cite{Weldon:2000fr}
\bibitem{Weldon:2000fr} 
  H.~A.~Weldon,
  %``Fermions without vierbeins in curved space-time,''
  Phys.\ Rev.\ D {\bf 63}, 104010 (2001)
  [gr-qc/0009086].
  %%CITATION = GR-QC/0009086;%%
  %10 citations counted in INSPIRE as of 14 Mar 2013


%\cite{Gies:2013noa}
\bibitem{Gies:2013noa} 
  H.~Gies and S.~Lippoldt,
  %``Fermions in gravity with local spin-base invariance,''
  Phys.\ Rev.\ D {\bf 89}, no. 6, 064040 (2014)
  [arXiv:1310.2509 [hep-th]].
  %%CITATION = ARXIV:1310.2509;%%
  %7 citations counted in INSPIRE as of 05 Feb 2015


%\cite{Kofink:1949}
\bibitem{Kofink:1949}
  W.~Kofink,
  % ``Zur Mathematik der Diracmatrizen: Die Bargmannsche Hermitisierungsmatrix A und die Paulische Transpositionsmatrix B''
  Math.\ Z.\ {\bf 51}, 702 (1949).


%\cite{Brill:1957fx}
\bibitem{Brill:1957fx} 
  D.~R.~Brill and J.~A.~Wheeler,
  %``Interaction of neutrinos and gravitational fields,''
  Rev.\ Mod.\ Phys.\  {\bf 29}, 465 (1957).
  %%CITATION = RMPHA,29,465;%%
  %308 citations counted in INSPIRE as of 23 gen 2015


%\cite{Unruh:1974bw}
\bibitem{Unruh:1974bw} 
  W.~G.~Unruh,
  %``Second quantization in the Kerr metric,''
  Phys.\ Rev.\ D {\bf 10}, 3194 (1974).
  %%CITATION = PHRVA,D10,3194;%%
  %152 citations counted in INSPIRE as of 29 Oct 2014


%\cite{Finster:1998ws}
\bibitem{Finster:1998ws} 
  F.~Finster, J.~Smoller and S.~-T.~Yau,
  %``Particle - like solutions of the Einstein-Dirac equations,''
  Phys.\ Rev.\ D {\bf 59}, 104020 (1999)
  [gr-qc/9801079].
  %%CITATION = GR-QC/9801079;%%
  %27 citations counted in INSPIRE as of 04 Apr 2013

 
%\cite{Casals:2012es}
\bibitem{Casals:2012es} 
  M.~Casals, S.~R.~Dolan, B.~C.~Nolan, A.~C.~Ottewill and E.~Winstanley,
  %``Kermions: quantization of fermions on Kerr space-time,''
  Phys.\  Rev.\ D {\bf 87}, 064027 (2013)
  [arXiv:1207.7089 [gr-qc]].
  %%CITATION = ARXIV:1207.7089;%%


%\cite{Gies:2013dca}
\bibitem{Gies:2013dca}
  H.~Gies and S.~Lippoldt,
  %``Renormalization flow towards gravitational catalysis in the 3d Gross-Neveu model,''
  Phys.\ Rev.\ D {\bf 87}, 104026 (2013)
  [arXiv:1303.4253 [hep-th]].
  %%CITATION = ARXIV:1303.4253;%%
  %1 citations counted in INSPIRE as of 06 Sep 2013


%\cite{Christiansen:2015}
\bibitem{Christiansen:2015}
  N.~Christiansen, K.~Falls, J.~Meibohm, J.~M.~Pawlowski, 
  M.~Reichert, in preparation.


%\cite{Watanabe:2004nt}
\bibitem{Watanabe:2004nt} 
  T.~Watanabe and M.~J.~Hayashi,
  %``General relativity with torsion,''
  gr-qc/0409029.
  %%CITATION = GR-QC/0409029;%%
  %12 citations counted in INSPIRE as of 12 Apr 2013


%\cite{Pauli:1936}
\bibitem{Pauli:1936}
  W.~Pauli,
  % ``Contributions mathématiques à la théorie des matrices de Dirac.''
  Ann.\ Inst.\ Henri Poincar\'{e} {\bf 6}, 109 (1936).


%\cite{Cornwell:1989bx}
\bibitem{Cornwell:1989bx} 
  J.~F.~Cornwell,
  ``Group Theory In Physics. Vol. 3: Supersymmetries And Infinite Dimensional Algebras,''
  London, UK: Academic (1989) 628 p. (Techniques of physics, 10).
  %1 citations counted in INSPIRE as of 07 Aug 2014


%\cite{Gies:2015cka}
\bibitem{Gies:2015cka} 
  H.~Gies and S.~Lippoldt,
  %``Global surpluses of spin-base invariant fermions,''
  arXiv:1502.00918 [hep-th].
  %%CITATION = ARXIV:1502.00918;%%


%\cite{Ogievetsky:1965ii}
\bibitem{Ogievetsky:1965ii} 
  V.~I.~Ogievetsky and I.~V.~Polubarinov,
  %``Spinors in gravitation theory,''
  Sov.\ Phys.\ JETP {\bf 21}, 1093 (1965)
  [Zh.\ Eksp.\ Teor.\ Fiz.\  {\bf 48}, 1625 (1965)].
  %%CITATION = SPHJA,21,1093;%%
  %42 citations counted in INSPIRE as of 23 Jan 2015


%\cite{Pitts:2011jv}
\bibitem{Pitts:2011jv} 
  J.~B.~Pitts,
  %``The Nontriviality of Trivial General Covariance: How Electrons Restrict 'Time' Coordinates, Spinors (Almost) Fit into Tensor Calculus, and 7/16 of a Tetrad Is Surplus Structure,''
  Stud.\ Hist.\ Philos.\ Mod.\ Phys.\  {\bf 43}, 1 (2012)
  [arXiv:1111.4586].
  %%CITATION = ARXIV:1111.4586;%%
  %2 citations counted in INSPIRE as of 23 Jan 2015


%\cite{Redlich:1983kn}
\bibitem{Redlich:1983kn} 
  A.~N.~Redlich,
  %``Gauge Noninvariance and Parity Violation of Three-Dimensional Fermions,''
  Phys.\ Rev.\ Lett.\  {\bf 52}, 18 (1984).
  %%CITATION = PRLTA,52,18;%%
  %452 citations counted in INSPIRE as of 10 Feb 2015


%\cite{Dunne:1998qy}
\bibitem{Dunne:1998qy} 
  G.~V.~Dunne,
  %``Aspects of Chern-Simons theory,''
  hep-th/9902115.
  %%CITATION = HEP-TH/9902115;%%
  %174 citations counted in INSPIRE as of 10 Feb 2015


%\cite{vanNieuwenhuizen:1981uf}
\bibitem{vanNieuwenhuizen:1981uf} 
  P.~van Nieuwenhuizen,
  %``Classical Gauge Fixing in Quantum Field Theory,''
  Phys.\ Rev.\ D {\bf 24}, 3315 (1981).
  %%CITATION = PHRVA,D24,3315;%%
  %12 citations counted in INSPIRE as of 25 Sep 2013


%\cite{Woodard:1984sj}
\bibitem{Woodard:1984sj} 
  R.~P.~Woodard,
  %``The Vierbein Is Irrelevant In Perturbation Theory,''
  Phys.\ Lett.\ B {\bf 148}, 440 (1984).
  %%CITATION = PHLTA,B148,440;%%
  %22 citations counted in INSPIRE as of 28 May 2013


%\cite{Pawlowski:2003sk}
\bibitem{Pawlowski:2003sk} %geometric FRG
  J.~M.~Pawlowski,
  %``Geometrical effective action and Wilsonian flows,''
  hep-th/0310018; %.
  %%CITATION = HEP-TH/0310018;%%
  %25 citations counted in INSPIRE as of 05 Feb 2015
%
%\cite{Donkin:2012ud}
%\bibitem{Donkin:2012ud} 
  I.~Donkin and J.~M.~Pawlowski,
  %``The phase diagram of quantum gravity from diffeomorphism-invariant RG-flows,''
  arXiv:1203.4207 [hep-th]; %.
  %%CITATION = ARXIV:1203.4207;%%
  %33 citations counted in INSPIRE as of 18 Feb 2015
%
%\cite{Demmel:2014hla}
%\bibitem{Demmel:2014hla} 
  M.~Demmel, F.~Saueressig and O.~Zanusso,
  %``RG flows of Quantum Einstein Gravity in the linear-geometric approximation,''
  arXiv:1412.7207 [hep-th].
  %%CITATION = ARXIV:1412.7207;%%
  %3 citations counted in INSPIRE as of 05 Feb 2015


%\cite{Manrique:2009uh}
\bibitem{Manrique:2009uh}  %bi gravity
  E.~Manrique and M.~Reuter,
  %``Bimetric Truncations for Quantum Einstein Gravity and Asymptotic Safety,''
  Annals Phys.\  {\bf 325}, 785 (2010)
  [arXiv:0907.2617 [gr-qc]].
  %%CITATION = ARXIV:0907.2617;%%
  %53 citations counted in INSPIRE as of 19 Feb 2015
%
%\cite{Codello:2013fpa}
\bibitem{Codello:2013fpa}
  A.~Codello, G.~D'Odorico and C.~Pagani,
  %``Consistent closure of renormalization group flow equations in quantum gravity,''
  Phys.\ Rev.\ D {\bf 89}, no. 8, 081701 (2014)
  [arXiv:1304.4777 [gr-qc]].
  %%CITATION = ARXIV:1304.4777;%%
  %15 citations counted in INSPIRE as of 18 Feb 2015
%
%\cite{Christiansen:2014raa}
\bibitem{Christiansen:2014raa} 
  N.~Christiansen, B.~Knorr, J.~M.~Pawlowski and A.~Rodigast,
  %``Global Flows in Quantum Gravity,''
  arXiv:1403.1232 [hep-th].
  %%CITATION = ARXIV:1403.1232;%%
  %13 citations counted in INSPIRE as of 19 Feb 2015
%
%\cite{Becker:2014qya}
\bibitem{Becker:2014qya}
  D.~Becker and M.~Reuter,
  %``En route to Background Independence: Broken split-symmetry, and how to restore it with bi-metric average actions,''
  Annals Phys.\  {\bf 350}, 225 (2014)
  [arXiv:1404.4537 [hep-th]]; %.
  %%CITATION = ARXIV:1404.4537;%%
  %11 citations counted in INSPIRE as of 05 Feb 2015
%
%\cite{Becker:2014jua}
%\bibitem{Becker:2014jua} 
  %D.~Becker and M.~Reuter,
  %``Propagating gravitons vs. 'dark matter` in asymptotically safe quantum gravity,''
  JHEP {\bf 1412}, 025 (2014)
  [arXiv:1407.5848 [hep-th]].
  %%CITATION = ARXIV:1407.5848;%%
  %3 citations counted in INSPIRE as of 05 Feb 2015


%\cite{Deffayet:2012zc}
\bibitem{Deffayet:2012zc} 
  C.~Deffayet, J.~Mourad and G.~Zahariade,
  %``A note on 'symmetric' vielbeins in bimetric, massive, perturbative and non perturbative gravities,''
  JHEP {\bf 1303}, 086 (2013)
  [arXiv:1208.4493 [gr-qc]].
  %%CITATION = ARXIV:1208.4493;%%
  %50 citations counted in INSPIRE as of 12 Feb 2015


%\cite{Kawai:1993fq}
\bibitem{Kawai:1993fq}  %exponential parametrization
  H.~Kawai, Y.~Kitazawa and M.~Ninomiya,
  %``Quantum gravity in (2+epsilon)-dimensions,''
  Prog.\ Theor.\ Phys.\ Suppl.\  {\bf 114}, 149 (1993).
  %%CITATION = PTPSA,114,149;%%
  %5 citations counted in INSPIRE as of 18 Feb 2015
%
%\cite{Eichhorn:2013xr}
\bibitem{Eichhorn:2013xr}
  A.~Eichhorn,
  %``On unimodular quantum gravity,''
  Class.\ Quant.\ Grav.\  {\bf 30}, 115016 (2013)
  [arXiv:1301.0879 [gr-qc]]; %.
  %%CITATION = ARXIV:1301.0879;%%
  %7 citations counted in INSPIRE as of 18 Feb 2015
%
%\cite{Eichhorn:2015bna}
%\bibitem{Eichhorn:2015bna} 
%  A.~Eichhorn,
  %``The Renormalization Group flow of unimodular f(R) gravity,''
  arXiv:1501.05848 [gr-qc]; %.
  %%CITATION = ARXIV:1501.05848;%%
  %1 citations counted in INSPIRE as of 05 Feb 2015
%
%\cite{Nink:2014yya}
\bibitem{Nink:2014yya} 
  A.~Nink,
  %``Field Parametrization Dependence in Asymptotically Safe Quantum Gravity,''
  arXiv:1410.7816 [hep-th]; %.
  %%CITATION = ARXIV:1410.7816;%%
  %4 citations counted in INSPIRE as of 05 Feb 2015
%
%\cite{Percacci:2015wwa}
\bibitem{Percacci:2015wwa} 
  R.~Percacci and G.~P.~Vacca,
  %``Search of scaling solutions in scalar-tensor gravity,''
  arXiv:1501.00888 [hep-th].
  %%CITATION = ARXIV:1501.00888;%%
  %2 citations counted in INSPIRE as of 18 Feb 2015


\end{thebibliography}
\end{document}